\documentclass[]{article}
\usepackage{arxiv}
\usepackage{amsmath}
\usepackage{graphicx}
\usepackage[procnames]{listings}
\usepackage{color}
\usepackage{siunitx}
\usepackage{placeins}
\usepackage{paralist}
\usepackage{balance}
\usepackage{graphics}
\usepackage{txfonts}
\usepackage{times}
\usepackage{hyperref}
\usepackage{url}
\usepackage{color}
\usepackage{textcomp}
\usepackage{booktabs}
\usepackage{ccicons}
\usepackage{todonotes}
\usepackage{float}
\usepackage[normalem]{ulem}
\usepackage{natbib}
\usepackage{threeparttable}
\usepackage[title]{appendix}

\newcommand{\ve}[1]{{\boldsymbol{#1}}}             
\newcommand{\ma}[1]{{\mathbf{#1}}}
\newcommand{\AAA}{\ma A}                           
\newcommand{\BBB}{\ma B}                           
\newcommand{\CCC}{\ma C}                           
\newcommand{\LLL}{\ma L}                           
\newcommand{\KKK}{\ma K}                           
\newcommand{\UUU}{\ma U}                           
\newcommand{\xx}{\ve x}                            
\newcommand{\xxx}{\ma x}                           
\newcommand{\dxxx}{\dot{\xxx}}                     
\newcommand{\yyy}{\ma y}
\newcommand{\uu}{\ve u}                            
\newcommand{\uuu}{\ma u}                           
\newcommand{\ddd}{\ma d}                           

\newcommand{\kkk}{\ma k}                           
\newcommand{\xxss}{\xx_{ss}}                       
\newcommand{\yyss}{\ve y_{ss}}                     
\newcommand{\uuss}{\uu_{ss}}                       
\newcommand{\Qc}{\ma Q_c}                          
\newcommand{\Qo}{\ma Q_o}                          
\newcommand{\Rc}{\ma R_c}                          
\newcommand{\Qt}{\ma Q_{t}}                        
\newcommand{\EEE}{\ma E}                           

\usepackage[printonlyused, nolist]{acronym}
\usepackage{tabularx}
\usepackage{multirow}
\usepackage{array}
\usepackage{caption}
\usepackage{subfigure}
\usepackage{pinlabel}
\usepackage{comment}

\makeatletter
\def\url@leostyle{%
  \@ifundefined{selectfont}{\def\UrlFont{\sf}}{\def\UrlFont{\small\bf\ttfamily}}}
\makeatother
\graphicspath{{simon_gp/}{matlab_plots/}{Figures/}}

\def\pprw{8.5in}
\def\pprh{11in}

\definecolor{linkColor}{RGB}{6,125,233}
\hypersetup{%
  pdftitle={Intermittent control as a model of mouse movements},
  pdfauthor={LaTeX},
  pdfkeywords={Intermittent control, pointing, HCI},
  bookmarksnumbered,
  pdfstartview={FitH},
  colorlinks,
  citecolor=black,
  filecolor=black,
  linkcolor=black,
  urlcolor=linkColor,
  breaklinks=true,
}

\title{intermittent control as a model of mouse movements}

\author{
  J. Alberto~\'Alvarez Mart\'in\\
  James Watt School of Engineering\\
  University of Glasgow, Scotland.\\
  alberto.alvarez-martin@glasgow.ac.uk\\
   \And
  Henrik~Gollee\\
  James Watt School of Engineering\\
  University of Glasgow, Scotland.\\
  hernik.gollee@glasgow.ac.uk\\
   \And
  J\"org~M\"uller\\
  Institute for Computer Science\\
  University of Bayreuth, Germany.\\
  joerg.mueller@uni-bayreuth.de\\
   \And
  Roderick~Murray-Smith\\
  School of Computing Science\\
  University of Glasgow, Scotland.\\
  roderick.murray-smith@glasgow.ac.uk
}

\begin{document}
\maketitle

\definecolor{keywords}{RGB}{255,0,90}
\definecolor{comments}{RGB}{0,0,113}
\definecolor{red}{RGB}{160,0,0}
\definecolor{green}{RGB}{0,150,0}

\lstset{language=Python,
        basicstyle=\ttfamily\small,
        keywordstyle=\color{keywords},
        commentstyle=\color{comments},
        stringstyle=\color{red},
        showstringspaces=false,
        identifierstyle=\color{green},
        procnamekeys={def,class}}

\begin{abstract}
  We present Intermittent Control (IC) models as a candidate framework for
  modelling human input movements in Human--Computer Interaction (HCI). IC
  differs from continuous control in that users are not assumed to use
  feedback to adjust their movements continuously, but only when the
  difference between the observed pointer position and predicted pointer
  positions become large.  We use a parameter optimisation approach to
  identify the parameters of an intermittent controller from experimental
  data, where users performed one-dimensional mouse movements in a
  reciprocal pointing task. Compared to previous published work with
  continuous control models, based on the Kullback-Leibler divergence from
  the experimental observations, IC is better able to generatively
  reproduce the distinctive dynamical features and variability of the
  pointing task across participants and over repeated tasks. IC is
  compatible with current physiological and psychological theory and
  provides insight into the source of variability in HCI tasks.
\end{abstract}

\keywords{Human-computer interaction, control theory, modelling, pointing, intermittent control}

\section{Introduction}
One of the most important areas of study of Human-Computer Interaction
(HCI) is how humans can interact with computers to input information into
computers or control their state. Most applied human--computer input is via
activity of the user's neuromuscular system causing movement of their body,
which is sensed by input devices and produces a change in the state of the
computer. The most widely-studied example of this in HCI is mouse movement,
where a user's hand moves a mouse to change the cursor position.  The goal
of this article is to provide a physiologically plausible model of such
user movements and the associated movement variability. While our long-term
objective is to understand movement during interaction with computers in
general, in this paper we focus on a simple core task: aimed movements of a
mouse cursor towards a static target of a certain width at a certain
distance.

Movement in interaction with computers is inherently dynamic and clearly
happens in a feedback loop.  Users observe the current state of the
computer (e.g., cursor position) and adjust their movements to change this
state into the state they desire -- {\it ``movements only make sense when
  they are precisely located in time and space that is, when they are part
  of an action, seeking to accomplish a goal. In this sense, simple
  goal-directed movements, such as pointing and grasping, can be considered
  to be the building blocks of more complex actions''} \citep{BooReiFer04}.
Control theory is the mathematical framework for systems with feedback
which are achieving specific objectives, and its relevance for HCI is
reviewed in \citep{Mur18} and \citep{MulOulMur17}.

In this paper we explore the benefits of interpreting interaction with
computers as intermittent control.  In intermittent control, open-loop
control trajectories are generated based on an internal estimate of the
state of interactions using a predictive model of how this state will
evolve over time. The open-loop trajectories are intermittently updated
with feedback information from continuous observations of the systems.
Only if the observed state deviates from the predicted state, do they
update their prediction and control accordingly. \citet{Loram2014} explain
why the intermittent control perspective is more physiologically plausible
as an explanation of human motor control than our previous continuous
control perspectives of interaction with computers (e.g.,
\citep{MulOulMur17}).  It has been shown that humans are generally not able
to adjust their movement continuously \citep{Navas1968}.  The switching and
intermittency inherent in Intermittent Control provides a powerful
explanation of the source of variability in human movement, and
potentially, it can also explain the emergence of phenomena that are
difficult to explain from a continuous control perspective, such as
submovements.

\subsection{Beyond Fitts' law}
Since the advent of graphical user interfaces, aimed movements towards a
spatially defined target have become a primary means of input to computers.
It is well known that all movements involve variability in their
performance.  For aimed movements, this has been described in the
``speed-accuracy tradeoff'', modelled through Fitts' law \citep{Fit54}.
Fitts' law allows the prediction of movement time $MT$ as a function of
distance $D$ to and width $W$ of the target, as $MT=a+b\log(D/W+1)$ in the
Shannon formulation \citep{mackenzie92}.  The application of Fitts' law to
understand interaction with computers is often viewed as one of the
greatest successes of the field of HCI \citep{guiard2004preface}.  Fitts'
law allows for the model-based evaluation of user interfaces, reducing the
number of necessary user tests, and serves in the automatic optimization of
user interfaces such as GUI layouts or keyboards.  Perhaps most
importantly, it has sharpened the intuition of generations of user
interface designers about the role of the users' motor control processes in
interaction with computers.

One of the strengths, but also limitations, of Fitts' law is that it
reduces the complexity of human movement to a single number, the movement
time, but it makes no testable predictions about the process of the
movement. The Fitts' law perspective offers no explanation of the causal
relationship that links conditions and outcomes in pointing. Neither the
biomechanical, perceptual and cognitive properties of the human nor the
properties of the computer interface such as input sensors, delays, jitter
or style of feedback can be understood from this perspective. It does not
allow any understanding of {\em why} a certain user interface, interaction
technique or input device is better than another. In particular, the
trajectory of the pointer, velocities and accelerations can not be
explained by Fitts' Law, so a richer understanding of the broader process
of movement is lost, and as \citep{BooReiFer04} observe, {\it ``while the
  duration of movement constitutes a particularly pertinent global measure
  of behavior, useful in many different contexts, it does not allow a full
  appreciation of the processes underlying behavioral organization. In the
  domain of perceptuomotor control, it is widely accepted that a more
  fine-grained window into these processes is available through the
  kinematics of movement''}.
Perhaps most critically, Fitts' law is limited to aimed movements. This may
make it more difficult for the field of HCI to think beyond aimed movements
and point-and-click as the foundation of interface design. In order to
develop a deeper understanding in HCI of how users create input to
computers by moving their bodies, we propose that we need to take a control
theoretic perspective, and we are exploring models which are appropriate
for the representation of purposeful human movement.

\subsection{Paper structure}

In this paper, we present a model of aimed mouse movement in the
interaction with computers, based on intermittent control. We build on an
existing dataset to fit the parameters of our event-driven intermittent
controller, allowing a comparison with the continuous implementations
proposed in \citep{MulOulMur17}.  The objectives of this paper are:
\begin{enumerate}
\item To identify the parameters of an intermittent controller from the
  experimental mouse movement data by using an optimisation approach.
\item Examine the ability of IC to generatively reproduce and explain
  distinctive dynamical features of the pointing task such as the velocity
  profile and the variability observed across participants.
\item Introduce IC as a plausible framework to understand and model user
  movement in the interaction with computers, as a novel analytic and
  practical tool for HCI research and practice.
\end{enumerate}

\noindent The content of the paper is organised as follows: After reviewing
related work in Section ~\ref{sec:relatedwork}, a general introduction to
the IC framework is given in Section~\ref{sec:overview}, focusing on its
role in pointing and human variability. We highlight that the model is
based on human physiological insight, and includes predictive properties
which are key to understanding and modelling human behaviour in interaction
contexts. Section~\ref{sec:formalIC} then provides a more mathematically
formal description of the theoretical details of
IC. Section~\ref{sec:experiment}-\ref{sec:optim} provide descriptions of
the experimental dataset and the models and optimisation process used for
the analysis.  Section~\ref{sec:modelling}-\ref{sec:densities} provide the
modelling results, demonstrating that Intermittent control models provide
improved modelling of the variability and dynamics of mouse movements
compared to previous publications.  In Sections~\ref{sec:discuss} and
\ref{sec:conclusions} we discuss the conceptual and practical advantages of
the model, and give an outlook on further developments.

\section{Related Work}
\label{sec:relatedwork}
\subsection{Fitts' Law and Information Theory}

Since the original work \citep{Fit54}, Fitts' relationship has been used in
a wide range of HCI-relevant publications. For more background,
\citet{GuiBea04} introduce a special issue celebrating 50 years of Fitt's
law and \citet{soukoreff2004towards} review 27 years of Fitts' law in HCI
and makes recommendations to HCI researchers wishing to construct Fitts'
law models for movement time prediction, or for the comparison of
conditions in an experiment. \citet{wobbrock2008error} show that Fitts' law
implies a predictive error rate model, and that the effect on error rate of
target size $W$ is greater than that of target distance $D$.

Dynamic aspects of Fitts' law task are explored in
\citep{billon2000dynamics,BooReiFer04,Gui93}, and \citep{Jag77,JagFla03}
link Fitts' law models to first and second order control models. While most
Fitts' law experiments are artificial lab studies, \citet{chapuis2007fitts}
collected kinematic data from 24 users over several months as part of their
normal interactions with a computer.  Some papers have explored dynamic
relationships in implementations of motor and display spaces, while not
taking an explicitly control-theoretic perspective. While pointing in the
physical world is constrained by physical laws, \citet{Bal04} observed that
pointing in the virtual world does not have to abide by the same
constraints, and compared approaches aiming to `beating' Fitts' law by
artificially reducing the target distance, increasing the target width, or
both, essentially adapting the dynamics of the control
task. \citet{BlaGuiBea04} introduced the concept of {\it Semantic
  pointing}, which manipulates control--display gain, using two independent
sizes for targets in motor space, adapted to its importance for the
manipulation, and in visual space, adapted to the amount of information it
conveys. Their experimental results suggested that the performance of
semantic pointing is given by Fitts' index of difficulty in motor rather
than visual space.

\subsection{Study of Movement from an intermittent perspective}
Control of body movements has been modelled as both discrete and continuous
processes, and there are strong arguments for either perspective:

In favour of the {\bf continuous perspective}, the human hand is a physical
inertial system, which changes its position smoothly over time. Although
difficult aimed movements can be segmented into submovements, identifying
the beginning and end of submovements is notoriously difficult. In some
movements, there do not seem to be submovements at all. The continuous
perspective can also explain other movements which are not aimed movements,
such as following a moving target, steering a cursor along some path,
controlling a character or vehicle in a video game, etc. Understanding
movement as a series of discrete submovements does not appear to concisely
capture these phenomena.

In favour of the {\bf discrete perspective}, data shows that humans are not
capable of continuously controlling their movements in
the 
same way as an engineered continuous control system. Craik first reported
this in \citep{Cra47,Cra48}. In particular, the {\em psychological
  refractory period} \citep{Telford1931} dictates that humans can not react
to further changes in the environment for a certain time after they reacted
to a change: The reaction onset to a second stimulus which follows shortly
after the first is delayed as the human is refractory during this
time. Crucially, a control paradigm based on continuous feedback is unable
to explain refractoriness in human motor control.

Interestingly, the perspectives of discrete movements vs. continuous
control may be less contradictory than it first seems.  Craik, proposed
that a series of discrete movements could appear to be continuous in
nature. From this perspective, the human hand moves continuously, but the
way the hand moves is only changed at certain points in time, or {\em
  intermittently}.  Craik's manual tracking experiments led to the
conclusion that the human operator uses an intermittent approach, dictated
by refractoriness, when tracking discrete reference steps, and that manual
control can be seen as the execution of discrete actions that are computed
based on the available sensory information and applied as individual
open-loop trajectories that are not modified by feedback until
completed. Since then, intermittency from the perspective of human motor
control has been studied extensively by the physiology community
\citep{Navas1968,Neilson1999,Loram2002,Oytam2005,Loram2006A}, and as
consequence, computational frameworks have been developed to model the
human operator as an intermittent controller using concepts of modern
control theory. On the engineering side, the intermittent controller has
its origins in the implementation of model-based predictive control (MPC)
in the presence of hard constraints \citep{Ronco1999} and the
observer-predictor-feedback architecture from \citet{Kleinman1969} and
\citep{kleinman1970optimal}.  It is important to note that, as explained
above, intermittency can masquerade as continuous control, in particular
when controlling an inertial system, and when the controller has a good
internal model of the system \citep{Gawthrop2011}.

Schmidt's Law \citep{SchZelSte78} addresses variability and dynamics by
manipulating amplitude and movement time and measuring the effective target
width $W_e$, leading to the relation $W_e = k\frac{W}{MT}.$ The insight
behind Schmidt's Law is that the human controls movement via discrete force
impulses and the overall variability is from the variation of the magnitude
and duration of the applied force.  \citet{Meyer1988} and \citet{MeyKeiKor90}
further developed this to the {\it optimised dual-submovement model} which
is consistent with both Schmidt's and Fitts' laws, with the variability of
submovements being proportional to the average velocity and that
variability leading to the requirement of multiple sub-movements optimised
to minimise the total movement time. The assumption being that subjects
move to the target region as quickly as possible, while maintaining a high
proportion of target hits. To achieve that, they need to cope with the
variability induced by motor noise.

The {\em Iterative Corrective Submovements} model
proposed by \citet{CroGoo83} is a model where an aimed movement is
understood as a series of individual submovements towards the target, each
with a constant error and constant duration. Crossman and Goodeve show how
Fitts' law can be derived from this model.

In HCI, movement is also often understood as a series of events, such as
submovements, and the Crossmann and Goodeve model is widely used. This
includes well-known HCI models, such as the work by \citet{Card1986}, who
introduced their `GOMS' framework where a Model Human Processor represented
movement as a series of discrete steps.  However, \citet{CroGoo83} also
presented an alternative continuous explanation of aimed movements, that
predicts the position of the hand at any point in time, during the
movement.\footnote{Interestingly, this first-order-lag model (1OL) of
  movement seems to be less well-known in HCI than the Iterative Corrective
  Submovements model presented in the same paper.  The 1OL assumes that the
  user is continuously observing the cursor position, and adjusts the
  velocity of the mouse as a constant fraction of the distance of the
  cursor to 
  the target. Crossman and Goodeve also show how Fitts' law can be derived
  from this different model.}

\begin{figure*}[!htbp]
\centering
\includegraphics{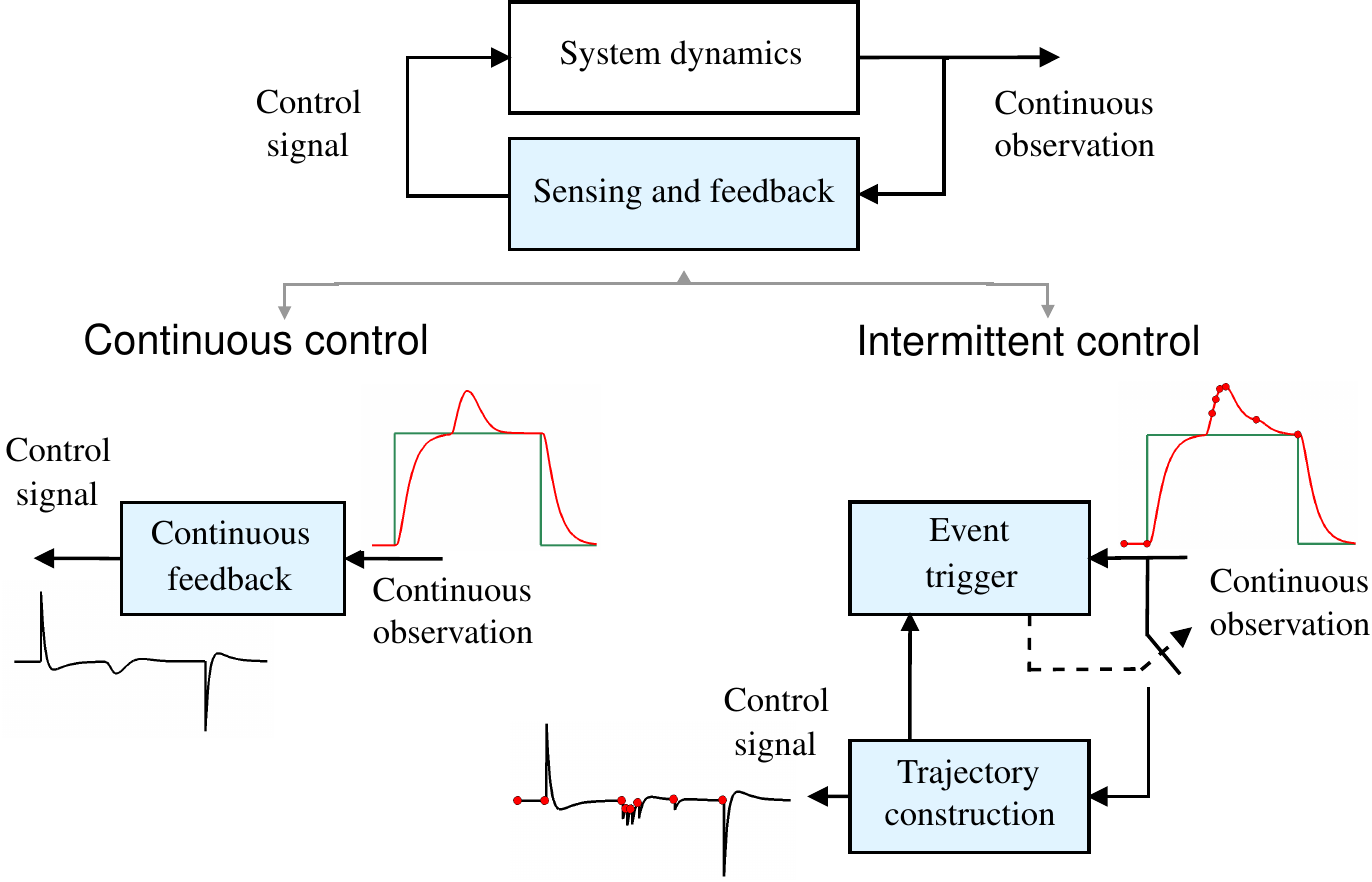}
\caption[Continuous and Intermittent control]{ Feedback control: continuous
  and intermittent control.  The top diagram shows a generic control loop:
  Feedback can be used to control the behaviour of a dynamic system, which
  involves ``sensing'' relevant variables to form a representation of the
  system's state, before generating a control signal based on a feedback
  control law. In continuous control (bottom left block diagram), the
  control signal (time-series in black) is generated continuously, based on
  observations of the system (time-series in red). In intermittent control
  (bottom right), feedback information from continuous observations is only
  used intermittently (determined by an \emph{event trigger}) to reset an
  open-loop control signal trajectory. The instances when feedback is used
  in intermittent control are indicated by circles on the red and black
  time-series, representing the system output and the control signal,
  respectively. The control signal trajectories between two consecutive red
  circles (or \emph{events}) are generated open-loop, i.e. without the use
  of feedback.}
\label{fig:ic-cc-simple}
\end{figure*}

\subsection{Intermittent Control}

Feedback control involves applying a control signal to a dynamic system,
based on observations which are combined into an estimate of the system's
state (``sensing''), and a feedback law,
cf. Fig.~\ref{fig:ic-cc-simple}. In the traditional continuous control
paradigm (bottom left in Fig.~\ref{fig:ic-cc-simple}), the control signal
is continuously adjusted using continuously observed information. In
contrast, intermittent control (IC), as introduced in
\citep{Gawthrop2009,Gawthrop2011,Gawthrop2015}, and shown in the bottom
right of Fig.~\ref{fig:ic-cc-simple}, employs an event-trigger to switch
between closed-loop and open-loop configurations, using feedback at only
specific moments in time to generate an control trajectory that is then
applied during the open-loop interval, where there is no feedback.

There are strong theoretical and practical modelling motivations for the
use of IC in human motor control. It has recently been shown that
refractoriness can also be observed in more complex human control tasks
\citep{KamGawGolLor13}, and that a control strategy which applies feedback
only intermittently can successfully model this phenomenon
\citep{LorKamGolGaw12}. The physiological basis for muscle activity in
human motor control is also a motivation for the use of intermittent
models, and provides insight into the source of variability in human
control of input devices \citep{Gollee2017}.  Intermittency can serve as a
mechanism to modulate the relationship between exploration and exploitation
or, in other words, the stability and plasticity trade-off that is critical
for adaptation and learning in human behaviour \citep{Loram2015}. When
humans learn to control something new, such as a mouse, they need to build
an internal model of the dynamics of what they are controlling. In
particular, they need to identify parameters of such an internal model from
interaction with the device, ie. in a closed-loop setting. In any situation
where we are trying to identify model parameters from noisy data acquired
in a closed-loop setting, there are problems of bias in parameters due to
the correlations between input and output induced by the feedback
\citep{Forssell1999a}. The intermittent breaking of the feedback loop in IC
allows model parameter fitting without the correlated noise from feedback
affecting the estimates.

Previous applications of IC in human motor control were mostly interested
in how humans can move their bodies from a physiological perspective, so
aimed movements were mostly investigated with the bare hand, or with
absolute control devices such as joysticks. This paper is the first
application of the formal IC framework to mouse movements in an HCI
context. The mouse provides only a relative mapping between mouse movements
and pointer movement, considerably complicating the control mapping. For
these reasons, pointing with the mouse has been investigated very rarely in
the field of human motor control. In contrast, mouse movements are at the
heart of the field of Human--Computer Interaction, looking at situations
where humans control the state of a computer system, via potentially
complicated mappings.

\subsection{Control of Computers}
Traditionally, HCI is often presented as {\it communication of information}
between the user and computer, and has used information theory to represent
the bandwidth of communication channels into and out of the computer via an
interface, but this does not provide an obvious way to measure the
communication, or whether the communication makes a difference.
Information theory {\it alone} is not sufficient as a framework for
modelling HCI.\footnote{Julien Gori has been updating the information
  theoretic analysis of Fitts' law tasks in \citep{gori2018modeling} and in
  \citep{gori2018speed} highlighted the potential benefits of including
  feedback in the analysis.}  Humans often want to control some aspect of
the world (e.g. the temperature in a room, or the volume of a music player)
via a computer, or change the state of a computer in its own right.  In
either case we have a dynamic feedback loop which passes through a
computer, where in order to communicate the simplest symbol of intent, we
typically require to move our bodies in some way that can be sensed by the
computer, often based on feedback while we are doing it. Our bodies move
smoothly through space and time, so any communication system is going to be
based on a foundation of continuous control. However, inferring the user's
intent is inherently complicated by the properties of the control loops
used to generate the information -- intention in the brain becomes
intertwined with the physiology of the human body and the physical dynamics
and transducing properties of the computer's input device, as well as the
brain's own predictions of these aspects. We therefore have sensitivity to
both variability of human motor control execution, and need to take user
predictions about their own behaviour into account, highlighting the role
of the model-predictive intermittent control methods explored in this
paper.\footnote{Note the similarity to issues in the {\it Joint cognitive
    systems} community, where \citep{Hol99,HolWoo05} argue that we need to
  focus on how the {\it joint} human--computer system performs, not on the
  communication between the parts.}  \citet{Mur18} gives an overview of the
role of control theory in the study of HCI and \citep{JagFla03} provide a
readable introduction to control and dynamics concepts for non-engineers.

Control theory provides an engineering framework which is well-suited for
analysis of closed-loop interactive systems.  This can include properties
such as feedback delays, sensor noise (see e.g. \citep{TreMur13}), or
interaction effects like `sticky mouse' dynamics, isometric joystick
dynamics \citep{BarSelRut95}, magnification effects, inertia, fisheye
lenses, speed-dependent zooming, all of which can be readily represented by
dynamic models.  The use of state space control methods was explored in
document zooming context in \citep{EslMur04,EslMur06,EslMur08,KraBroRoh10}
and \citet{QuiMalCoc13} reviewed the challenge of optimising touch
scrolling transfer functions and used a robot arm to identify the dynamics
of commercial products. Examples of the use of dynamic models in
interactive dynamic scrolling systems are now widespread in commercial
systems, and examples in the academic literature, include
\citep{WilMurHug07,ChoMurKim07}. Highlighting the relevance of knowledge of
variability in user movements, \citet{quinn2016} used control models to
understand how input trajectories associated with words entered into
gesture keyboards are likely to vary.

The control perspective can also inspire unusual approaches to interaction,
such as {\it motion pointing} interfaces which infer the user's intent
based on detection of control behaviour, as originally developed by
\citet{WilMur04b} and built on by \citet{FekElmGui09} and, via eye
tracking, in a number of applications described in
\citep{VelCarEst17}. Improved estimation of variability in closed-loop
control behaviour would increase the performance of many of these
approaches significantly, as they rely on measurement of the divergence of
distributions. Motion pointing requires a good model of the user's ability
to perform certain actions. Recent interest in the relevance of detailed
inverse biomechanical simulation in HCI includes
\citep{bachynskyi2015informing,bachynskyi2016biomechanical}, while
\citet{fischer2020reinforcement} explores forward biomechanical simulation
to simulate pointing movements.

In \citep{MulOulMur17} we presented and compared several manual control
models of mouse pointing identified from the same mouse movement data used
in this paper. These models are generative, estimating not only movement
time, but also pointer position, velocity, and acceleration on a
moment-to-moment basis. The continuous control models tested captured some
of the important dynamic characteristics, but did not explain the
variability shown in the data, which led to a poorer fit for low index of
difficulty (ID) targets. The models included continuous second order
dynamics models, and Costello's surge model, which has a switching
characteristic.

The switching characteristic separates the movement into an open-loop
initial surge phase and a later continuous current control phase. This
switching behavior has also been modelled by \citet{aranovskiy16cdc} and
further in \citep{aranovskiy20journal}. Based on the optimised
dual-submovement model and the surge model they develop a two-phase model
that takes into account nonlinear pointing transfer functions (PTF) which
map from mouse movement to cursor movement. While the ballistic movement
phase only has access to proprioceptive feedback for control, the following
corrective movement phase is guided visually from the pointer position
which is mediated by the PTF. The switch between ballistic and corrective
movement phase happens at a predefined distance from the target after a
transition period with zero acceleration. Aranovskiy et al. derive the
exponential stability of the model under some mild assumptions about the
PTF, and validate the model against experimental data of reciprocal
pointing in the case of a constant gain function. In contrast to the
intermittent control model presented in this paper, the model of Aranovskiy
explicitly models pointing with a possibly non-constant PTF. On the other
hand, the intermittent control model explicitly models intermittency of the
control process, the observer and predictive capability of the human, the
dynamics of the neuromuscular system, as well as noise and delay in the
nervous system. Being based on physiological principles of motor control,
the intermittent control model has the potential advantage that it is
applicable to a range of movement tasks other than mouse movement.

The {\it Intermittent Click Planning Model} (ICP) of
\citet{park2020intermittent} describes the process by which users plan and
execute optimal click actions, from which the model predicts the pointing
error rates for target tracking tasks. Their ICP model assumes that the
user is an intermittent controller, following on from the intermittent BUMP
model described in \citep{bye2008bump}. The main difference between our
approach and the BUMP model is that in this paper we use an event-trigger,
leading to a variable distribution of open-loop intervals, whereas the BUMP
model are based on a constant intermittent interval. Physiological evidence
suggests that the open-loop interval is not constant, so an event trigger
should be the more reasonable explanation from \citet{GawLorGolLak14}. The
ICP model also assumes that the user is a statistical encoder that makes
optimal use of the externally provided information that allows for
estimation of click timings. More complex threshold functions for IC are an
interesting opportunity for future research.

\section{Overview of the intermittent control Model}
\label{sec:overview}

This section provides an explanation of the main elements in IC with an
emphasis on human motor control and integrates them in the form of a
unified control model. The purpose is to relate the function of fundamental
features of human control in pointing and tracking tasks with classical
control theory concepts, while highlighting the differences between IC and a
continuous control point of view.

In this context, the flow of information is represented by \emph{signals}
(such as the pointer position or motor control signals within the user)
that are used as \emph{inputs} to specific \emph{systems} (such as the
computer or the biomechanics of the user), and depending on the role of the
system, different \emph{outputs} are generated. In human motor control and
HCI there are many operations involved in classic tasks that have a direct
representation in control theory; for instance, for pointing tasks where
the goal is to reach a target using a pointer device of some sort, the
target can be seen as a \emph{reference} signal that should be introduced
to the model \citep{MulOulMur17}. This signal allows the calculation in
real-time of the \emph{error} between the target position and the pointer
position. In most systems, the goal is to reduce this error as much as
possible and a common way to achieve this is by using \emph{feedback},
which means that the output signal delivered by the system we are trying to
control is used as an input to a \emph{controller} that computes and
applies a correcting signal to the same system. For human in-the-loop
systems, this process happens when the system output, such as the display
content, is sensed, processed and integrated via our sensory stream, to
form an internal representation of the task in the brain
\citep{Wolpert1998,Shadmehr2012}, meaning that the central nervous system
(CNS) uses a combination of prior and present sensory information to
estimate internal and external states that are relevant to the task
\citep{Bays2007}, such as pointer position and velocity. This internal
representation provides the capability to predict how the system output
will behave in the future based on its current state, while cancelling the
effects of common delays that affect the transmission of these signals
through our neural pathways \citep{Miall1993,Gawthrop2011} or within the
computer, to then combine all the available information in order to
generate control commands, which cause the muscles to contract in a way to
achieve the desired effect.

\begin{figure*}[!htbp]
\centering
\includegraphics{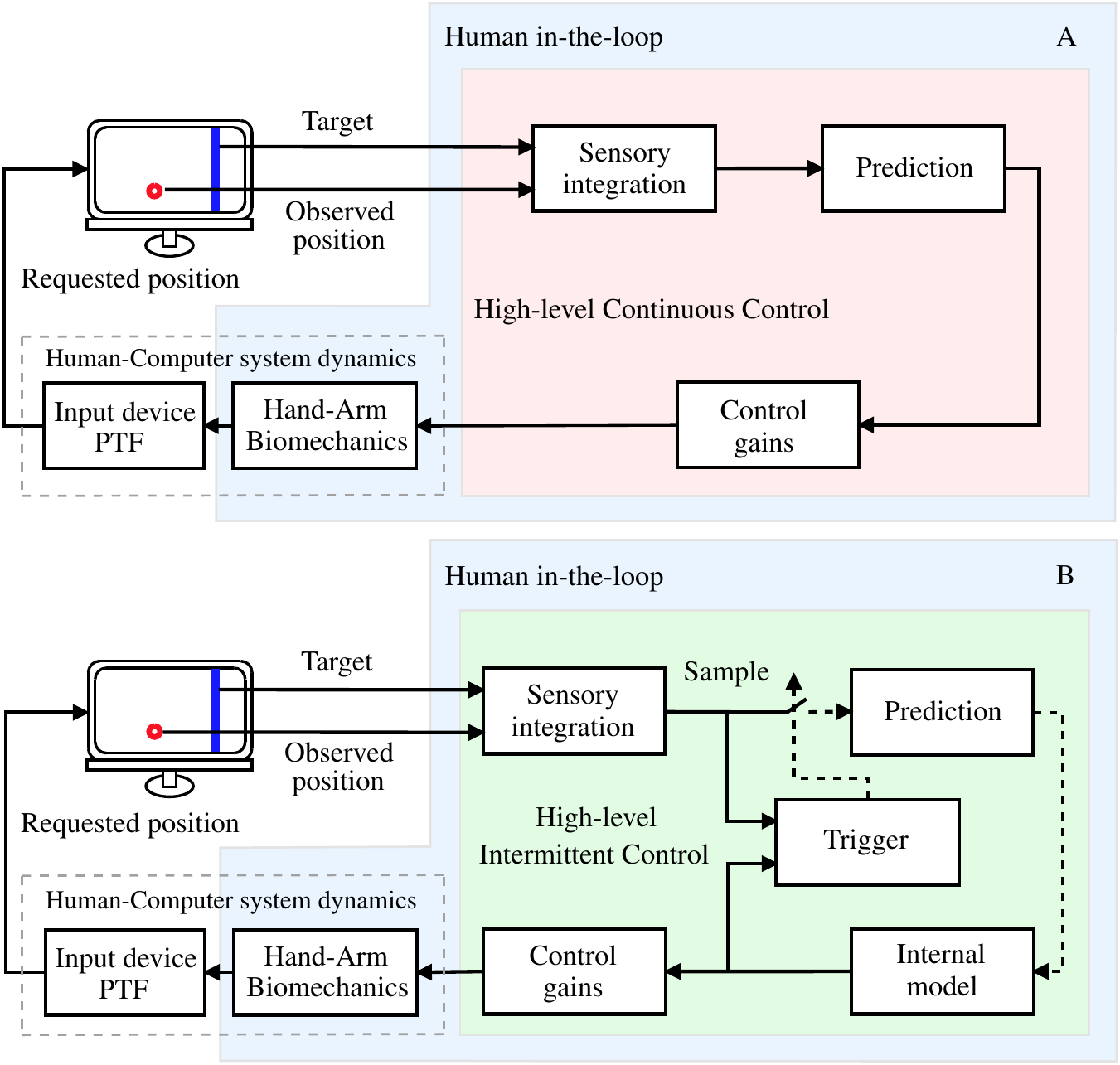}
\caption[Continuous and Intermittent control models]{The continuous control
  (CC, above) and intermittent control (IC, below) frameworks in the
  context of a general HCI task. The black solid lines represent signals
  going in and out the different systems continuously, whereas the dashed
  lines indicate that they are active intermittently. (A) Continuous
  control performs three tasks: Sensory integration, Prediction, and input
  generation modulated by a set of Control gains. (B) Intermittent control
  (green box), includes an Internal reference model, a Trigger, and
  Sampling mechanism. Both models take into account the Human-Computer
  system dynamics (grey dashed-line box) as the combination of two separate
  systems: The Hand-Arm Biomechanics and the Input Device Pointing Transfer
  Function (PTF).  The pointing task is to bring the displayed pointer (red
  circle) to the Target (blue vertical bar). The user is represented in
  blue.}
\label{fig:ic-cc-paradigm}
\end{figure*}

The aforementioned process can be defined by three main components: 1) the
system that describes the evolution in time (time-series) of the controlled
variable or output, known simply as the {\it system}, which includes the
computer hard- and software, 2) a high-level controller in charge of
sensing outputs, predicting future states and generating control inputs
which is implemented in the users' CNS and finally, 3) the neuromuscular
system (NMS) of the user which provides the necessary correction, normally
as a force or torque applied to a pointing mechanism, such as a joystick or
a computer mouse.

A continuous flow of signals between a system and its user through any
given interface is an example of a continuous interaction model where the
system output is continuously being sensed and consequently, control
commands are generated as a result of this continuous stream. Although this
continuous model has been the dominant framework to explain motor control,
there are situations where this closed-loop system is not always
continuous, for instance when the contact between the hand and the input
device to the system is interrupted \citep{Loram2011}, or when the system
output is not available for measurement. These situations yield a system
that is intermittently-continuous. However, a third source of intermittent
behaviour arises as a result of the refractory period that is observed
frequently in reaction tasks \citep{Telford1931}, which implies that even
with a continuous flow of information in the form of feedback, humans
respond to a succession of sufficiently rapid stimuli in a serial way,
starting the response for new stimuli only after finishing the processing
and execution of the previous one, thus evolving for short periods of time
in an open-loop configuration \citep{Cra47,Navas1968}. This serial-ballistic
operation mode is at the core of the intermittent control model that is
presented in this paper for HCI tasks, providing a flexible architecture
that includes continuous interaction as a special case of a more general
framework.

\subsection{The intermittent control framework}

Let us introduce the general intermittent control model by presenting the
continuous control counterpart first using a block diagram, which is a
common graphical representation of the structure and purpose of a control
system. In Fig.~\ref{fig:ic-cc-paradigm}A (above), a continuous interaction
model is shown in the context of target tracking using a pointer device
(e.g., computer mouse or joystick). The arrows represent signals travelling
through the loop that act as inputs and outputs to the different
\emph{systems} in the model (depicted as white boxes with solid black
borders). The user is represented by the light blue coloured box containing
two subsystems, the \emph{Human-Computer system dynamics} box (in grey
dashed lines) which includes the dynamics of the arm and hand in
relationship to the pointer device, and a high-level continuous controller
(CC) shown in a red coloured box comprised by three blocks: Sensory
integration, Prediction, and Control gains. The \emph{Human-Computer system
  dynamics} box contains the group of variables that the user attempts to
control and it receives a correcting input signal that the CC system
generates. The mouse velocity measured by the mouse sensor is transferred
to the pointer velocity via a possibly non-linear pointing transfer
function (PTF) \citep{casiez08cdgain,casiez11bricolage}. The output of this
block (e.g., pointer position on a display) is then sensed via visual
feedback and used to generate a new control action.  The CC system as a
whole is in charge of reducing the error between the target signal (desired
pointer position) and the output (current pointer position) by combining
sensory streams and intended targets to form an internal model of the task
at hand. It is argued that the brain uses some form of model based
\emph{Prediction} to better understand the consequences of our commands
during movement and to diminish the effect of slow transmission pathways
\citep{Miall1993,Bhushan1999,Gawthrop2008,Gawthrop2011}, this concept is
represented by the block that follows the system integration process. The
last stage involves the selection and computation of an appropriate
high-level \emph{Control} signal which is applied continuously.

The IC model, shown in Fig.~\ref{fig:ic-cc-paradigm}B (below), incorporates
the same elements of the continuous control model with the addition of a
\emph{Trigger} mechanism within the user's CNS, that has the fundamental
role of opening or closing the loop right after the output of the sensory
integration process; this effectively \emph{samples} the output and
relevant states of the system. Regulating the flow of information at this
point allows the user to achieve continuous measurement of the output
variables while applying control commands intermittently. This mechanism
can replicate refractoriness by enforcing a minimum amount of time in which
the system would evolve in an open-loop configuration. In addition to this,
the controller could extend these open-loop intervals beyond this minimum
time by comparing the states of an internal dynamical model which
represents the task and the sensory outputs coming in the form of
feedback. This concept is discussed in the following section.

\subsection{Event generation based on an internal model}

The generation of control commands by the user in IC (shown in
Fig.~\ref{fig:ic-cc-paradigm}B) can be seen as the interplay of an
\emph{Internal model} and a set of \emph{Control gains} within the CNS of
the user that modulate the input signal that is applied to the system.  The
internal model is a central feature of IC and serves as reference in terms
of performance, using a dynamical model to mimic the behaviour of a
continuous delay-free version of the system \citep{Gawthrop2015};
therefore, in the ideal scenario of control inputs and system outputs that
are not corrupted by motor or observation noise respectively, or affected
by external \emph{disturbances}, the states of this internal model should
match the ones generated by the sensory integration process.

On the other hand, if the observations predicted by the user differ from
the actual observations, it means that there are unaccounted disturbances
which produce a deviation from the reference internal model. It is possible
to use this source of error to decide if the sampling mechanism should stay
open or to close it, which would update the Prediction stage and
consequently the internal model itself with measured states. During the
time in which the sampling mechanism is open, the internal model is
evolving on its own and producing states \emph{continuously} without using
available feedback information, the fact that this process happens
continuously ensures that a control signal is always going to be
computed. In the case of a discrete movement, such as the initial ballistic
submovement towards the target (surge), the motor signals in charge of the
execution are planned in advance and executed ballistically.  In order to
plan such a ballistic movement, the internal model is necessary.  One way
to determine the mode in which the overall control loop should operate
(open vs closed-loop) is to compare the error between the sensed output and
the version of the output generated by the internal model against a fixed
value, establishing a triggering \emph{threshold}. If the error is larger
than the threshold then the controller should rely on feedback to reduce it
(trusting sensed information more). This constant comparison carried out by
the trigger mechanism generates \emph{events}, which define the moment in
time when the error overcomes the threshold value. This particular
implementation of IC is often referred as Event-Based or Event-Driven IC
\citep{Gawthrop2009}. When an event is created, the feedback loop is closed
only for a small instant and then reopened, which forces the controller to
stay open-loop for a minimum amount of time before feedback can be used
again due to new events.

The IC model shown in Fig.~\ref{fig:ic-cc-paradigm}B was conceived from a
computational perspective as an additional layer of an existing continuous
model proposed by Kleinman, which used a linear optimal controller to
approximate the response of a human operator in manual control tasks
\citep{Kleinman1969,kleinman1970optimal}. The previously mentioned elements
of sensory integration, prediction, and control generation formed the basis
for Kleinman's model, linking physiology concepts with control theory,
while providing solid basis for IC.

\subsection{Human variability in motion}

Variability is a distinctive element of human motion and human-in-the-loop
systems. A human generating a control action repeatedly over a system with
a defined set of external perturbations, would generate output responses
that are slightly different from each other. Physiologically, variability
is generated at any level of the motor pathway and it is normally
attributed to a combination of factors: 1) neural spike
initiation-propagation, as well as synaptic transmission and muscle
activation, all rely on biophysical and chemical events which are
stochastic in nature
\citep{jonesSourcesSignalDependentNoise2002,dec.hamiltonScalingMotorNoise2004,faisalStochasticSimulationsReliability2007,faisalNoiseNervousSystem2008}
and, 2) variation in movement planning and decision making processes
happening in the CNS \citep{dhawaleRoleVariabilityMotor2017,Gollee2017}.

From a computational point of view, the variability of human movements has
been explained by adding motor and observation noise to a continuous
representation of the overall closed-loop system
\citep{Levison1969,Kooij2007,Kooij2011}, implying that these differences are
the result of a linear process in combination with a stochastic component,
introduced by the random noise. However, there is experimental evidence
that shows how manual control is subject to a refractory process
\citep{Navas1968,Miall1986,Loram2011,Loram2014}, strengthening the view of
sustained sensorimotor control as a discrete sequence of open-loop
intervals combined with feedback instances of sensory information. It has
also been proposed that a triggering mechanism is behind the decision of
closing or opening the feedback loop, where the timing of each event is
decided by error signals crossing a threshold value.

Results from a visuo-manual tracking task \citep{Loram2011,Gollee2017}, in
which the participants used a joystick to control the position of an
unstable dynamical system (represented by a dot on a screen), showed that
an intermittent-predictive controller is able to explain both the linear
and nonlinear components of the output response at both excited and
non-excited frequencies, without adding any coloured noise. Whereas a
continuous predictive controller requires the addition of a separate motor
noise profile for each of the conditions in the experiment to adequately
describe the output response. These findings suggest not only that the
aperiodic triggering mechanism is a plausible hypothesis to explain manual
tracking but also that the intermittency is directly related to the
observed variability.

The constant switching between open and closed-loop evolution introduces
other benefits, such as the possibility to clearly distinguish the effects
of the applied input commands, the external disturbances and the natural
dynamics of the system \citep{Loram2011}. Moreover, the variability that is
introduced by the aperiodic triggering process of IC plays an important
role in adaptation and learning by regulating the trade-off between
exploration and exploitation \citep{Loram2015}. It has been shown that
higher rates of movement-to-movement variability in motor output predict
faster learning rates for specific motor control tasks. This suggests that
regulating variability according to the environment and the task at hand
might be an essential feature of the CNS to promote either fine control
when precision is needed or broad exploration to accelerate adaptation and
learning \citep{Wu2014}.

\section{The intermittent control model}
\label{sec:formalIC}

In this section, a more detailed explanation of IC is given, with an
emphasis on describing the main elements of the framework more formally,
from a control theory perspective, to support a reader wishing more insight
into the model structures, rationale and behaviour. While the theory of IC
in this section is fundamental to understand the details of the framework
as a control methodology, the main results of the paper are largely
accessible if skipped at a first reading. The high-level IC described in
Fig.~\ref{fig:ic-cc-paradigm}B in the context of a pointing task, is
reintroduced using a more general approach in Fig.~\ref{fig:ic}, which is
in turn based on the versions presented in
\citep{Gawthrop2007,Gawthrop2011,Gawthrop2015}. Together, the
\emph{Hand-Arm Biomechanics} and \emph{Input device PTF} blocks from
Fig.~\ref{fig:ic} represent a mathematical model of the neuromuscular
dynamics of the user and the user interface. Although user interface
dynamics can, in principle, be highly nonlinear, e.g., via nonlinear PTFs,
for simplicity in the following we assume that the overall dynamics can be
expressed as a linear state-space model of order~$n$, that corresponds to
the case of constant gain in pointing, and which can be expressed as
follows:

\begin{figure*}[!htbp]
\centering
\includegraphics{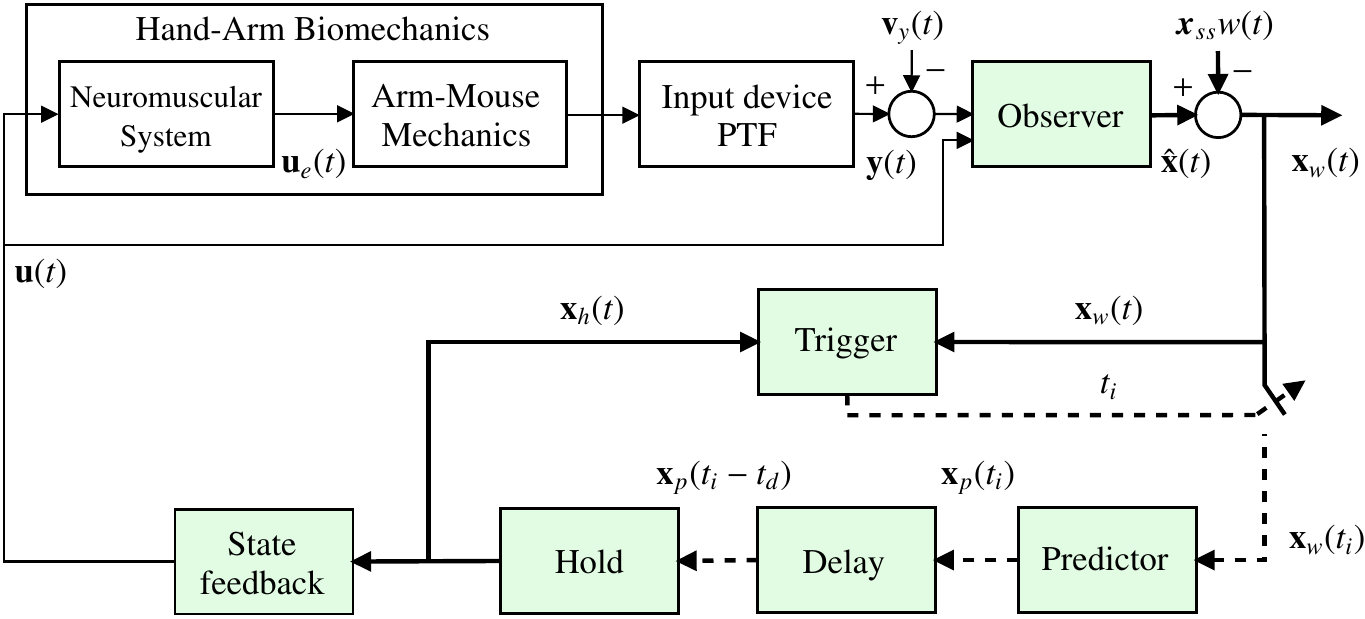}
\caption[The intermittent control model]{Diagram of the intermittent
  controller within the CNS of the user. The hold states $\xxx_h$ are
  compared with the state-estimates $\xxx_w$ in the trigger block. If the
  difference exceeds a predefined threshold, then the block creates events
  at times denoted by $t_i$. The hold states $\xxx_h$ are used to generate
  the control signal $\uuu$ during the open-loop period, and it is reset
  only at times $t_i$ by the predictor block. The dashed lines represent
  signals that are defined only at $t_i$. The thick lines/arrows of the
  diagram represent vector signals, whereas the thin ones represent scalar
  versions, for the single-input single-output case. This figure is based
  on the representation given in \citep{Gawthrop2011}. The Hand-Arm
  Biomechanics block contains the Neuromuscular System which generates the
  force that is applied to the input device and the Arm-Mouse Mechanics
  block that translates it into input to the pointing transfer function.}
\label{fig:ic}
\end{figure*}

\begin{equation}\label{eq:ssmodel}
\begin{aligned}
  \dxxx (t) & = \AAA \xxx (t)+\BBB \uuu (t) \\
  \yyy (t) & = \CCC \xxx (t)\:,
\end{aligned}
\end{equation}
with $\xxx (0) = \xxx_{0}$ as initial condition and where
$\xxx \in\mathbb{R}^{n}$ corresponds to the system state such as pointer
position and velocity, and muscle activation, $\yyy \in\mathbb{R}^{n_{y}}$
corresponds to the output such as pointer position on the display, and
$\uuu \in\mathbb{R}^{n_{u}}$ corresponds to the input such as muscle
excitation.  $t$ represents continuous time. $\AAA$ is a $n \times n$
matrix representing how the system evolves without control input, $\BBB$ is
a $n \times n_{u}$ matrix, representing the impact of control signals such
as muscle excitation on the system, and $\CCC$ is a $ n_{y} \times n$
matrix, representing how the system state maps to the users' observation,
such as pointer position on the display.

Two concepts from control theory are potentially generally useful for HCI
models. These are the {\it observer} and the {\it predictor}. The control
goal of the user is to bring the system output $\yyy(t)$ as close as
possible to the target signal $w(t)$ (e.g., the pointing target). The
system integration process mentioned in Fig.~\ref{fig:ic-cc-paradigm}B
which combines multiple sensory channels, can be interpreted as an
estimation problem. The central issue is that the control problem might not
be solvable from the observations alone, if the true state is unknown. For
example, in pointing, the user can observe only the pointer position. If
the pointer velocity, muscle activation etc. can not be observed directly,
they need to be estimated in order to enable the user to control the
system.

{\bf Observer:} To solve this issue, the input $\uuu(t)$ and the output
$\yyy(t)$ are fed to a linear \emph{observer}, which is a dynamical model
based on~(\ref{eq:ssmodel}), that reconstructs the full state of the system
$\hat{\xxx}(t)$ using only output measurements and information on the input
that is being applied. Once the target signal $w(t)$ has been introduced as
$\xxx_{w}(t) = \hat{\xxx}(t) - \xxss w(t)$, the resulting observed states
$\xxx_{w}(t)$ get sampled, this updates the rest of the structures in the
feedback loop, only at discrete points in time $t_{i}$.

{\bf Predictor: } In this implementation, the role of the \emph{predictor}
is to compensate for transmission time-delays which might be present in the
feedback loop both from the users neural system and from the computer
system, this is done by predicting the future states of the system
$\xxx_{p}(t_{i})$ based on the state estimates $\xxx_{w}(t_{i})$ which are
defined only when a sample is taken.

The concept of an internal model mentioned earlier in
Fig.~\ref{fig:ic-cc-paradigm}B, is represented in control terms as a
generalised \emph{hold}, which is a dynamical model that mimics the
behaviour of the overall closed-loop system by continuously generating the
state vector $\xxx_{h}(t)$. The \emph{trigger} compares the hold states
$\xxx_{h}(t)$ and the observed states $\xxx_{w}(t)$ to generate a
prediction error $e_p$. An event is generated by the trigger at $t_i$ once
$e_p$ exceeds a predefined threshold $q$. The hold states $\xxx_{h}(t)$ are
used to compute a \emph{state feedback} control input $\uuu(t)$
continuously, which is the signal that drives the neuromuscular system
producing the final input to the system $\uuu_{e}(t)$. Notice that both the
input $\uuu(t)$ and the output $\yyy(t)$ can be affected by motor noise
$\ma v_{u}(t)$ and sensory noise $\ma v_{y}(t)$ respectively.

Before introducing definitions for the hold, the predictor and the trigger
mechanism, it is important to define the different time frames that are
used in IC to accurately distinguish between what happens during the
open-loop intervals and the instances of feedback.

\subsection{Intermittent control time-frames}
IC combines the use of the following time-frames:

\begin{enumerate}
  \item \textit{\bf{Continuous-time} ($t$)}, represents the time in which the
    system defined in equation~(\ref{eq:ssmodel}) evolves.
  \item \textit{\bf{Discrete-time} ($t_i$)}, time instants at which
    an event is generated, indexed by \textit{i}. The time between event
    instants is known as the \emph{intermittent interval} $\Delta_{i}$. The
    \textit{i}th intermittent interval can be defined as
    \begin{equation} \label{eq:deltai}
    \Delta_{ol}=\Delta_{i}=t_{i+1}-t_{i} \:.
    \end{equation}
    A \emph{sampling delay} $\Delta_{s}$ can be used to sample the observed
    states $\xxx_{w}(t)$ once a fixed-time interval has elapsed after the
    detection of an event at $t_i$. If $\Delta_{s} = 0$ then the sample is
    taken at $t_i$.
  \item \textit{\bf{Intermittent-time} ($\tau$)}, a continuous variable
    that is restarted every $\Delta_{i}$ according to
    \begin{equation} \label{eq:tau}
    \tau=t-t_i \:.
    \end{equation}
    A lower limit $\Delta_{ol}^{\text{min}}$ can be
    specified within a given intermittent interval as
    \begin{equation} \label{eq:minol}
    \Delta_{i}>\Delta_{ol}^{\text{min}}>0 \:.
    \end{equation}
    The lower limit $\Delta_{ol}^{\text{min}}$, also known as \textit{minimum open-loop
    interval}, has been used before to model the \textit{psychological
    refractory period} observed in human motor control \citep{Gawthrop2011}.

\end{enumerate}
Based on these definitions, it is now possible to define the model used as
the generalised hold, which is a continuous closed-loop approximation of
the overall system that relies on the following underlying controller.

\subsection{Underlying continuous controller}
A state-feedback controller for the system in equation~(\ref{eq:ssmodel}),
with gain $\kkk$, can be established by implementing the following
control-law
\begin{equation} \label{eq:control-law}
  \uuu (t) = -\kkk \xxx (t) \:.
\end{equation}
Standard procedures such as pole-placement or the linear quadratic
regulator (LQR) approach \citep{Goodwin2001} can be used to obtain the gain
$\kkk$. The resulting closed-loop system, defined by the state vector
$\xxx_{c}$ is
\begin{equation}\label{eq:closed-loop}
  \begin{aligned}
  \dxxx_{c} (t) & = \AAA_{c} \xxx_{c} (t)\:,
  \end{aligned}
\end{equation}
which depends on $\xxx_{c} (0) = \xxx_{0}$ as the initial condition and a
closed-loop matrix $\AAA_{c}$ defined as follows
\begin{equation} \label{eq:Ac}
  \AAA_{c} = \AAA - \BBB \kkk \:.
\end{equation}
The target signal $w(t)$ can be introduced by defining the
following system
\begin{equation}\label{eq:ss-steady-state}
\begin{aligned}
 \ma 0_{n\times1} & = \AAA \xxss(t)+\BBB \uuss(t)\\
  \yyss(t) & = \CCC \xxss(t)  \:,
\end{aligned}
\end{equation}
where $\xxss$, $\uuss$, and $\yyss$ correspond to the steady-state versions
of the states, inputs, and outputs respectively. Solving for $\xxss$ and
$\uuss$ can be done by writing the system as follows
\begin{equation} \label{eq:ic-ss}
  \left[\begin{array}{c} \xx_{ss}(t) \\
\uu_{ss}(t) \end{array}\right] = \left[\begin{array}{cc} \AAA & \BBB
\\ \CCC & \ma 0
      \end{array}\right]^{-1}\left[\begin{array}{c} \ma 0_{n\times1}\\
1 \end{array}\right] \:.
    \end{equation}
Thus, the overall control input from (\ref{eq:control-law}) can be redefined as
\begin{equation}\label{eq:full-control-law}
\begin{aligned}
  \uuu(t) & = -\kkk\left( \xxx(t)- \xxss  w(t) \right) + \uuss w(t) \\
          & = -\kkk \xxx(t) + \left(\uuss + \kkk \xxss \right) w(t) \:.
\end{aligned}
\end{equation}
By defining $\ma r = \uuss + \kkk \xxss$, a simplified expression is
obtained
\begin{equation} \label{eq:final-control-law}
\begin{split}
  \uuu(t) = -\kkk \xxx(t) + \ma r w(t) \:.
\end{split}
\end{equation}
The expression in~(\ref{eq:final-control-law}) reduces the error between
the output $\yyy(t)$ and the target signal $w(t)$, depending on the value
of the gain $\kkk$. This assumes that the matrices $\AAA$ and $\BBB$ are
such that the model in~(\ref{eq:ssmodel}) is \emph{controllable} with
respect to $\uuu(t)$, which means that the control input $\uuu(t)$ has
the ability to drive the state-vector $\xxx(t)$ from its initial condition to
any final value in a finite amount of time.

\subsection{State observer}
In many cases, the system state $\xxx(t)$ in (\ref{eq:ssmodel}) is not
fully available for direct measurement. For example, usually only the
pointer position is shown by the computer, but not the pointer
velocity. One way to overcome this is to implement a state observer that
estimates the portion of $\xxx(t)$ that is not known using only available
information such as the output $\yyy(t)$, known states, and the current
input $\uuu(t)$. An observer for the system described in (\ref{eq:ssmodel})
can be designed by defining a vector $\hat{\yyy}(t)$ which contains all the
available signals as follows
\begin{equation} \label{eq:observer}
  \dot{\hat{\xxx}} (t) = \hat{\AAA} \hat{\xxx} (t)+\BBB \uuu (t)+\ma L \left[\hat{\yyy}(t) - \ma v_y(t) \right] \\
\end{equation}
where $\hat{\AAA} = \AAA - \ma L \hat{\CCC}$ and $\hat{\yyy}(t)$ is defined as
\begin{equation} \label{eq:observation-matrix}
  \hat{\yyy} (t) = \hat{\CCC} \xxx(t) \:.
\end{equation}
Therefore, the matrix $\hat{\CCC}$ defines which elements from $\xxx(t)$ are
used as inputs for the observer in order to reconstruct the full state. The
matrix $\ma L$ is a design parameter similar to the feedback gain $\kkk$ in
that both of them can be designed via pole-placement or LQR methods.

\subsection{Generalised hold}

The generalised hold uses the dynamics imposed by~(\ref{eq:Ac}) to define
the open-loop behaviour in IC. This is achieved by implementing the
following state-feedback control-law
\begin{equation} \label{eq:u_openloop}
\uuu(t)=\uuu(t_i+\tau)=-\kkk \xxx_h(\tau) \:,
\end{equation}
where the hold states $\xxx_h$ evolve in the intermittent time $\tau$
according to the following autonomous system
\begin{equation} \label{eq:hold_system}
  \dxxx_{h}(\tau)=\AAA_{h} \xxx_{h}(\tau) \:.
\end{equation}
When the observed states are sampled at $t = t_i$, the hold states
$\xxx_{h}$ are reinitialised using vector $\UUU_{i}$ as follows
\begin{equation} \label{eq:U_IC}
\UUU_i = \KKK_h \xxx_p(t_i-t_d) \:.
\end{equation}
Expression (\ref{eq:U_IC}) takes the predicted states $\xxx_{p}(t_i-t_d)$,
which cancel the effect of the time-delay $t_d$, to reset the hold state
$\xxx_{h}$ at the start of each intermittent interval
\begin{equation} \label{eq:reset_U_IC}
\xxx_h(t_i) = \UUU_i \:.
\end{equation}
The square matrix $\KKK_h = \ma I_{n \times n}$ in (\ref{eq:U_IC}) is
defined as the \emph{intermittent control gain}. The overall dynamics of
the autonomous system in~(\ref{eq:hold_system}) are then determined by
matrix $\AAA_h$. If the closed-loop matrix $\AAA_{c}$ in~(\ref{eq:Ac}) is
used as follows
\begin{equation} \label{eq:smh-Ac}
\AAA_h = \AAA_c \:,
\end{equation}
then the generalised hold becomes a \emph{system-matched hold}
\citep{Gawthrop2011A}, that effectively approximates the dynamics of the
closed-loop system defined in \eqref{eq:closed-loop}. The assignment in
(\ref{eq:smh-Ac}) provides a common ground to compare the estimated states
provided by the observer and to those coming from the hold, which can be
seen as reference states for the IC framework.

\subsection{Intermittent prediction}
A predictor can be implemented as a compensation strategy in the presence
of time-delays. The following dynamical system can be established during
the intermittent time frame $\tau$
\begin{equation} \label{eq:dynamics_predictor}
\dxxx_{p}(\tau)=\AAA \xxx_{p}(\tau)+\BBB \uuu(\tau) \:,
\end{equation}
with $\xxx_{p}(0)=\xxx_{w}(t_{i})$ and evaluated at
$\tau = t_d$. Combining~(\ref{eq:dynamics_predictor})
and~(\ref{eq:hold_system}) yields the following extended system
\begin{equation} \label{eq:comb_dynamics_predictor}
\frac{d}{d\tau}\ma X(\tau)=\AAA_{ph}\ma X(\tau) \:,
\end{equation}
subject to $\ma X(0)=\ma X_{i}$ as the initial condition,
and where
\begin{equation}
\AAA_{ph}=\left[\begin{array}{cc}
\AAA & -\BBB \kkk\\
0_{n\times n} & \AAA_{h}
\end{array}\right] \:.
\end{equation}

The extended state vector $\ma X$
from~(\ref{eq:comb_dynamics_predictor})
is defined as
$\ma X(\tau)=\left[\begin{array}{cc}\xxx_{p}(\tau) &
    \xxx_{h}(\tau)\end{array}\right]^{T}$ during the open-loop interval. At
$t_i$, $\ma X$ takes the following form
\begin{equation} \label{eq:state_i_predictor}
\ma X_{i}=\left[\begin{array}{c}
\xxx_{w}(t_{i})\\
\xxx_{p}(t_{i}-t_{d})
\end{array}\right] \:.
\end{equation}
The solution of~(\ref{eq:dynamics_predictor}) at $\tau = t_d$ yields
\begin{equation} \label{eq:solution_predictor}
\ma X(t_{d})=e^{\AAA_{ph} t_{d}}\ma X_{i}\:.
\end{equation}
From~(\ref{eq:solution_predictor}), the predicted states $\xxx_p$ can be
obtained every intermittent interval using the following expression
\begin{equation} \label{eq:predicted_states}
\xxx_{p}(t_{i})=\EEE_{pp}\xxx_{w}(t_{i})+\EEE_{ph}\xxx_{h}(t_{i}) \:,
\end{equation}
where the matrices $\EEE_{pp}$ and $\EEE_{ph}$ of dimension $n\times n$,
are partitions of the $2n \times 2n$ matrix $\EEE$ defined as
\begin{equation} \label{eq:E_matrix_partition}
\EEE = e^{\AAA_{ph}t_{d}} =\left[\begin{array}{cc}
\EEE_{pp} & \EEE_{ph}\\
\EEE_{hp} & \EEE_{hh}
\end{array}\right] \:.
\end{equation}
Both $\EEE_{pp}$ and $\EEE_{ph}$ can be obtained offline.

\subsection{Event detection}
IC produces an aperiodic sequence of events, determined by the error
between the hold states $\xxx_{h}$ and the closed-loop observer state
$\xxx_w$ as follows
\begin{equation} \label{eq:trig-prediction-error}
   e_x = \xxx_h(t) - \xxx_w(t) \:.
\end{equation}
An event is generated when $e_{x}$ is greater than a predefined threshold
$q$, according to the following quadratic switching function
\begin{equation} \label{eq:quadraticerror}
   e_{x}^{T}(t) \, \Qt e_{x}(t) > 1 \:,
\end{equation}
where $\Qt$ is a positive semi-definite matrix that defines which states
are considered in order to detect events and thus trigger the use of
feedback. For instance, an IC triggering on the error of a two dimensional
state vector would implement the next switching function
\begin{equation} \label{eq:quadraticerror_ex}
   \bigg(\frac{e_{x_{1}}}{q_1}\bigg)^{2} + \bigg(\frac{e_{x_{2}}}{q_2}\bigg)^{2} > 1 \:,
\end{equation}
with a matrix $\Qt$ containing the following values in its diagonal
\begin{equation} \label{eq:Qt_detail}
\Qt =\left[\begin{array}{cc}
\frac{1}{(q_{1})^2} & 0\\
0 & \frac{1}{(q_{2})^2}
\end{array}\right] \:.
\end{equation}
This particular choice of the event detection mechanism allows to set
different threshold values for specific states and to use different
combinations of them.

\section{Experiment Setup and Dataset}
\label{sec:experiment}

The results shown in this paper are based on the \emph{Pointing Dynamics
  Dataset} that was collected from a mouse pointing experiment described in
\citep{MulOulMur17}. The IC framework from the previous sections is used to
identify the parameters of a standard intermittent controller. The full
dataset is publicly
available.\footnote{\url{http://joergmueller.info/controlpointing/}} In the
aforementioned experiment, 12 participants were asked to control a white
pointer that was shown on a screen. The pointer reflected the changes in
the $x$-dimension of the mouse exclusively and it was restricted to move
only horizontally.

The task consisted of clicking a number of one-dimensional targets that
were displayed in sequence. During a block of trials, the active target was
presented in a different color compared to the previous target, and the
distance between them, as well as the width of the target, stayed constant
throughout the block. Each block uses a specific combination of distance
and width; a total of 8 combinations were applied for each subject: 1)
distance of 212 mm and widths of 0.83, 3.32, 14.1 and 70.6 mm, 2) distance
of 353 mm and widths of 1.38, 5.54, 23.5, 118 mm. Fitts' law \citep{Fit54}
states that the time required to move to a specific target area $MT$ is a
function of both the distance to the target $D$ and the size of the target
$W$ as follows
\begin{equation} \label{eq:Fitts}
MT = a + b ID,
\end{equation}
where the Index of Difficulty (ID) is
\begin{equation} \label{eq:ID}
ID = \log_2 \left( \frac{D}{W} + 1 \right)\:,
\end{equation}
which is known as the Shannon formulation \citep{mackenzie92}. Based on
(\ref{eq:ID}) and the selected combinations of distance between targets and
target width, the resulting ID for both 212 mm and 353 mm conditions are 8,
6, 4 and 2.  Therefore, each participant had to complete a full block of
trials per ID for the two distance conditions. The experiment blocks were
designed to have 102 trials each, divided into 22 trials for training and
80 for the rest of the task.

The participants were asked to stay below a 5\% error rate while clicking
on every target as quickly as possible. A beep sound would indicate the
user if the click was made outside the target area, triggering the
appearance of the next target on screen for that particular block. The
users were not asked to repeat failed trials (missed targets). All
participants completed a training phase of 22 trials for each block before
starting the experiment and had a short break between to rest between
blocks.

\section{Model implementation and controller design}
\label{sec:modelimp}
The intermittent controller shown in Fig.~\ref{fig:ic} was implemented in
Matlab\footnote{Matlab release version R2019a. The MathWorks, Inc., Natick,
  Massachusetts, United States.} using the Control Systems Toolbox to
generate appropriate transfer function and state-space representations of
the human dynamics and the system to be controlled. The block labelled as
\emph{Neuromuscular system} in Fig.~\ref{fig:ic} was implemented as a
second order system with time-constants of 50~ms
\citep{milner1973contractile},
\begin{equation} \label{eq:transfer_nms}
    G_{nms}(s) = \frac{1}{\left(0.05s + 1\right)^{2}} \:,
\end{equation}
where $G(s)$ is the transfer function in the Laplace domain. This transfer
function was converted to its equivalent state-space system in the form of
equation~\eqref{eq:ssmodel} with
\begin{equation} \label{eq:nms}
  \AAA_{nms}=\left[\begin{array}{cc}
    -20 & 20 \\
    0 & -20
    \end{array}\right] \:, \quad \BBB_{nms}=\left[\begin{array}{c}
    0 \\
    20
    \end{array}\right] \:, \quad \CCC_{nms}=\left[\begin{array}{cc}
    1 & 0
    \end{array}\right] \:.
\end{equation}
Similarly, the block shown in Fig.~\ref{fig:ic} as \emph{Arm-Mouse Mechanics}
was established as a double integrator in the state-space
representation of \eqref{eq:ssmodel} with
\begin{equation} \label{eq:sys}
  \AAA_{sys}=\left[\begin{array}{cc}
    0 & 1 \\
    0 & 0
    \end{array}\right] \:, \quad \BBB_{sys}=\left[\begin{array}{c}
    0 \\
    1
    \end{array}\right] \:, \quad \CCC_{sys}=\left[\begin{array}{cc}
    1 & 0
    \end{array}\right] \:,
\end{equation}
where the associated state vector is
$\xxx_{sys}(t)=\left[\begin{array}{cc}\xxx_{pos}(t) &
    \xxx_{vel}(t)\end{array}\right]^{T}$, comprised of the pointer position
$\xxx_{pos}(t)$ and velocity $\xxx_{vel}(t)$. This assumes that the output
of the neuromuscular system, depicted as $\uuu_{e}(t)$ in
Fig.~\ref{fig:ic}, is a force generated by the participant, applied to a
unit mass in a low friction environment, and that the output of the overall
system, $\yyy(t)$, is the position of the mass, which in this case is
equivalent to the position of the pointer on the screen. The combination of
(\ref{eq:nms}) and (\ref{eq:sys}) yields a fourth order system based on the
full state vector $\xxx(t)$, which is then used to design the overall
controller.

The feedback gain vector $\kkk$ shown in (\ref{eq:control-law}) was
designed using optimal control via the LQR design method
\citep{Goodwin2001}, to ensure that the closed-loop matrix $\AAA_c$,
which defines the behaviour of the generalised hold in IC
(\ref{eq:smh-Ac}), has stable eigenvalues. This involves the
minimisation of the LQR cost function
\begin{equation} \label{eq:lqr}
J_{LQR}=\int_{0}^{\infty}\left[\xxx(t)^{T}\Qc\xxx(t)+\uuu(t)^{T}\Rc\uuu(t)\right]dt
\:,
\end{equation}
and the solution of its associated algebraic Ricatti equation. Both
$\xxx(t)$ and $\uuu(t)$ in (\ref{eq:lqr}) are weighted by design matrices
$\Qc$ (an $n \times n$ diagonal matrix that must be positive semi-definite)
and $\Rc$ (an $n_{u} \times n_{u}$ diagonal positive definite matrix, where
$n_u$ corresponds to the number of inputs in the system). The diagonal
nature of $\Qc$ is in fact a design choice, which is based on the
assumption of not having crossed-term effects between the states. Using the
same LQR method on the dual problem of state estimation, the closed-loop
observer gain matrix $\LLL$, can be computed to force the estimation error
to vanish asymptotically, by choosing a design matrix $\Qo$. The
observation matrix in (\ref{eq:observation-matrix}) was defined as
$\hat{\CCC} = \left[\begin{array}{cccc} 1 & 0 & 0 & 0 \end{array}\right]$,
which means that only the first state of the state vector $\xxx(t)$,
corresponding to the pointer position $\xxx_{pos}(t)$, is used by the
observer to estimate the rest to the unknown states. In the context of this
work, matrices $\Qc$ and $\Qo$ were defined as parameters to be optimised,
whereas $\Rc$ was fixed for the entire process with a value of 1 (in this
case it is a scalar since there is only one input to the system,
$n_{u} = 1$). The time-delay $t_{d}$ remained constant at $0.01$ sec for
all participants and conditions. Note that the effective time-delay is a
combination of the constant $t_d$ and the optimised threshold $q$ and
minimum open-loop interval $\Delta_{ol}^{\text{min}}$.

To detect events, a threshold $q$ was applied only to
$\xxx_{pos}(t)$. However, the actual value of $q$ was obtained via the
optimisation process. As suggested in \citep{Gawthrop2009}, the use of a
\emph{disturbance observer} is recommended to reduce the effect of zero
mean constant disturbances $\ddd(t)$ at the input; therefore, an integrator
is normally used to account for them and compensate accordingly.

\section{Optimisation approach}
\label{sec:optim}

Based on the systems (\ref{eq:nms}) and (\ref{eq:sys}), an optimisation
procedure was used to fit a set of controller parameters to the
experimental data. The following parameters were identified as a result of
the optimisation process: The LQR design matrices $\Qc$ and $\Qo$, the
prediction error threshold $q$ for triggering purposes, the minimum
open-loop interval $\Delta_{ol}^{\text{min}}$ and a mismatch gain
$\ma A_p = 1 - p$.

The purpose of $\ma A_p$ is to model the cases where the control input that
is applied to system is different by a fixed amount from what it should be
by design. This is implemented as follows: from Fig.~\ref{fig:ic}, the
output of the neuromuscular system $\ma u_e(t)$, which serves as the input
to the system in (\ref{eq:sys}), is multiplied by a quantity $p$, resulting
in
\begin{equation}
\dxxx_{sys}(t) = \AAA_{sys} \xxx_{sys}(t)+\BBB_{sys} \uuu_{sys}(t) \:,
\end{equation}
where $\uuu_{sys}(t) = p\uuu_e(t)$. When $p = 1$, the full input is
applied. A mismatch is generated when $p$ is differnt than 1. The
optimisation process identified the value of $p$ directly; however, the
mismatch gain $\ma A_p$ is reported in the following sections as it gives a
better insight on the difference between $\uuu_e(t)$ and
$\uuu_{sys}(t)$. This discrepancy has a significant effect in the overall
performance, specially when the states get close to the target position.

As observed in \citep{MulOulMur17}, the initial set of parameters used to
start the optimisation process had a negligible effect on the resulting
optimised parameters, converging consistently to the same set. As a
consequence, a common starting point was selected for all conditions and
subjects, which is shown in Table~\ref{tab:initial-parameters} including
the lower and upper limits for each of them.

A pattern-search method was
used for all subjects and conditions as implemented in Matlab's
Optimisation Toolbox. Pattern-search methods are numerical optimisation
routines that avoid the calculation or approximation of derivatives to
minimise an objective function \citep{torczon1997convergence}. Based on a
starting point, the method tries to find a new point that has a lower
objective function value. The selection of the points to test is done based
on pattern vectors which define points around the initial guess. If a lower
objective function value is obtained for a particular point, then the point
gets selected and a new set of scaled pattern vectors are applied
(increasing the effective search area). If there is no point with a lower
objective function value, then the area covered by the pattern vectors is
reduced by a predefined factor. The process is repeated until a minimum is
found.  In general, these methods are useful to speed up repetitive
optimisation tasks.

\begin{table}[ht]
\centering
\begin{threeparttable}
\caption{Initial optimisation values\tnote{*}}\label{tab:initial-parameters}
\begin{tabular}{cccccc}
\hline
\toprule
 & \bfseries $Parameter$ & \bfseries $Initial$ $value$ & \bfseries $Min.$ $value$ & \bfseries $Max.$ $value$ & \bfseries $Units$  \\
\midrule
$LQR$ & \texttt{$\ma Q_o$}  & $10$ & $0.00001$ & $100000$ & - \\
  & \texttt{$\ma Q_{c,diag}$} & $\left[\begin{array}{cccc} 1 & 1 & 1 & 1 \\ \end{array}\right]$ & $0.00001$ & $100000$ & - \\
\midrule
$Timing$ & \texttt{$q$} & $0.03$ & $0.001$ & $1$ & - \\
 & \texttt{$\Delta_{ol}^{min}$}& $0.05$ & $0.03$ & $1$ & $sec$ \\
\midrule
$Mismatch$ & \texttt{$p$} & $0.9$ & $0.1$ & $2$ & - \\
\bottomrule
\hline
\end{tabular}
\begin{tablenotes}
\item[*]\textit{The values are shown in three different categories}:
  $\ma Q_c$ and $\ma Q_o$ correspond to the \emph{LQR} design parameters,
  the threshold $q$ and the minimum open-loop interval
  $\Delta_{ol}^{\text{min}}$ as \emph{timing} parameters, and finally the
  value of $p$ which determines the \emph{mismatch} gain $\ma A_p = 1 -
  p$. The minimum and maximum values for each parameter used as limits
  during the optimisation process are also shown.
\end{tablenotes}
\end{threeparttable}
\end{table}

\subsection{Data partitioning}

\begin{figure*}[!htbp]
\centering
\includegraphics{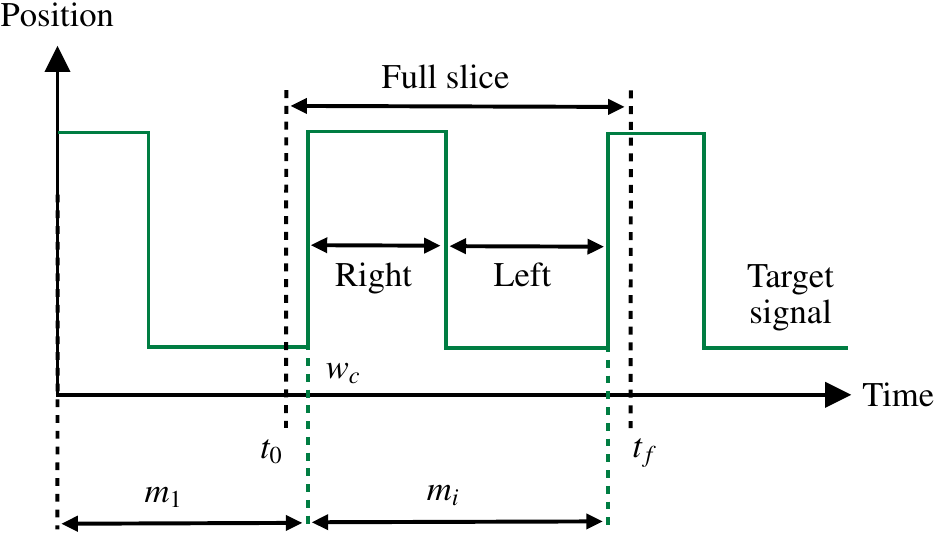}
\caption[Full slice diagram]{Data partitioning diagram. The figure shows a
  hypothetical target signal $w(t)$ (in green) to illustrate how the slices
  were delimited. The initial point of the slice $t_0$ starts 10 samples
  before a change in reference $w_c$ where the participant has to move the
  pointer position to the target on the right side of the screen. The
  ending point $t_f$ is 10 samples after the next change in reference in
  the same direction. The second portion of the slice (shown as
  \emph{Left}) corresponds to the movement from the target on the right to
  the target on the left. As described in section \ref{controller-switching},
  $m_1$ represents the best controller of the bank $m$, which is used to
  start a simulation and $m_i$ is a randomly selected controller from the
  same bank.}
\label{fig:slice-diagram}
\end{figure*}

The amount of data used for each optimisation differs from
\citep{MulOulMur17}. Instead of fitting an intermittent controller to the
entire block of trials, the data was partitioned according to the following
criteria: the time elapsed between two successive changes in the target
position would constitute a full \emph{slice}. Essentially, each slice is
composed of the pointer trajectories made by the subject to reach
consecutive targets (two consecutive trials). An individual IC is then
fitted to each slice. In Fig.~\ref{fig:slice-diagram}, a prototypical
target signal is used to illustrate how the data was partitioned into
individual slices. Each block had 80 trials, this resulted in 40 full
slices per block. The optimisation was carried out using the information of
the last 20 slices in the block, leaving the rest for evaluation.

\subsection{Criteria for model fitness via cost functions}
To establish how well the intermittent controller fits the experimental
data, we use a general cost function $J$ measuring the difference between
pointer movement predicted by the model and actual pointer movement, which
also includes the difference in terms of the pointer velocity. The
optimiser considers the root mean squared error (RMSE) of these quantities
as follows
\begin{equation} \label{eq:opt_cost}
J=c_{p}\sqrt{\frac{{\sum_{i=1}^{n_j}}\left(\hat{y}_i-y_i\right)^{2}}{n_j}}+c_{v}\sqrt{\frac{{\sum_{i=1}^{n_j}}\left(\hat{v}_i-v_i\right)^{2}}{n_j}}\:,
\end{equation}
where $\hat{y}$ and $\hat{v}$ are the simulated pointer position and
velocity respectively. The number of samples considered for each slice is
shown as $n_j$ and the constants $c_p$ and $c_v$ denote values that act as
weights for each of the errors. The errors in (\ref{eq:opt_cost}) between
the positions and velocities are squared, the result is averaged over the
number of samples in each trial and then the square root of the obtained
quantity is computed. This formulation provides a clear basis to evaluate
the fitness of the resulting intermittent controllers. For the optimisation
procedure, the values of $c_p$ and $c_v$ were set to 0.5, resulting in
equal weights for each of the error terms in (\ref{eq:opt_cost}).

\subsection{Controller switching} \label{controller-switching} The
resulting models of the optimisation procedure can be used to carry out a
simulation over a collection of trials for each participant. This can be
done by selecting the best model available according to different criteria;
for instance, selecting the model with the lowest RMSE score or the one
that produced the lowest value of the cost function defined in
(\ref{eq:opt_cost}). A simulation using this criterion would yield a
response that is acceptable in terms of accuracy and steady-state error;
however, it would not be able to capture some of the inherent variability
of human control. To create a controller that exhibits an appropriately
wide variety of responses, we took a multiple-model approach, instead of
just applying a single model for all trials. This collection of parameters
which are compatible with individual realisations of the behaviour helps
represent the behavioural variability.

For all participants, a bank of models, which we will call $m$, was
generated for each of the conditions in the experiment. This bank is
essentially a ranked list based on the value of (\ref{eq:opt_cost}), that
contains the intermittent controllers that were derived using the optimised
parameters for each slice. This amounts to 20 controllers per condition. To
start a simulation using this multiple model approach, the best controller
in the bank, or $m_1$, is used against the first trial, which is comprised
from an initial pointer movement towards the right target followed by a
second movement to the left target. This is illustrated in
Fig.~\ref{fig:slice-diagram}, where $w_c$ shows the moment in time when the
target changes. At $w_c$, a new controller $m_i$ is selected from the model
bank and applied for the duration of the next trial only, where $i$
represents a random integer number between 1 and 20. This random selection
of controllers from the bank $m$ starts only after the first trial has
finished and it is maintained until all the trials are completed. This
effectively ensures that different controllers are used for a pair of back
and forth movements, producing a trial response that is slightly different
compared to the previous trial, resulting in a more variable trajectory
overall.

The following section introduces the results obtained when this multiple-model
approach is used to generate dynamic responses in order to compare them to
the experimental data.

\section{Modelling Results}
\label{sec:modelling}
The results obtained from the optimisation process applied to all
participants and conditions are presented in this section. First, a
visualisation of the observed variability in some of the responses is
introduced, then the optimisation parameters obtained in each condition are
shown including some of the dynamic responses showing the pointer position
evolution, phase planes where position and velocity are compared, and Hooke
plots which show the acceleration profiles through time.

\subsection{Response variability}
\begin{figure*}[!htbp]
\centering
\includegraphics{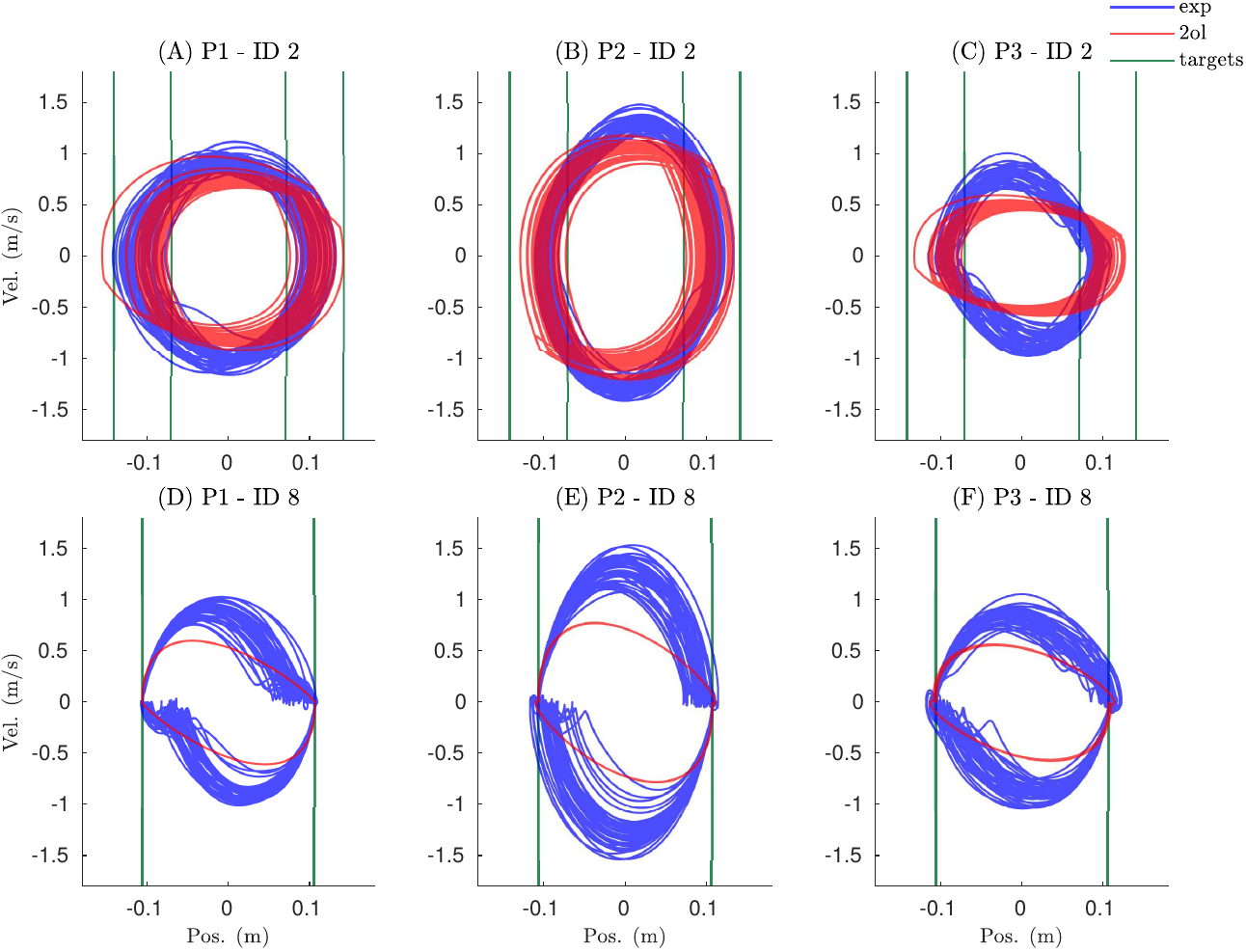}
\caption[Phase planes for participants 1, 2 and 3]{Phase planes showing
  pointer position vs. velocity for participants 1, 2 and 3. Experimental
  trajectories for each participant are shown in blue, left and right
  target limits are represented by green vertical lines, and the
  trajectories in red correspond to the second order lag model (2ol) from
  \citep{MulOulMur17}. A, B, and C, show data for a distance of 212 mm with
  a target width of 70.6 mm, corresponding to an ID of 2. D, E, and F, show
  data for the same distance between targets but with a different target
  width of 0.83 mm (ID 8). Participants 1, 2 and 3 are represented by each
  of the columns from left to right. From the simulated trajectories in
  red, it is clear that the 2ol model provides some degree of variability
  for ID 2 but when the ID increases, it does not reproduce the observed
  variability of the experimental trajectories.}
\label{fig:phase123}
\end{figure*}

The ID of the task has a significant impact on the overall behaviour. In
Fig.~\ref{fig:phase123}, the phase planes (pointer velocity in the vertical
axis vs pointer position in the horizontal axis) for participants 1, 2 and
3 are shown to illustrate how the ID influences the control strategy and
the human response. The bottom row (D, E, F), shows the experimental data
for an ID of 8, which is regarded as the most difficult task in the
experiment. A clear indicator of this is how small the target width is
(0.83 mm), compared to a target width of 70.6 mm (ID of 2) presented in A,
B and C. The target width is delimited by vertical lines on each side. For
each condition, the experimental phase planes are shown in blue, whereas
the ones in red correspond to the continuous second order lag controller
(2ol) described in \citep{MulOulMur17}. The 2ol controller is different in
many levels when compared to IC, but its most fundamental difference is the
fact that it uses feedback at all times to generate the control input that
is applied to the system.

The bottom row in Fig.~\ref{fig:phase123} (D, E, F), shows responses where
an initial burst in speed and position is made, commonly known as the
surge, followed by a series of small position corrections as the pointer
approaches the target. It is clear for instance that the landing position
for the surge movement differs in every trial -- a consequence of human
variability. The top row of figures (A, B, C) exhibits a drastically
different behaviour since the target width is greater, there is variability
present but the participants seem to engage in an almost constant motion
towards the targets without making corrections as they approach. The 2ol
controller does a relatively good job tracking the easy conditions (ID 2);
however; when applied to the most difficult condition (ID 8), the phase
plane does not show the same level of variability seen in the experimental
data. The parameters of the 2ol controller were also obtained via
optimisation as described in \citep{MulOulMur17} resulting in an individual
parameter set for each condition, which was subsequently applied in
simulation.

\subsection{Optimised parameters and their interpretation}
\label{sec:optimised-parameters}
\begin{figure*}[!htbp]
\centering
\includegraphics[width=1\linewidth]{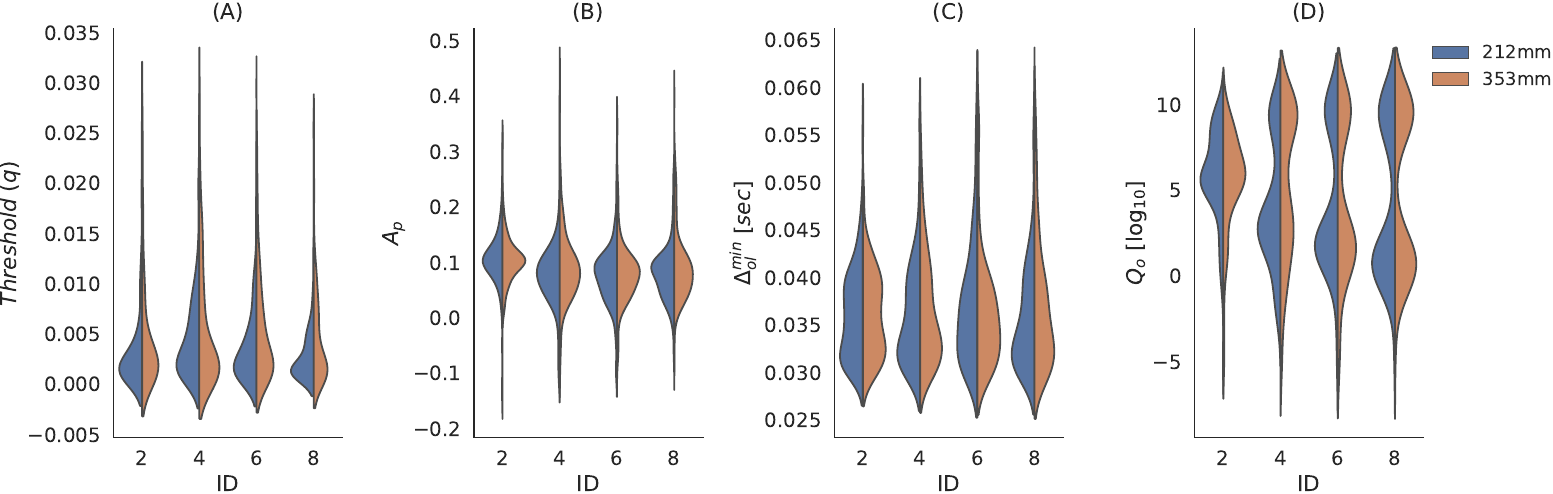}
\caption[Optimised controller parameters for each condition]{Optimised
  controller parameters for each condition. (A) the threshold $q$, (B)
  mismatch gain $\ma A_p$, (C) the minimum open-loop interval
  $\Delta_{ol}^{\text{min}}$ , and (D) the observer gain $\ma Q_o$ are
  shown as violin plots including data of all subjects and categorised by
  ID (horizontal axis). The results are also grouped according
  to the two values of distance between targets used in the experiment
  (left: 212 mm and right: 353 mm). The shape of the distributions
  for ID 2 is slightly different compared to the rest of the conditions.}
 \label{fig:control_params_vs_seaborn}
\end{figure*}

The optimisation process had 4 free scalar parameters $\ma Q_o$, $\ma A_p$,
$q$ and $\Delta_{ol}^{\text{min}}$, as well as the matrix $\ma Q_c$, where
only its diagonal elements are of interest (the off diagonal entries are
zero), thus adding 4 extra parameters for a total of 8. In
Fig.~\ref{fig:control_params_vs_seaborn}, the optimised parameters for all
subjects are shown when grouped by ID. This includes the threshold $q$ (A),
the gain $\ma A_p$ (B), the minimum open-loop interval
$\Delta_{ol}^{\text{min}}$ (C) and finally the observer gain $\ma Q_o$ (D,
shown in log scale). For each parameter, the data distribution is shown as
a violin plot for the two different distances of 212 mm (left side) and 353
mm (right side) respectively, and grouped according to the ID.

Fig.~\ref{fig:control_params_vs_seaborn}A, shows that the threshold $q$ has
slightly larger values for ID 2, which is consistent for the distributions
of both distances. The data are more concentrated for the cases of ID 4, 6,
and 8 as shown by the narrow distributions. All conditions exhibit long
tails, indicating the presence of higher thresholds in some of the
optimised models. The mismatch gain $\ma A_p$ in
Fig.~\ref{fig:control_params_vs_seaborn}B also exhibits a clear trend. For
ID 4, 6, and 8, the data distribution is wider for these three cases,
indicating that there was less of a mismatch compared to ID2, which shows
most of its values around $\ma A_p = 0.1$. The minimum open-loop interval
$\Delta_{ol}^{\text{min}}$ in Fig.~\ref{fig:control_params_vs_seaborn}C is
comparable for the different levels of ID, showing that most of the data
lie within a 0.03 to 0.05 sec range.

\begin{figure*}[!htbp]
\centering
\includegraphics[width=1\linewidth]{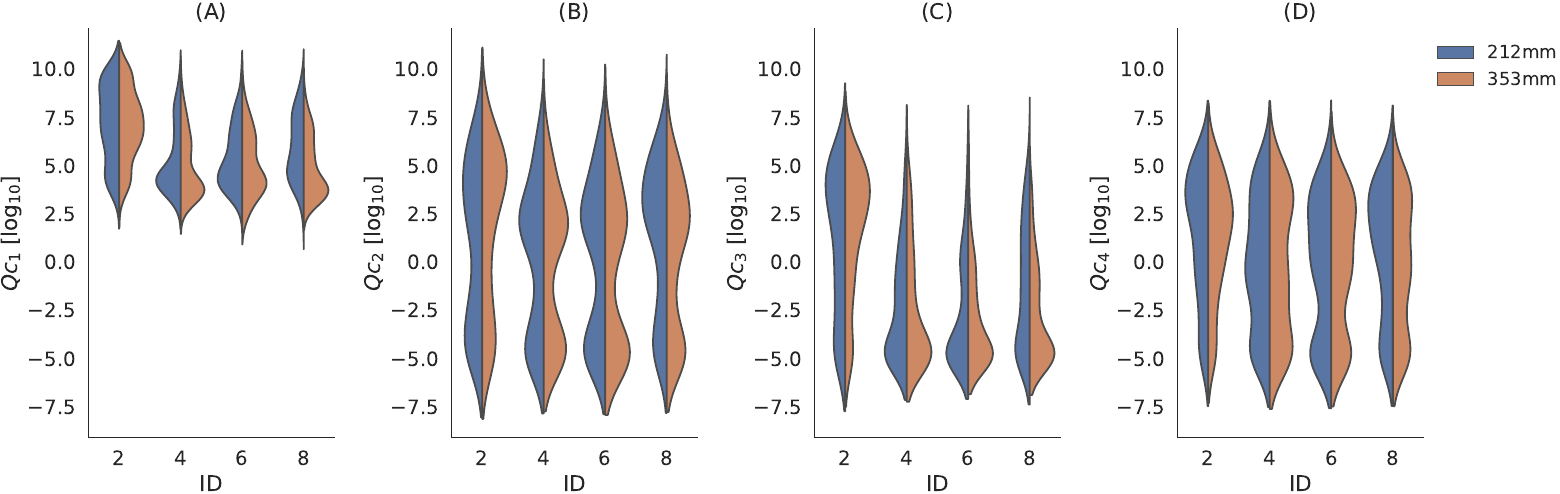}
\caption[Optimised $\ma Q_c$ parameters for each condition]{Optimised
  $\ma Q_c$ parameters for each condition. This figure shows the values of
  the four elements in the diagonal of matrix $\ma Q_c$, for all subjects
  and segmented by ID (horizontal axis). The results for $\ma Q_{c1}$,
  $\ma Q_{c2}$, $\ma Q_{c3}$ and $\ma Q_{c4}$ are shown in A, B, C, and D
  respectively. The violin plots show the data for the two distances
  between targets (left, in blue: 212 mm and right, in brown: 353 mm).
  $\ma Q_{c1}$ has consistently higher values compared to the rest of the
  elements in $\ma Q_{c}$ and ID 2 has a different distribution shape than
  the rest of the conditions}.
 \label{fig:Qc_vs_seaborn}
\end{figure*}

The values for the design matrix $\ma Q_c$ are presented in
Fig.~\ref{fig:Qc_vs_seaborn} using a log scale, where
\ref{fig:Qc_vs_seaborn}A, B, C and D show the data of the four elements in
the diagonal of matrix $\ma Q_c$. The first and second elements of
$\ma Q_c$ shown in Fig.~\ref{fig:Qc_vs_seaborn}A and
Fig.~\ref{fig:Qc_vs_seaborn}B are of particular importance: $\ma Q_{c1}$,
since it is associated to the position state which is the defined output of
the controller, and $\ma Q_{c2}$ is related to the pointer velocity
state. These two values have a distinct impact in the overall transient
behaviour. $\ma Q_{c1}$ shows higher values across all conditions compared
to the other elements in $\ma Q_c$, which means that the controller is
trying to put more emphasis on the position state in order to reduce the
error with respect to the reference. However, comparing the distributions
for each ID in Fig.~\ref{fig:Qc_vs_seaborn}A, it can be seen that the
optimisation yields controllers with higher values of $\ma Q_{c1}$ for ID 2
in general, whereas the rest of the IDs have distributions that are similar
in shape. This seems to be related to the result shown in
Fig.~\ref{fig:control_params_vs_seaborn}B where ID 2 exhibits a larger
model mismatch $\ma A_p$. A plausible explanation is that the LQR method
tries to compensate for the mismatch by assigning a higher weight on the
position state, which in the end results in a higher controller gain. For
$\ma Q_{c2}$ in Fig.~\ref{fig:Qc_vs_seaborn}B, the distributions are
bimodal for all ID showing lower values in general compared to
$\ma Q_{c1}$. Similarly, the distribution of ID 2 presents more data
grouped towards the positive range of values, indicating that the
optimisation procedure penalised the velocity state more heavily. This is
also in agreement with the result for $\ma A_p$ in
Fig.~\ref{fig:control_params_vs_seaborn}B. Overall, the difference in
behaviour of ID 2 against the rest is evident, and the higher weights for
$\ma Q_{c1}$ and $\ma Q_{c2}$ reveal that as the ID increases, a higher
degree of precision is needed and therefore the controller becomes more
cautious by producing smaller gains $\kkk$.

\begin{figure*}[!htbp]
\centering
\includegraphics[width=1\linewidth]{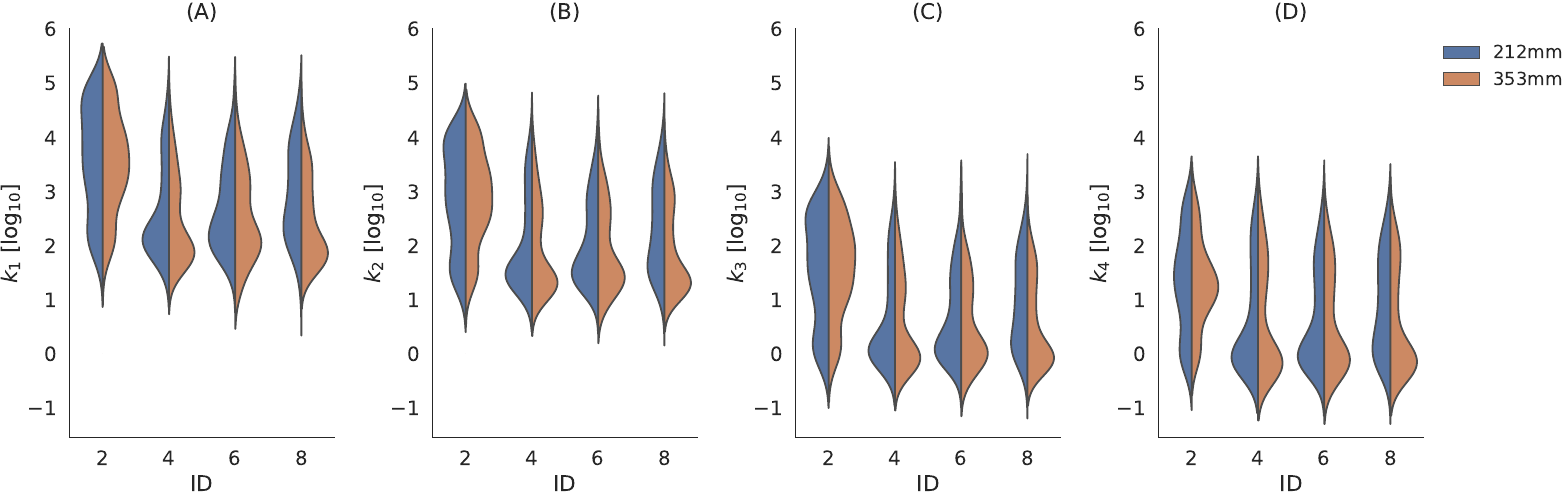}
\caption[Resulting controller gains $\kkk$ for each condition]{Resulting
  controller gains $\kkk$ for each condition. This figure shows the four
  elements of the gain vector $\kkk$ for all participants and segmented by
  ID (horizontal axis). The results for $\kkk_{1}$, $\kkk_{2}$, $\kkk_{3}$
  and $\kkk_{4}$ are shown in A, B, C, and D respectively. Each violin plot
  presents the results for the two distances between targets (left, in
  blue: 212 mm and right, in brown: 353 mm) side by side. The controller
  gains follow a similar trend compared to the design matrix $\ma Q_c$ from
  Fig.~\ref{fig:Qc_vs_seaborn}, with ID 2 showing some differences in shape
  against the rest of the conditions.}
 \label{fig:k_vs_seaborn}
\end{figure*}

As a result of the LQR method that involves the design matrix $\ma Q_c$, as
shown in (\ref{eq:lqr}), a set of controller gains $\kkk$ can be obtained
to implement the control law described in
(\ref{eq:control-law}). Fig.~\ref{fig:k_vs_seaborn} presents the values of
each element in $\kkk$ in a log scale, grouped by ID, and displaying side
by side the distribution for each target distance in the experiment. It can
be seen, by comparison with Fig.~\ref{fig:Qc_vs_seaborn}, that ID 2 still
shows higher values for all the elements in $\kkk$. However, the bimodal
distributions of $\ma Q_{c2}$ in Fig.~\ref{fig:Qc_vs_seaborn}B were
converted to a single low peak with long tails towards the higher values of
$\kkk$.

\subsection{Dynamic responses}
\begin{figure*}[!htbp]
\centering
\includegraphics{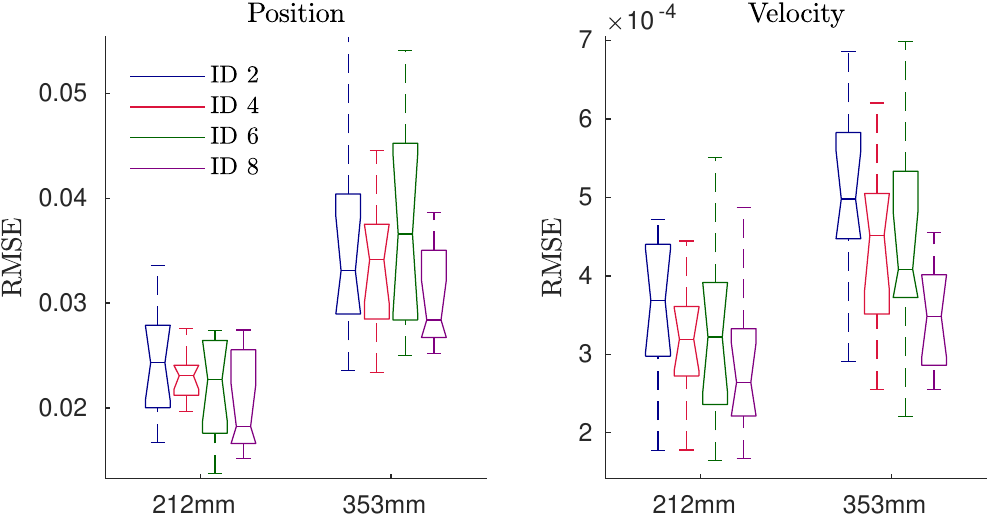}
\caption[RMSE for position and velocity.]{RMSE comparison for the position
  and the velocity profiles. The left and right figures show position and
  velocity RMSE respectively. The results in both cases are grouped
  according to the distance between left and right targets (212 mm and 353
  mm, horizontal axis) and also according to the ID of the task, where each
  box plot represents one of the four conditions of the experiment.}
\label{fig:rmse_vs}
\end{figure*}

The position and velocity RMSE is calculated to give a measure of
performance for the optimised controller in each condition according to
(\ref{eq:opt_cost}). The results across all participants are shown in
Fig.~\ref{fig:rmse_vs}. The left figure is the position RMSE for both
distances (212 and 353 mm) and the one on the right side corresponds to the
velocity RMSE, with the data being grouped according to the ID. In both
cases, the error when the distance between left and right targets is 353
mm, is larger compared to the errors registered for 212 mm. The position
error is larger in ID 2 for both distances compared to the rest of the
conditions, suggesting that the more highly variable behaviour in low
constraint cases is harder to model with this particular model. In terms of
velocity error, when the task is difficult (i.e. ID 8), the values are
smaller compared to the rest of the conditions. This suggests that since
more precision is needed due to the target width being reduced, a more
cautious regulation is required by the user, in terms of velocity. A
similar argument can be made for the position error. In general, the easier
conditions (ID 2 in particular) allow a greater degree of error because the
targets are wider.

\begin{figure}[!htbp]
\centering
\includegraphics[width=1\linewidth]{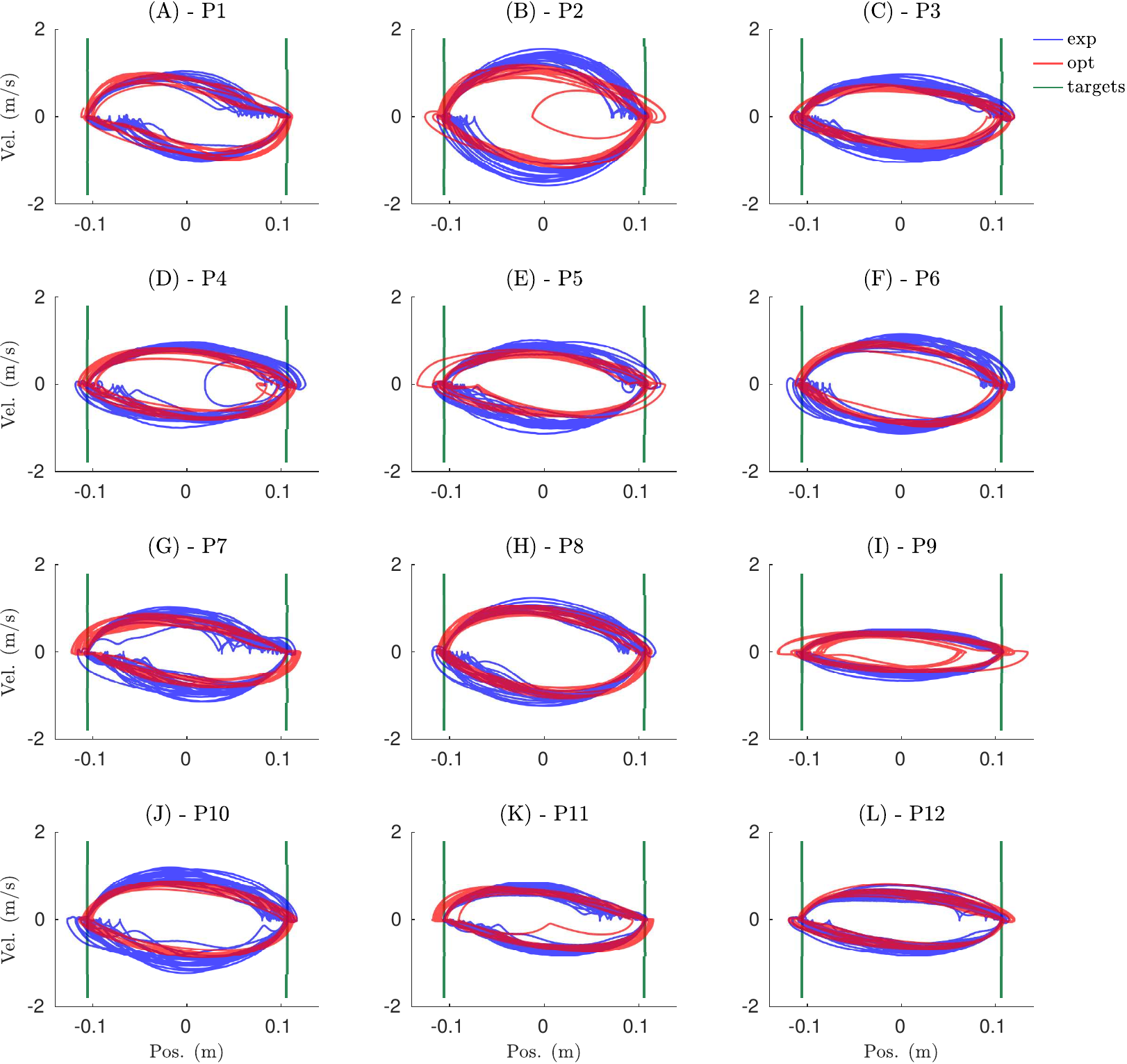}
\caption[Phase planes for all participants when the distance between
targets is 212 mm and the ID of the task is 8.]{Phase planes for all
  participants, a distance between targets of 212 mm and an ID of 8. The
  phase planes compare the pointer position (horizontal axis) against the
  pointer velocity (vertical axis).  The trajectories in blue correspond to
  the experimental data and the ones in red represent the resulting
  trajectories when the optimised IC is used. The location of the targets
  is shown by the vertical lines in green. The optimised IC trajectories
  capture more of the observed variability when compared with the second
  order lag model (2ol) responses shown in Fig.~\ref{fig:phase123} for this
  condition.}
\label{fig:phase_all_765_3}
\end{figure}

The following figures show the time-series, phase planes and Hooke plots
for participant 10, as well as some comparisons made for all
participants. Fig.~\ref{fig:phase_all_765_3} shows the phase planes for ID
8 when the distance between the left and right targets is the shortest (212
mm). This is the most difficult task since the targets are not only close
to each other but the target width is also small (0.83 mm). The human
response (in blue) tends to fall short of the target (vertical lines in
green) and subsequently approaches it via a series of small corrections, as
well as overshooting it. The controller switching strategy described
in~\ref{controller-switching} yields trajectories that capture the
variability in some regions of the phase planes and it is particularly good
at reproducing the trajectories that overshoot the target. This can be seen
more evidently in participants 3, 4, 8, 10, and 12. In
Fig.~\ref{fig:phase_all_765_3}H, the trajectories that start from the left
target depart from small area to then approach the right target, widening
and spreading as they get closer.

\begin{figure}[!htbp]
\centering
\includegraphics[width=1\linewidth]{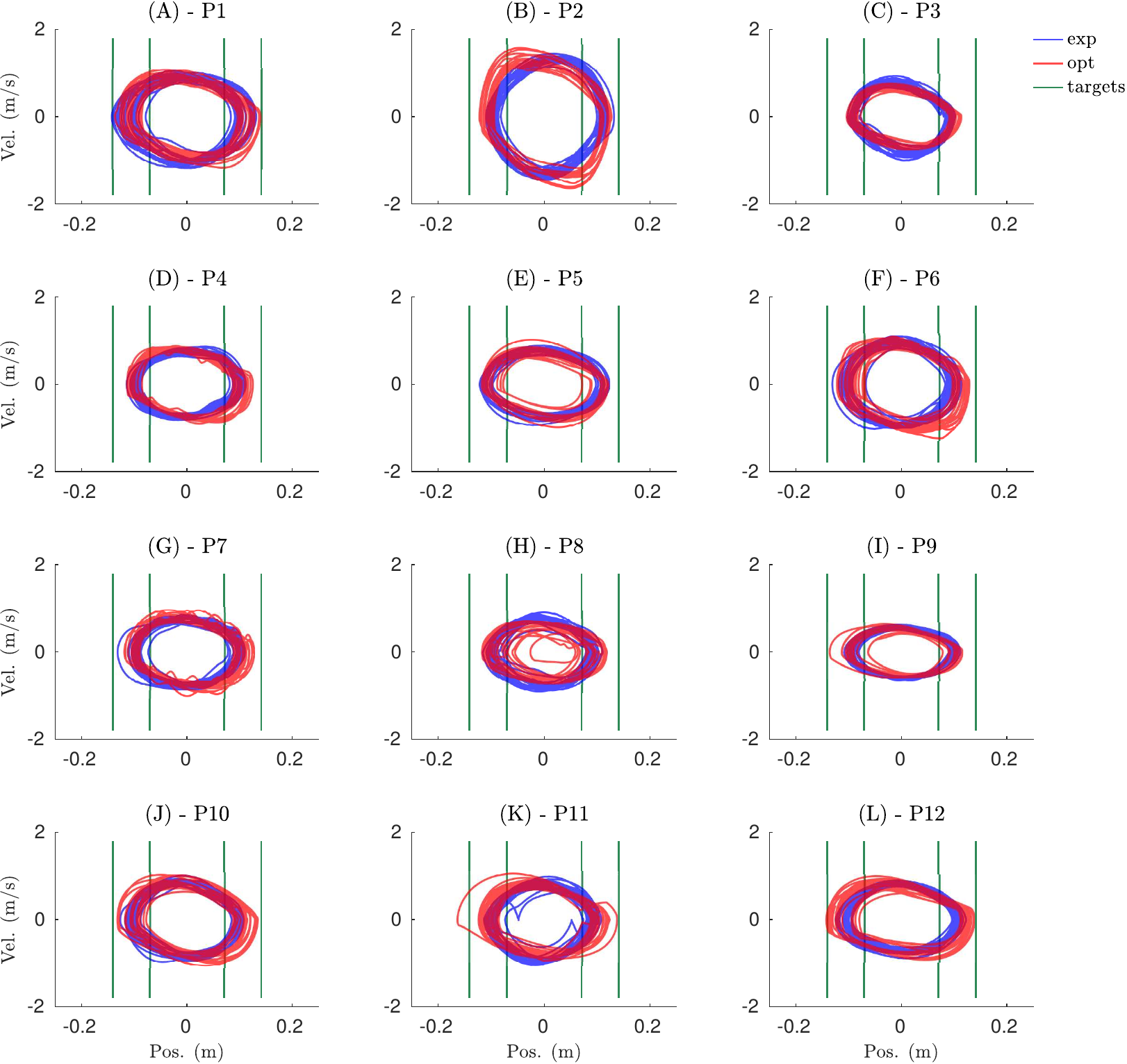}
\caption[Phase planes for all participants when the distance between
targets is 212 mm and the ID of the task is 2.]{ Phase planes for all
  participants, a distance between targets of 212 mm and an ID of 2. The
  phase planes compare the pointer position (horizontal axis) against the
  pointer velocity (vertical axis).  The trajectories in blue correspond to
  the experimental data and the ones in red represent the resulting
  trajectories when the optimised IC is used. The location of the targets
  is shown by the vertical lines in green. The optimised phase planes in
  red follow the shape of the experimental ones, displaying the variability
  of the trajectory.}
\label{fig:phase_all_765_255}
\end{figure}

In Fig.~\ref{fig:phase_all_765_255}, the phase planes for a distance of 212
mm are shown when the ID is 2 (target width is the largest). The response
of the controller is capable of capturing the overall behaviour and some of
its associated variability, as the shape of the response changes from a
sequence of corrections (Fig.~\ref{fig:phase_all_765_3}) to a more
continuous circular movement. In this case, since the difficulty is low,
the participants can afford a greater degree of error in the initial
portion of the movement leading to trajectories with greater variability
and with no correction as they land in the target area. The controller, in
most cases, exhibits trajectories that follow closely the ones generated by
each participant. ID 2 is the condition that is substantially different in
terms of behaviour than all the rest, as described in
Fig.~\ref{fig:Qc_vs_seaborn}A and Fig.~\ref{fig:Qc_vs_seaborn}B.

\begin{figure*}[!htbp]
\centering
\includegraphics[width=1\linewidth]{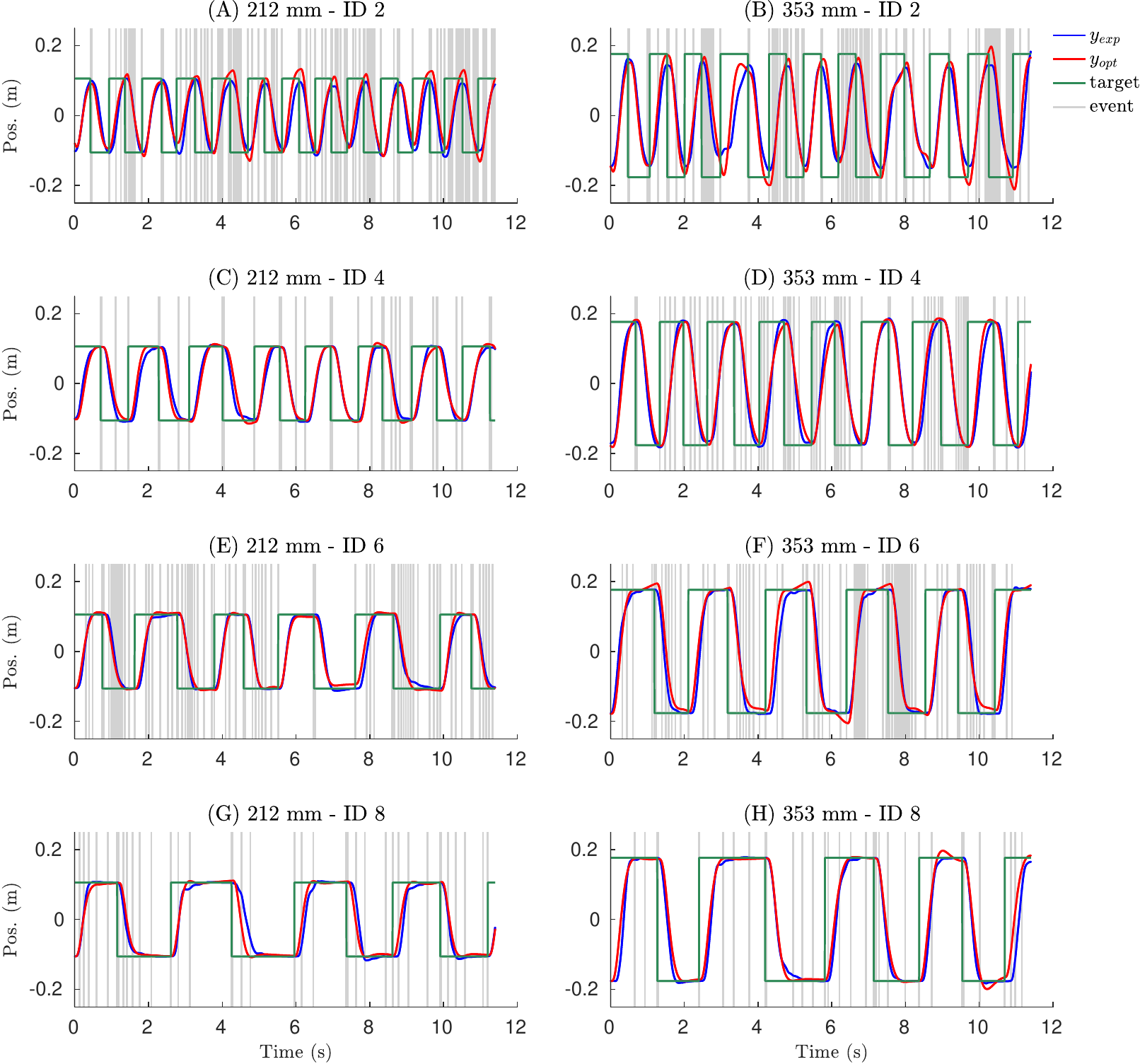}
\caption[Pointer position for participant 10 in all conditions.]{Pointer
  position for participant 10 in all conditions. Each row corresponds to a
  particular ID, starting with 2 for the top row and ending with 8 at the
  bottom. The column on the left shows the pointer position (vertical axis)
  for a distance between targets of 212 mm (A, C, E, G). The right column
  shows the same quantities for a distance of 353 mm (B, D, F, H). The
  horizontal axis shows around 11 seconds of each condition. The
  participant response $y_{exp}$ is shown in blue, while the trajectories
  generated by the IC are shown in red as $y_{opt}$. The target or
  reference signal is shown in green. The vertical lines in grey show the
  instances in time when an event was triggered by the optimised IC,
  signalling the use of feedback.  The simulated IC trajectories follow
  closely the ones generated by the participant; for each change in
  reference, subtle differences can be seen in the IC trajectories as a
  consequence of the controller switching approach explained in
  section~\ref{controller-switching}.}
\label{fig:Pos_P10}
\end{figure*}

In Fig.~\ref{fig:Pos_P10}, the time-series associated with the pointer
position of participant 10 is shown for the initial part of each trial,
where the position recorded from the experiment is in blue, the trajectory
generated by the IC is in red, and the target signal (reference) is shown
in green. Also, the moments in time when the IC generated an event to
trigger the use of feedback are represented by vertical lines in gray.

The left column in Fig.~\ref{fig:Pos_P10} shows data for a distance between
left and right targets of 212 mm (A, C, E, G), and the right column shows
it for 353 mm (B, D, F, H). Each row in the figure corresponds to a
different ID, starting from 2 at the top and ending with 8 at the
bottom. It is possible to see how in all conditions, both the experimental
and IC trajectories follow the reference closely. In particular, the
trajectory generated by the IC exhibits a small degree of error compared to
the experimental result, especially as the response approaches the
target. Also, for the most difficult conditions, such as ID 6 or 8 (E, F,
G, and H), the position follows a more traditional step response with a
small steady-state error. However, the top row shows how the response
becomes more similar to a sinusoidal signal due to the quick changes in the
target signal in green.

IC generates events when the comparison between the hold and the observed
states exceeds the specified threshold value as shown in
(\ref{eq:trig-prediction-error}). Due to the controller switching strategy
described in section~\ref{controller-switching}, a different threshold is
applied at every change in reference $w_c$ (Fig.~\ref{fig:slice-diagram}),
which creates different trigger patterns depending on the trial and the
model that was selected. This is evident in Fig.~\ref{fig:Pos_P10}B and
Fig.~\ref{fig:Pos_P10}G where the trials around 8 secs exhibit events that
are close in temporal proximity; however, when the next trial starts, the
open-loop intervals $\Delta_{ol}$ become larger due to the different
parameters being applied including a new threshold $q$ value, which results
in a slower triggering rate compared to the previous trial. During the time
between events, the IC is evolving in an open-loop configuration, using
only states that are generated internally by Eq.~(\ref{eq:u_openloop}).

\begin{figure*}[!htbp]
\centering
\includegraphics[width=1\linewidth]{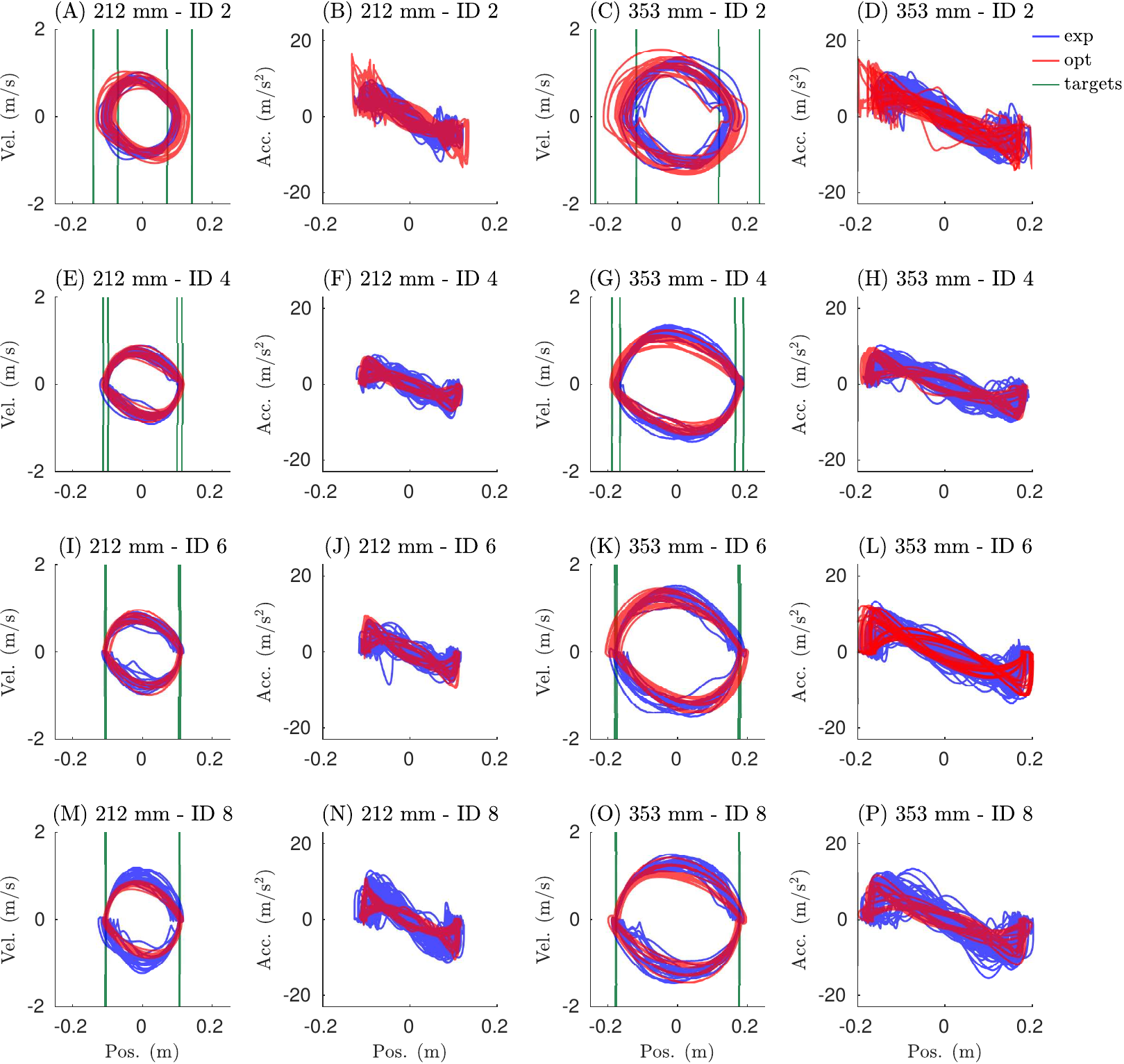}
\caption[Phase planes and Hooke plots for participant 10 in all
conditions.]{Phase planes and Hooke plots for participant 10 in all
  conditions. Each row corresponds to a particular ID, starting with 2 for
  the top row and ending with 8 at the bottom. The two columns on the left
  show the phase planes (pointer position vs velocity) and Hooke plots
  (pointer position vs acceleration) for a distance between targets of
  212 mm (A, B, E, F, I, J, M, N). The two columns on the right show the same
  quantities for a distance of 353 mm (C, D, G, H, K, L, O, P). The participant's
  response is shown in blue, while the trajectories generated by the IC are
  shown in red. The target or reference signal is shown in green using
  vertical lines.}
\label{fig:Phase_P10}
\end{figure*}

Fig.~\ref{fig:Phase_P10} presents the phase planes and Hooke plots of
participant 10 for all conditions. The first two
columns from left to right contain data for a distance between targets of
212, whereas the third and fourth columns represent a distance of 353
mm. Each row corresponds to a specific ID, starting from ID 2 in the top
row and ending with ID 8 at the bottom. In general, the phase planes follow
the overall behaviour of the human response (in blue); however, it is
possible to see that the IC response (in red) captures the variability seen
in the experimental result in all conditions, being slightly less accurate
as the difficulty increases.

The trajectories in Fig.~\ref{fig:Phase_P10}E and
Fig.~\ref{fig:Phase_P10}O, provide some insight into the type of control
applied by the participants in these conditions. Once the trajectory
approaches the target, the control policy that is used results in a
reduction of the pointer velocity accompanied by a sequence of small
corrections to reduce the error (more frequent in difficult conditions), or
in an error in the opposite direction in the form of overshoot (when the
response passes the intended target). For these two cases, the IC output
not only covers almost the entire range of possible trajectories generated
by this participant, but also reproduces the overshoot behaviour that was
previously mentioned. The Hooke plots (second and fourth columns) compare
how the pointer acceleration changes with time. In
Fig.~\ref{fig:Phase_P10}F and Fig.~\ref{fig:Phase_P10}P, the acceleration
trajectory generated by the IC is smooth (in red), and follows the
experimental result (blue) in terms of its overall shape.

\subsection{Open-loop intervals}

\begin{figure*}[!htbp]
\centering
\includegraphics{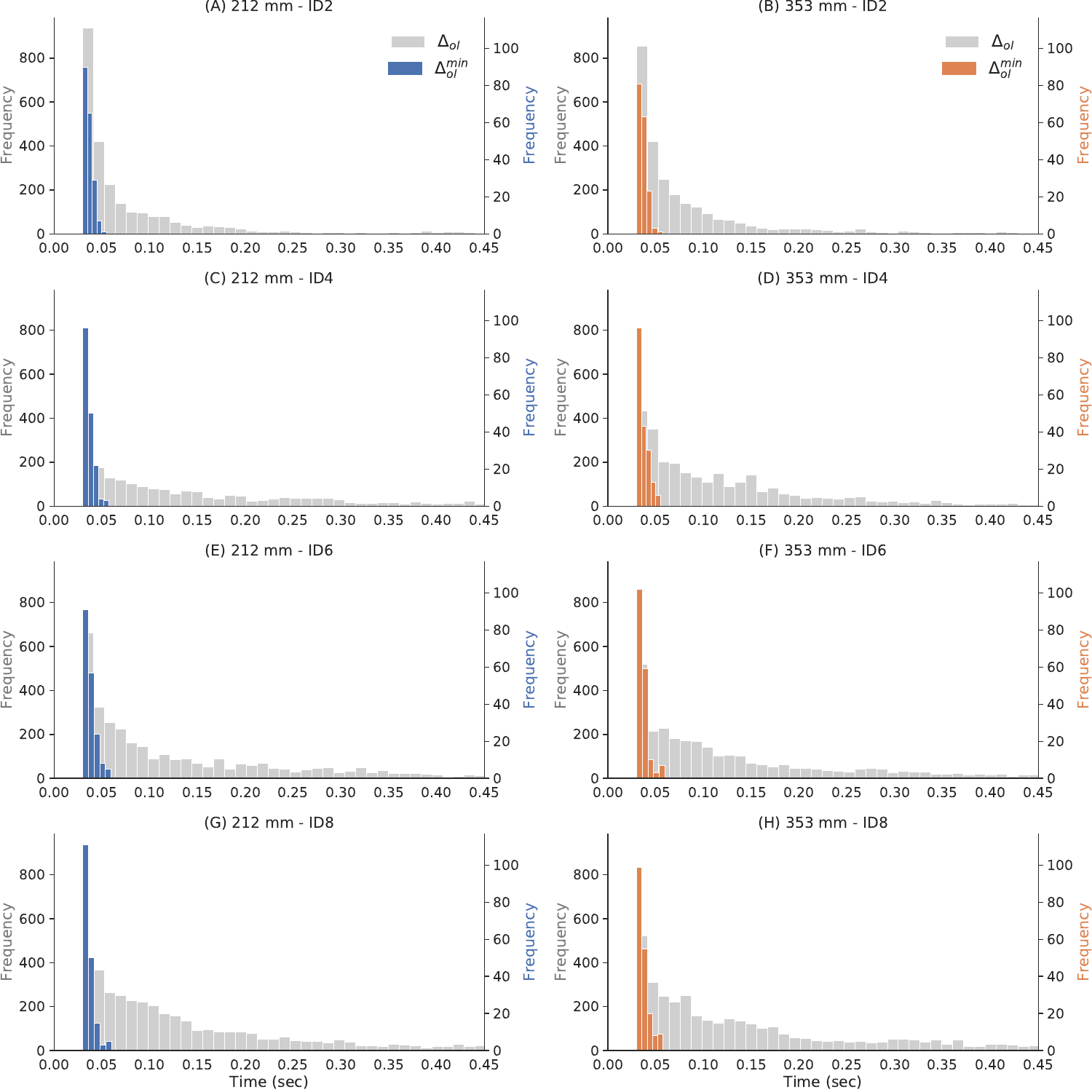}
\caption[Minimum open-loop intervals and the actual open-loop
intervals.]{Open-loop interval distributions for all participants. The
  actual open-loop interval $\Delta_{ol}$, generated by IC, is shown as a
  grey histogram summarising the data of all participants for a particular
  condition. Similarly, blue and brown histograms corresponding to the
  minimum open-loop interval $\Delta_{ol}^{\text{min}}$ are shown for a
  distance between targets of 212 mm and 353 mm respectively. The vertical
  axis corresponds to the frequency or bin count of a particular value of
  $\Delta_{ol}$ (left axis) and $\Delta_{ol}^{\text{min}}$ (right
  axis). The horizontal axis displays the time in seconds.}
\label{fig:hist_kde}
\end{figure*}

The histograms of the open-loop intervals $\Delta_{ol}$, generated by the
intermittent controller, is presented in Fig.~\ref{fig:hist_kde}, where all
the $\Delta_{ol}$ values for a specific condition where considered for all
the participants in the experiment.  These values are the actual open-loop
intervals recorded from simulation, which can be greater than the minimum
open-loop interval $\Delta_{ol}^{\text{min}}$ for a particular slice. The
minimum open-loop interval $\Delta_{ol}^{\text{min}}$ belongs to the set of
parameters used in the optimisation approach described in
section~\ref{sec:optimised-parameters}, with distributions shown in
Fig.~\ref{fig:control_params_vs_seaborn}. In Fig.~\ref{fig:hist_kde}, we
decided to overlap both $\Delta_{ol}$ and $\Delta_{ol}^{\text{min}}$ to
show the differences between the two and their relationship with specific
conditions of the experiment.

Fig.~\ref{fig:hist_kde}A, C, E, and G, represent the trials where the
distance between targets is 212 mm. Similarly, Fig.~\ref{fig:hist_kde}B, D,
F, and H, correspond to 353 mm. From top to bottom, each ID is shown as a
row, starting with ID 2 at the top (A, B) and ending in ID 8. The values of
$\Delta_{ol}$ were encapsulated as a histogram (in grey) which shows the
frequency of a particular value for a single condition or ID, and its scale
is shown on the left $y$ axis of each sub-plot. The values of
$\Delta_{ol}^{\text{min}}$ are shown as either blue or brown histograms,
corresponding to distances between targets of 212 mm and 353 mm
respectively.

The results in Fig.~\ref{fig:hist_kde} show that the majority of the
intervals are short in general, having higher frequencies at lower values
of $\Delta_{ol}$. Considering Fig.~\ref{fig:hist_kde}A and
Fig.~\ref{fig:hist_kde}B, it is clear that histograms in grey decay at a
faster rate compared to the rest of the conditions for both distances,
showing very few instances of $\Delta_{ol}$ beyond 0.25 sec. For ID 4, 6,
and 8, the tail of the histogram extends beyond the 0.25 sec consistently,
meaning that the open-loop intervals are longer in general compared to the
easiest condition of ID 2.

The small values of $\Delta_{ol}$ for ID 2 seem to be caused by the effects
of the identified threshold $q$ and the mismatch gain $\AAA_p$ (shown in
Fig.~\ref{fig:control_params_vs_seaborn}) and the fact that the IC model
had more trouble fitting the response for this particular condition in
terms of the target position and velocity, which is displayed in
Fig.~\ref{fig:rmse_vs}. The RMSE for ID 2 in both distances is larger that
than the corresponding error in the other conditions, probably as a result
of the model trying to capture the variability observed in these trials;
however, the poor fit results in low threshold $q$ values which indicates
that prediction errors would be addressed by using feedback more often. If
the threshold $q$ is low, the triggering patterns of IC would most likely
become close to the imposed minimum open-loop interval
$\Delta_{ol}^{\text{min}}$, and in some cases triggering would be as fast
as the minimum open-loop interval for that particular model would allow.

This behaviour is also affected by the mismatch gain $\AAA_p$. Having large
values of $\AAA_p$ means that the control input that is applied to the
system is different from the input that would generate the desired
performance. Since the input is affected, the states of the system might
not reach the specified targets and this would eventually lead to higher
triggering rates.

The minimum open-loop intervals $\Delta_{ol}^{\text{min}}$, for both
distances, have similar histogram distributions that extend to 0.05 sec
approximately.  However, the most difficult conditions, i.e., ID 6 and 8,
seem to have slightly more values of $\Delta_{ol}^{\text{min}}$ around 0.05
than the rest of the conditions. Fig.~\ref{fig:hist_kde} provides evidence
of the interplay between the threshold $q$ and $\Delta_{ol}^{\text{min}}$
as well as for the event-driven nature of IC, since even when IC could
trigger as fast as $\Delta_{ol}^{\text{min}}$ all the time (blue and brown
histograms), it does it only after the threshold has been exceeded, leading
to the longer open-loop intervals shown by the grey histograms.

\section{Densities based on repeated simulations}
\label{sec:densities}
Using the controller switching strategy from
section~\ref{controller-switching}, multiple simulations were carried out
using the optimised parameters and the associated models in order to create
a probability distribution of the IC simulation results, in the phase
space.  The most basic non-parametric approach to visualise probability
densities is simply to binning the data into discrete blocks to create 2D
and 3D histograms to expose the variability observed in the data.  The
availability of such probability density estimates would allow us to
predict the likelihood of a given trajectory that a specific participant
might take for a particular task. In addition to simple histograms, we can
estimate continuous density functions. In this case, the densities are
estimated by {\it Kernel Density estimates (KDE)}. KDE is a non-parametric
approach to estimate the probability density function of a random variable,
and is essentially a data smoothing problem where continuous inferences
about the population are made, based on a finite data sample
\citep{rosenblatt1956remarks,parzen1962estimation}.  To do this, the first
step was to create densities based on the recorded time-series of all the
simulations (200 simulations in total), in particular the pointer position
and its velocity. With this information, a phase-plane density was created.

\begin{figure*}[!htbp]
  \centering
  \includegraphics[width=1\linewidth]{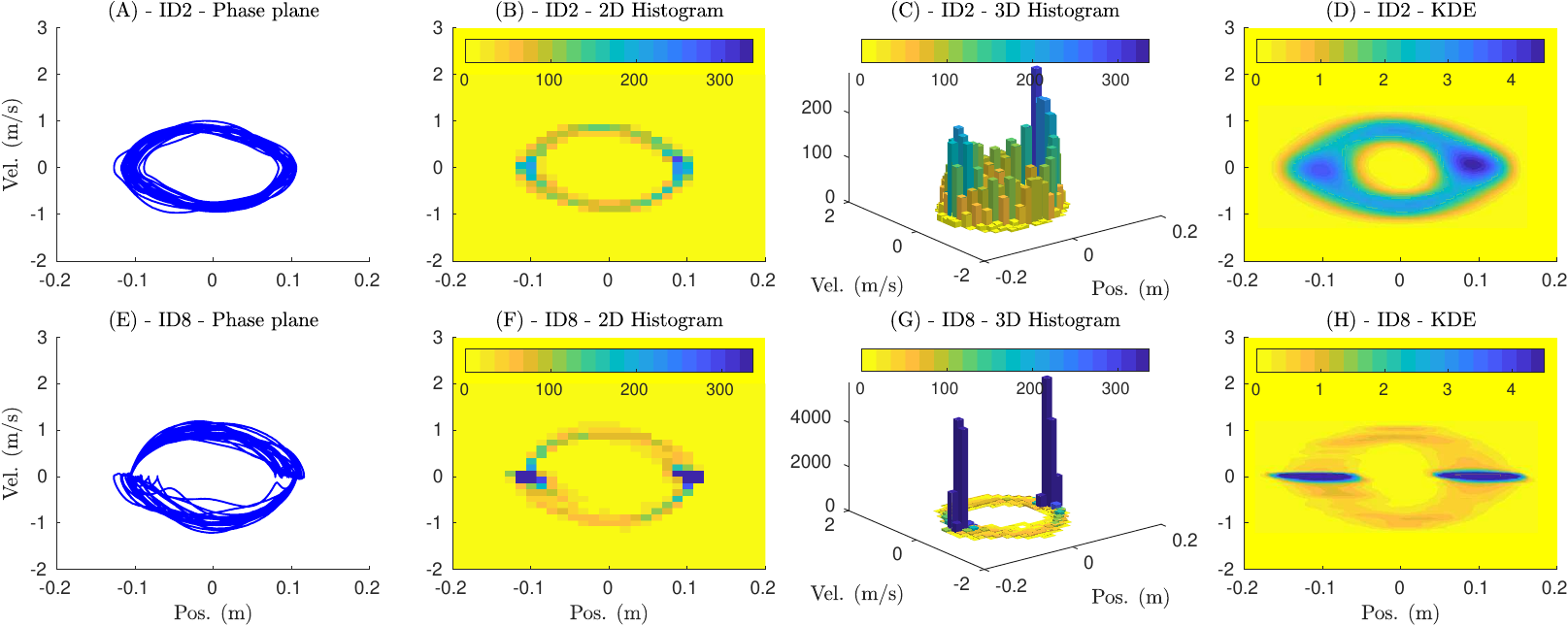}
  \caption[P10 - densities - experimental - 765 - ID2 and ID8]{Experimental
    phase planes, histograms and densities for participant 10 and a
    distance of 212 mm. (A) shows the recorded phase plane trajectory. (B)
    is a 2D histogram representation of the phase plane, generated with a
    grid of 30 by 30 bins. The scale represents the number of data points
    in each bin. (C) is a 3D histogram where the height of the bars
    corresponds to bin count, as in the 2D version. (D) is the kernel
    density estimate of the probability density of the corresponding
    data. The top row (A, B, C, D) shows data for ID 2, whereas ID 8 is
    displayed at the bottom (E, F, G, H). The horizontal and vertical axes
    correspond to pointer position and velocity respectively. For C and G,
    these two axes appear now in the horizontal plane and the height of the
    bins in the vertical one.}
  \label{fig:P10_summary_density_exp}
\end{figure*}

To establish a comparison, histograms and KDE densities generated from the
experimental data from Participant 10 are presented first, followed by
similar visualisations generated from the IC models that were
identified. In Fig.~\ref{fig:P10_summary_density_exp}, a summary of
different visualisations of the experimental data is shown, for ID 2 and 8,
when a distance between targets is 212 mm. The phase plane is shown for
reference (Fig.~\ref{fig:P10_summary_density_exp}A, E), followed by a 2D
histogram representation that was generated based on the phase plane
time-series (Fig.~\ref{fig:P10_summary_density_exp}B, F). These
heatmap-like figures in 2D can be extended to obtain a 3D representation as
shown in Fig.~\ref{fig:P10_summary_density_exp}C and
Fig.~\ref{fig:P10_summary_density_exp}G, where the height of each histogram
bar represents the bin count in a $30 \times 30$ grid. It is possible to
observe how, as the phase plane trajectory approaches the targets, the
count increases showing areas and bars in a dark blue colour. This is a
clear indication of the participant slowing down to land on the target
which results in more data points around these areas. Finally, in
Fig.~\ref{fig:P10_summary_density_exp}D and
Fig.~\ref{fig:P10_summary_density_exp}H, a 2D kernel density estimate is
generated to have a probabilistic representation of the data.

\begin{figure*}[!htbp]
  \centering
  \includegraphics[width=1\linewidth]{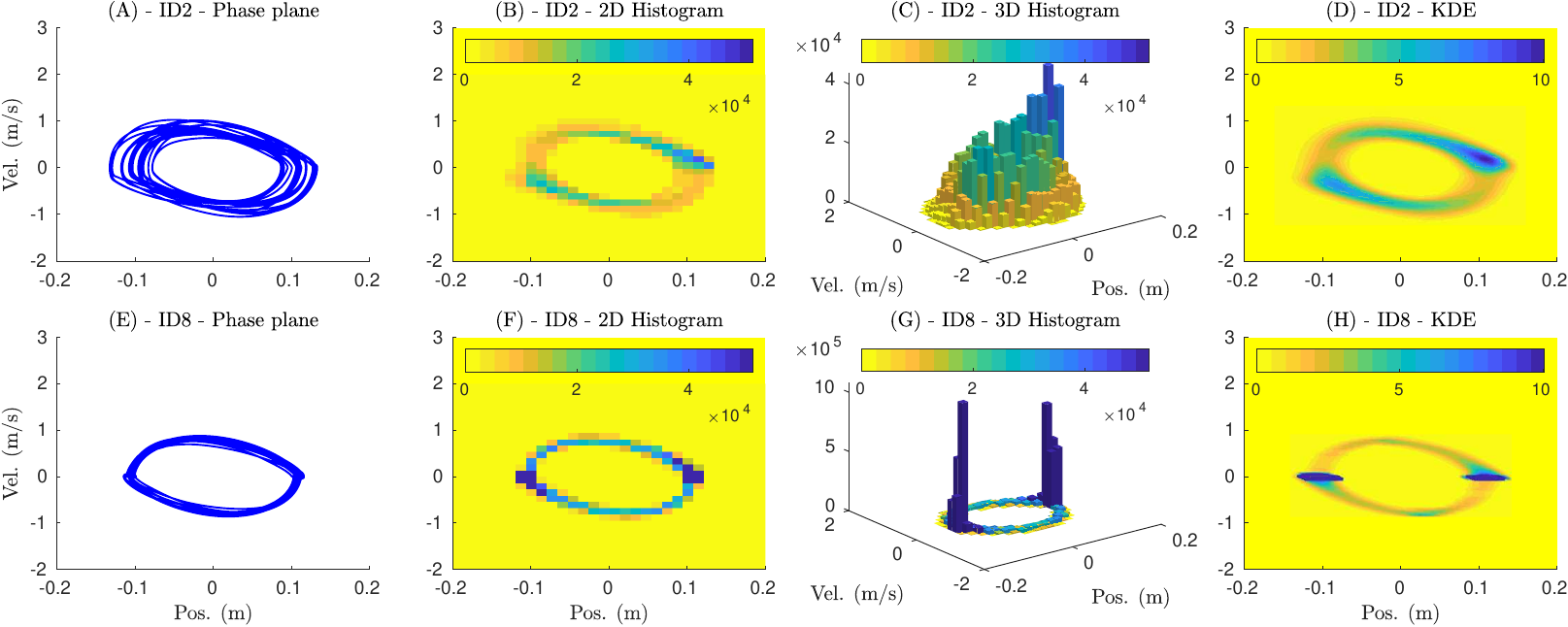}
  \caption[P10 - densities - simulation - 765 - ID2 and ID8]{Simulated
    phase planes, histograms and densities for participant 10 and a
    distance of 212 mm. A total of 200 simulations, using the multiple
    controller approach, were used to generate the same type of
    visualisations as in Fig.~\ref{fig:P10_summary_density_exp}. The
    simulated densities have similar shapes to the experimental ones in
    Fig.~\ref{fig:P10_summary_density_exp}, showing also higher higher
    density values around the targets.}
  \label{fig:P10_summary_density_sim}
\end{figure*}

\begin{figure*}[!htbp]
  \centering
  \includegraphics{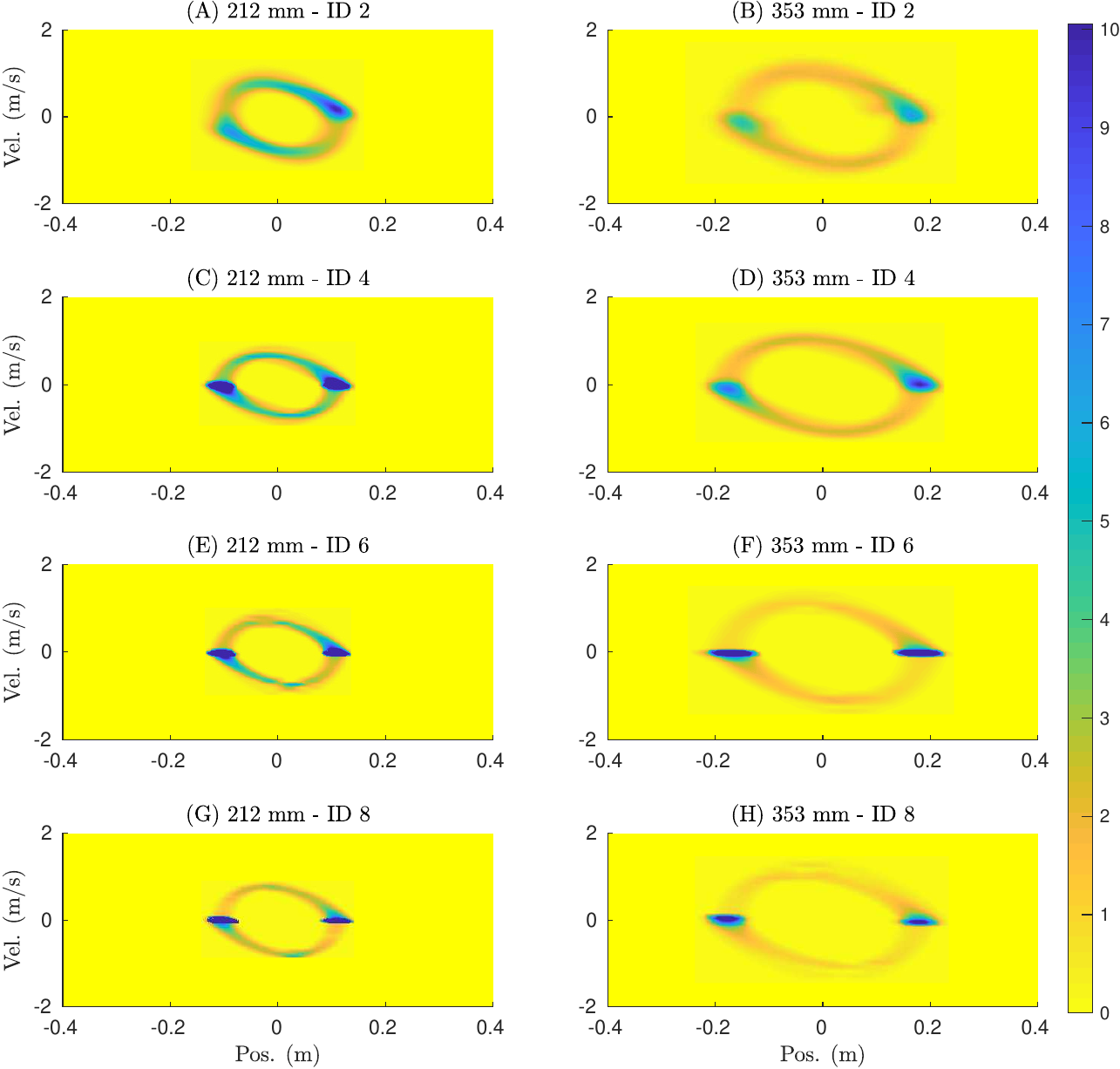}
  \caption[P10 - densities - simulation]{Simulated kernel density estimates
    for participant 10 in all conditions and distances. (A), (C), (E) and
    (G) correspond to a distance of 212 mm whereas (B), (D), (F), and (H)
    are for 353 mm. The different conditions are represented by each row of
    plots, with ID 2 shown at the top and finishing with ID 8 at the
    bottom. The horizontal and vertical axes in the figure represent
    pointer position and velocity respectively. The simulated KDEs have
    lower values for ID 2 in both distances (A and B) since the participant
    uses a more cautious approach to hit the target, slowing down
    considerably, which leads to more data points in these regions.}
  \label{fig:P10_density_sim}
\end{figure*}

The identified IC models were used to run 200 simulations using the
multiple controller approach to build a density representation. First we
show a summary with the corresponding phase plane, 2D-3D histograms and
densities for ID 2 and ID 8 for a distance of 212 mm in
Fig.~\ref{fig:P10_summary_density_sim}, which is comparable to
Fig.~\ref{fig:P10_summary_density_exp}, followed by the simulated densities
for all distances and conditions (Fig.~\ref{fig:P10_density_sim}).

The simulated densities show a similar shape to the experimental
counterparts, with the variability also present for low ID values such as 2
(A, B). Similarly, the difficult conditions (G and H) also show high
density values around the targets showing how the IC model trajectories
also spend more time in the vicinity of the target before accelerating in
the opposite direction.

\subsection{Kullback-Leibler divergence for repeated simulations}
\label{subsec:kullback-leibler}
The information from the repeated simulations can be used to calculate a
distance measure that evaluates how much of the observed experimental
distributions is captured by the simulated realisations of the phase
planes. To measure the similarity between these, we used a Kullback-Leibler
(KL) divergence measure (also called relative entropy). The KL divergence
is a measure of how one probability distribution differs from a second,
reference probability distribution. For discrete probability distributions
$P$ and $Q$ defined on the same probability space, ${\mathcal {X}}$, the KL
divergence from $Q$ to $P$ is
\begin{equation}
    D_{KL} (P || Q) = \int p(x) \log \left( \frac{p(x)}{q(x)} \right)  dx,
\end{equation}
where $p$ and $q$ denote the probability densities of $P$ and $Q$. We took
the approach described in
\citep{wangDivergenceEstimationMultidimensional2009}, using the associated
software\footnote{\url{https://github.com/slaypni/universal-divergence}} to
quantify the information loss when compared against the experimental
data. We also quantified the loss if the continuous second order lag
controller (2ol) \citep{MulOulMur17} is used to generate an individual set
of simulated trajectories.

\begin{figure*}[!htbp]
  \centering
  \includegraphics[width=1\linewidth]{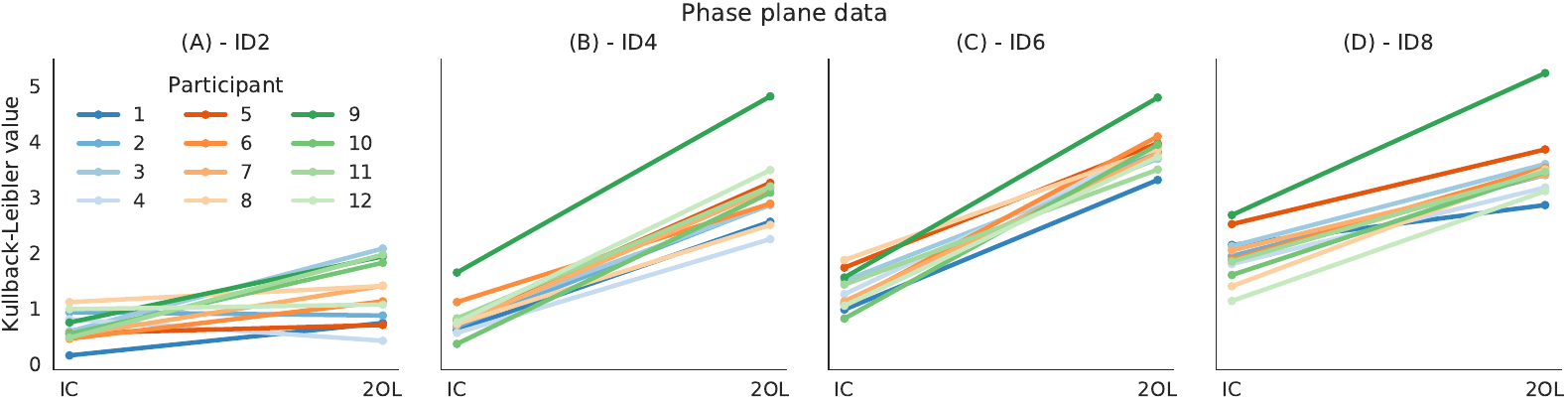}
  \caption[Kullback-Leibler divergence]{Kullback-Leibler divergence values
    of IC and 2ol controllers. (A), (B), (C), and (D) show the KL value, in
    the vertical axis, for all participants in conditions ID 2, ID 4, ID 6,
    and ID 8, respectively. Each line corresponds to a participant, where
    the left end of the line is the value for IC and the right end point
    represents the value of the 2ol controller. For IC, individual KL
    values were registered from 20 different simulation runs, from which a
    mean value was calculated. The mean is reported in this figure. The
    positive slopes of most of the lines indicate that IC provides a better
    fit to the observed distribution of the experimental data.}
  \label{fig:kl-values}
\end{figure*}

The KL divergence values were computed for 20 simulations of the IC, for a
distance between targets of 212 mm and for all participants and
conditions. To obtain each value, two vectors of samples coming from the
simulated time-series of the pointer position and velocity, i.e., the phase
plane information, were compared against the corresponding experimental
time-series. The reported values for IC represent the mean of the 20 KL
values of each simulation run. The mean of the resulting 20 KL values for
each participant was computed. This is the reported value IC in
Fig.~\ref{fig:kl-values}, which shows a comparison of the mean KL values
over 20 simulation runs of IC and a simulation of the 2ol continuous
controller. Each subplot corresponds to a specific ID, starting with ID 2
on the left side (Fig.~\ref{fig:kl-values}A), ending with ID 8 on the
right. The subplots show the KL values on the vertical axis and the two
controllers are represented on the horizontal axis (IC on the left, 2ol on
the right). The slope of each line (one per participant), shows how
different the KL value is for a specific condition. A larger number means
that the information loss is greater, therefore a small number indicates
that the distribution from the simulation is closer to the observed
distribution of the experimental data.

\begin{figure}[htbp!]
  \centering
  \includegraphics[width=0.9\linewidth]{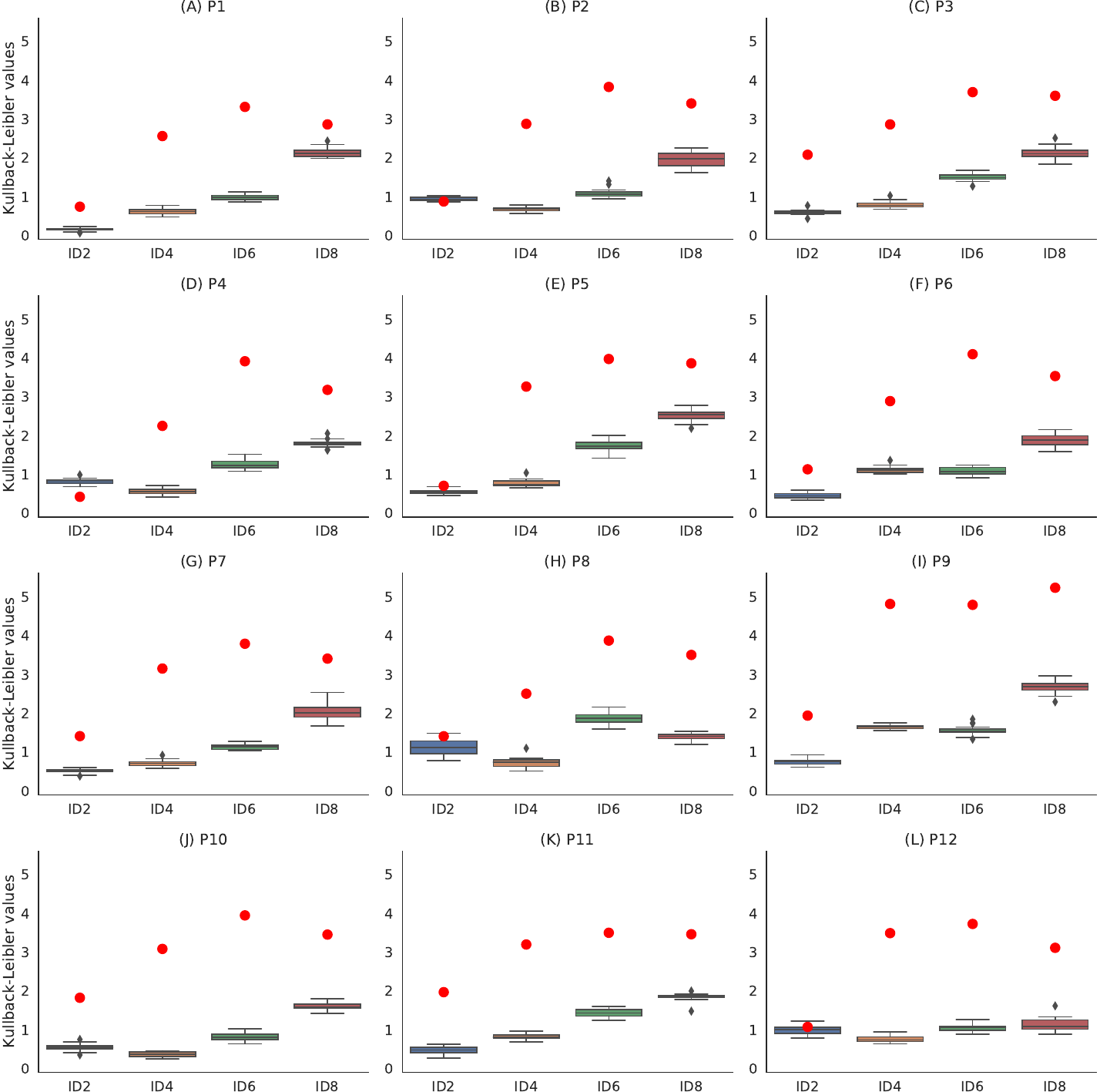}
  \caption[Boxplots of the KL divergence of each user and
  condition.]{Boxplots of the Kullback-Leibler divergence values for all
    participants and conditions. The results for each participant are
    grouped according to the ID of the task (horizontal axis). Overall, the
    easier tasks result in lower KL values for all participants. Red
    circles associated with each boxplot represent the value of the 2OL
    result for a particular condition. If the red dot is above the upper
    whiskers of the boxplot, this means that it is above the distribution
    of the IC results, indicating the significance of the difference
    between the conditions.}
  \label{fig:kl_boxplots_20iter_10sub_with_cc}
\end{figure}

Most lines in Fig.~\ref{fig:kl-values} have a positive slope except for two
participants (P2 and P4) in ID 2 (Fig.~\ref{fig:kl-values}A). To gain more
insight into the KL-divergence measure, view the phase plane responses for
these participants in Fig.~\ref{fig:phase_all_765_255}, which does show
repeated divergence in the model behaviour. The positive slopes show the KL
divergence for IC are typically lower than for the 2ol controller,
indicating a better fit to the distribution. Overall, the divergences for
both controllers in ID 2 are smaller compared to the rest of the conditions
(Fig.~\ref{fig:kl-values}B, C, D) in part because although ID 2 results are
{\it more} variable than others, the variability itself is relatively
consistent, making it easier to model.\footnote{Table~\ref{tab:kl-values}
  in the supplementary material contains all the values that were used to
  obtain Fig.~\ref{fig:kl-values}.} The variability of individual
participants can be seen in
Fig.~\ref{fig:kl_boxplots_20iter_10sub_with_cc}, again compare the size and
variability of divergence with the phase portrait plots for the same
participants in Fig.~\ref{fig:phase_all_765_255}. The boxplots give a clear
indication that overall the KL divergence tends to be lower for lower IDs,
and that more difficult tasks have a higher KL divergence, and a larger
variance in that divergence. Overall, the easier tasks result in lower KL
values for IC than 2ol for all participants, apart from ID 2 for P2, P4,
P5, P8 and P12.

\section{Low-dimensional visualisations of model structure}\label{sec:umap}

While the models may appear complex at first sight, it is possible to
explore a simplified model space in two ways: 1. experimentation with
optimisation of a reduced subset of model parameters, and 2. use of
low-dimensional visualisation techniques to map the full parameter space to
one or two dimensions to investigate the range of model behaviour for the
different tasks. We combine both approaches in this section.

\subsection{Optimisation of a reduced parameter set}
Analysis of the variation in model parameters suggested that we could
potentially fix some of the values for all users, and restrict the
identification process to a subset of parameters. This helps us understand
the key factors in the resulting models. We therefore repeated the
modelling process described earlier, but restricted the optimisation
process to a reduced parameter set, where only $\ma Q_{c1}$, $\ma Q_{c2}$,
$q$, and $\ma A_p$ were optimised, leaving out the minimum-open loop
interval $\Delta_{ol}^{\text{min}}$, the observer gain $Q_o$ and the rest
of the diagonal entries in $\ma Q_c$. The parameters that were left out
from the optimisation took the following fixed values: $\ma Q_o = 10$,
$\ma Q_{c3} = \ma Q_{c4} = 1$, $\Delta_{ol}^{\text{min}} = 0.3$ sec.

\begin{figure*}[!htbp]
\centering
\includegraphics[width=1\linewidth]{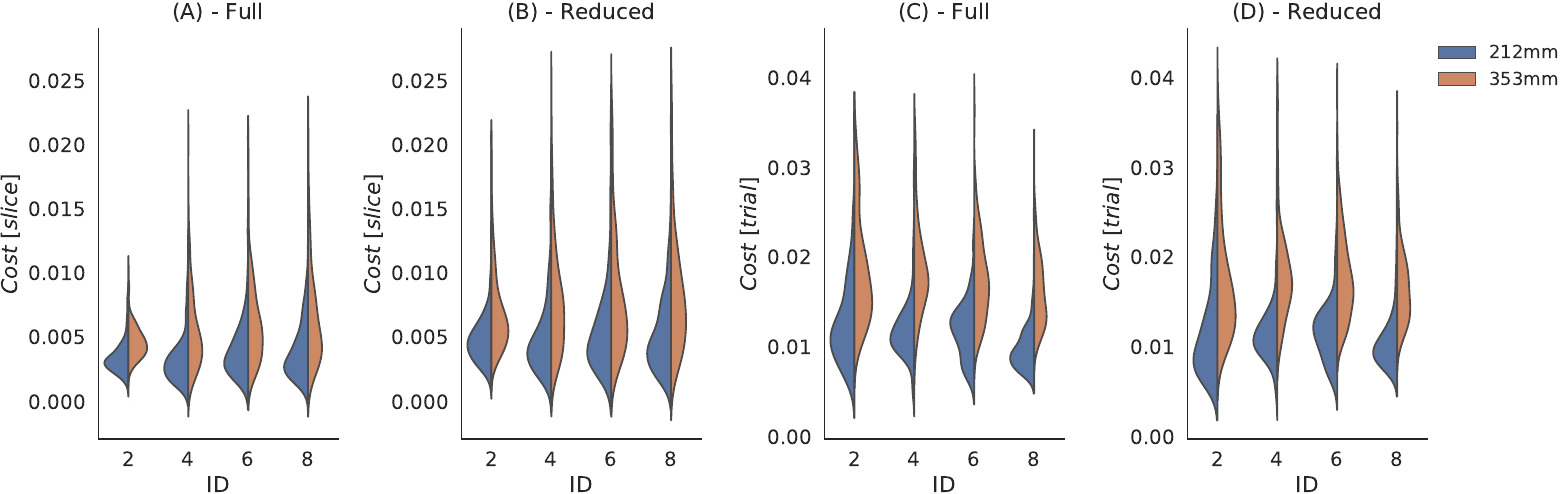}
\caption[Full vs. Reduced set of parameters comparison]{Slice and trial
  cost for the reduced and full parameter sets in each condition. (A) The
  cost per slice for the full parameter set, (B) cost per slice for the
  reduced parameter set, (C) cost over all trials for the full parameter
  set, and (D) cost over all trials for the reduced parameter set, are
  shown as violin plots for all participants and categorised by ID on the
  horizontal axis. The results are also grouped according to the two values
  of distance between targets used in the experiment (left: 212 mm and
  right: 353 mm).}
\label{fig:costs_vs_combined_full_vs_min}
\end{figure*}

To measure how well the optimisation using a reduced parameter set would
compare to the full set optimisation in terms of quality of fit, we decided
to evaluate the cost function in (\ref{eq:opt_cost}) for each individual
slice. In addition to this, we evaluated the cost function when the
controller obtained from each of the slices was simulated against all
trials of the condition. Fig.~\ref{fig:costs_vs_combined_full_vs_min} shows
the results of this comparison, where
Fig.~\ref{fig:costs_vs_combined_full_vs_min}A and B show the cost per slice
for all participants for the full and reduced parameter sets respectively;
similarly, Fig.~\ref{fig:costs_vs_combined_full_vs_min}C and D show the
cost over all trials for the two sets.

From Fig.~\ref{fig:costs_vs_combined_full_vs_min}A and B, we can see that
the reduced parameter optimisation does not represent too much of a trade
off in quality of fit, since the cost per slice values are very similar
with the most noticeable difference in ID 2, where the reduced set shows
slightly higher cost values. The cost over the entire trials
(Fig.~\ref{fig:costs_vs_combined_full_vs_min}C, D) for the two sets is very
similar in both value and shape of the distributions; however, when
compared to the cost per slice, we can see that it is generally higher
across all conditions. This is an expected consequence of simulating a
single controller against all trials in a condition and the main reason to
use the switching control strategy introduced in
section~\ref{controller-switching}.

\begin{figure*}[!htbp]
\centering
\includegraphics[width=1\linewidth]{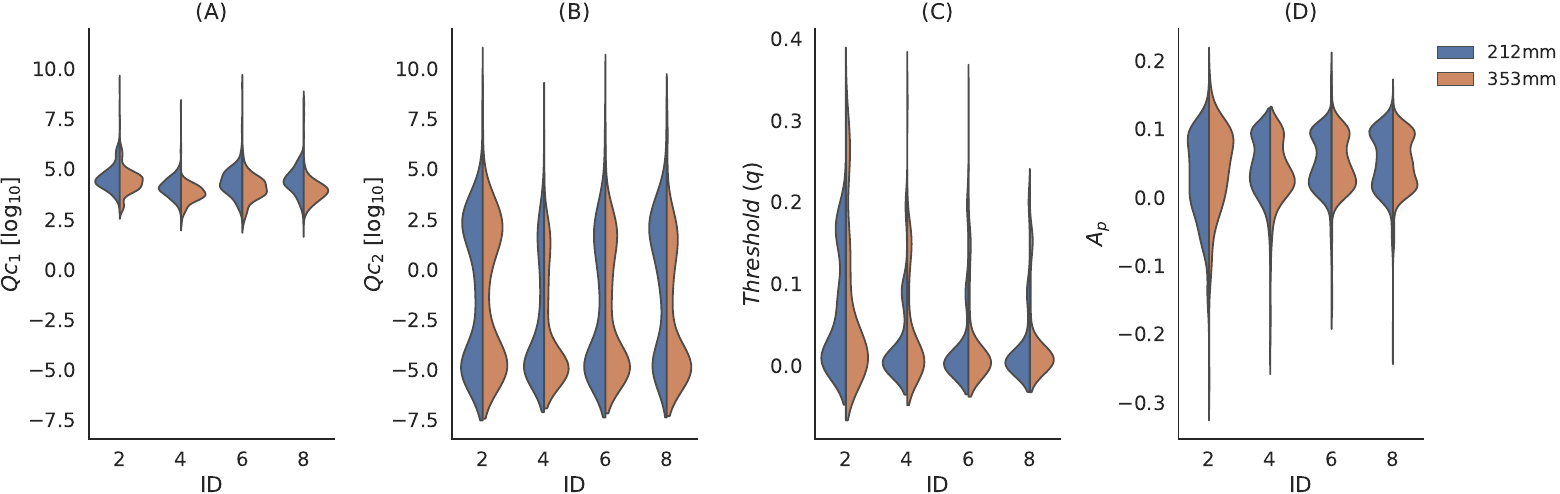}
\caption[Reduced set of optimised controller parameters for each
condition]{Optimised controller parameters (for a reduced set) for each
  condition. (A) $\ma Q_{c1}$, (B) $\ma Q_{c2}$, (C) the threshold $q$, and
  (D) the mismatch gain $\ma A_p$ are shown as violin plots including data
  of all participants and categorised by ID (horizontal axis). The results
  are also grouped according to the two values of distance between targets
  used in the experiment (left: 212 mm and right: 353 mm).}
 \label{fig:control_params_vs_seaborn_combined_min}
\end{figure*}

The ability to successfully represent the data with only 4 parameters
identified from the experimental data, may help reassure readers that the
models can capture general properties of pointing behaviours, even in
variable conditions such as ID 2, and is not `overfitting' the experimental
data. The distributions of these 4 parameters are shown in
Fig.~\ref{fig:control_params_vs_seaborn_combined_min}.

In Fig.~\ref{fig:control_params_vs_seaborn_combined_min}A, we can see how
$\ma Q_{c1}$ has more compact distributions compared to the same figure for
the full set of parameters (Fig.~\ref{fig:Qc_vs_seaborn}A), whereas
$\ma Q_{c2}$ in Fig.~\ref{fig:control_params_vs_seaborn_combined_min}B
still shows elongated bi-modal distributions that are comparable in shape
to those in Fig.~\ref{fig:Qc_vs_seaborn}B. The threshold $q$ and the
mismatch gain $\ma A_p$ are shown in
Fig.~\ref{fig:control_params_vs_seaborn_combined_min}C and
Fig.~\ref{fig:control_params_vs_seaborn_combined_min}D respectively. The
mismatch gain $\ma A_p$ in the reduced parameter set has now a longer
distribution that spans from -0.1 to 0.1 for ID 2 when compared to its
distribution in the full parameter set
(Fig.~\ref{fig:control_params_vs_seaborn}B) which is more compact and
centered around 0.1.

\begin{figure*}[!htbp]
\centering
\includegraphics[width=1\linewidth]{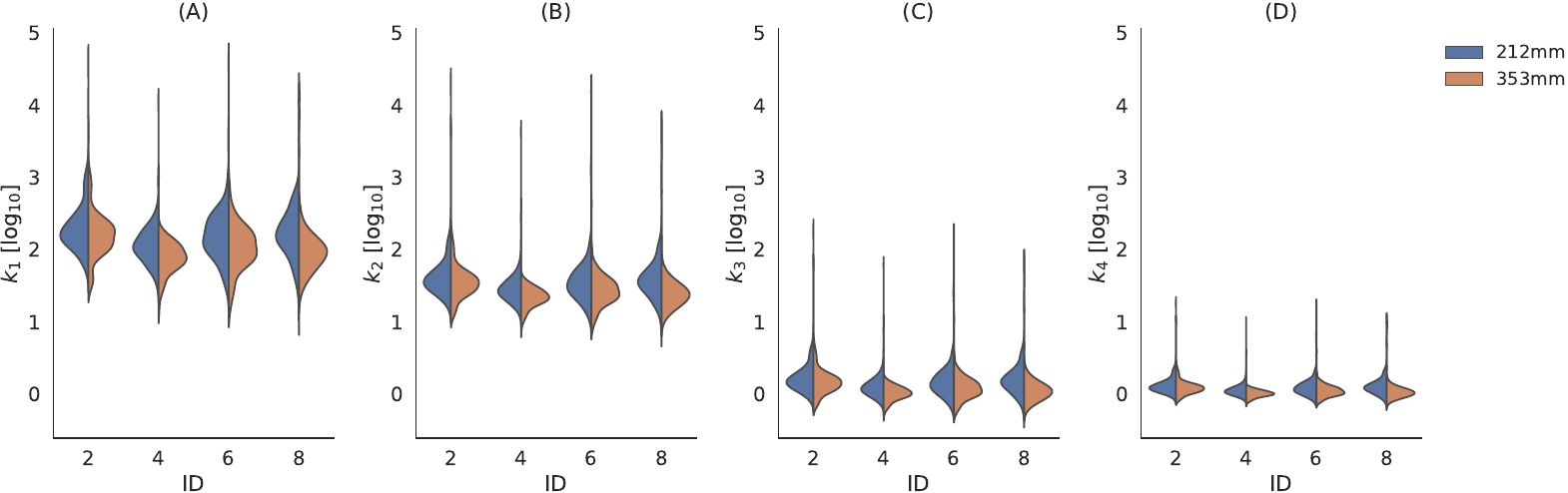}
\caption[Resulting controller gains $\kkk$ for each condition]{Resulting
  controller gains $\kkk$ (for a reduced set) for each condition. This
  figure shows the four elements of the gain vector $\kkk$ for all
  participants and segmented by ID (horizontal axis). The results for
  $\kkk_{1}$, $\kkk_{2}$, $\kkk_{3}$ and $\kkk_{4}$ are shown from left to
  right in A, B, C and D respectively. Each violin plot introduces the
  results for the two distances between targets (left, in blue: 212 mm and
  right, in brown: 353 mm) side by side.}
 \label{fig:k_vs_seaborn_min}
\end{figure*}

The impact of the optimised $\ma Q_c$ values is on the gain parameters,
therefore in Fig.~\ref{fig:k_vs_seaborn_min} we show all elements of the
state-feedback gain vector $\kkk$ which is computed via LQR using the
matrix $\ma Q_c$ as a design parameter.

The distributions of the state-feedback gain vector $\kkk$ for the reduced
parameter set, shown from left to right in Fig.~\ref{fig:k_vs_seaborn_min},
are more compact when compared to the corresponding gains of the full
parameter set optimisation in Fig.~\ref{fig:k_vs_seaborn}; on the other
hand, the overall trend is similar for both parameter sets where $\kkk_1$
and $\kkk_2$ are both higher than $\kkk_3$ and $\kkk_4$, and $\kkk_2$ being
lower in general than $\kkk_1$.

\subsection{Visualisation of low-dimensional embeddings with UMAP}
\begin{figure*}[!htbp]
\centering
\includegraphics[width=1\linewidth]{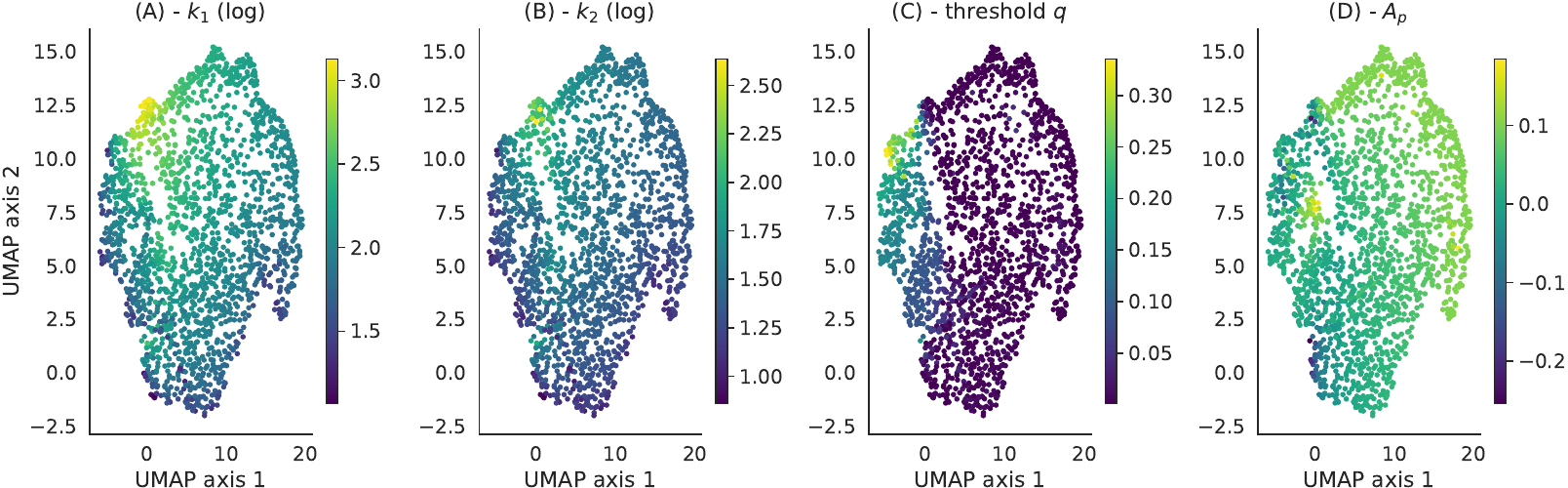}
\caption[UMAP 2D embedding]{Uniform Manifold Approximation using a 2D
  embedding, coloured using the value of each parameter. From left to right
  we have the state-feedback gains $\kkk_1$ and $\kkk_2$ (A, B), followed
  by the threshold $q$ and the mismatch gain $A_p$ (C, D). Both axes
  represent the two dimensions of the UMAP embedding, which contains the
  data for all participants, IDs and distances between targets. This gives
  an indication of the typical ranges and combinations of parameters
  obtained from the reduced set optimisation. $\kkk_1$ and $\kkk_2$ are
  higher for the samples in the top left region of the embedding. Low
  levels of the threshold $q$ are quite common except for a small patch of
  samples located on the left side of the embedding. The mismatch gain
  $A_p$ has higher values towards the right side, with a small patch of
  negative values at the bottom.}
\label{fig:umap-min-1}
\end{figure*}

\begin{figure*}[!htbp]
\centering
\includegraphics[width=1\linewidth]{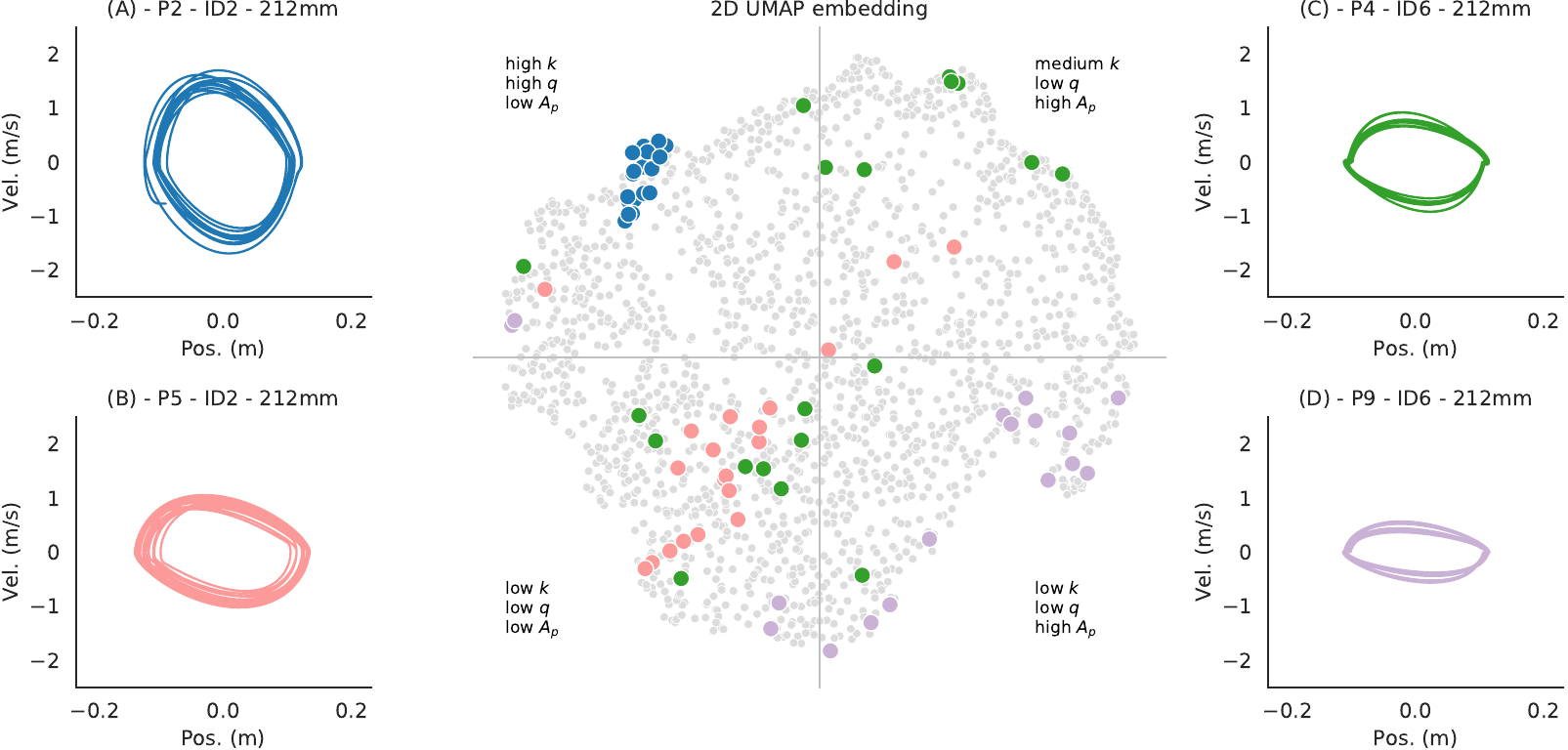}
\caption[UMAP 2D embedding for 212 mm]{Some samples of individual users and
  conditions for the 212 mm distance between targets and the associated
  phase planes. Participants 2 (A, blue) and 5 (B, pink) for an ID of 2 are
  shown in the left column, while participants 4 (C, green) and 9 (D, light
  purple) are displayed on the right. The corresponding samples in the 2D
  UMAP embedding are highlighted using the same colours. The embedding is
  classified roughly in four regions with approximate features of the
  optimised parameters, such as high and low values of $\kkk$, $q$ and
  $\ma A_p$. Note how the high variability ID 2 case is in the high gain
  region, while e.g., the P5 data is spread more evenly in the diagonal
  axis from low $k$, low $A_p$ to medium $k$, high $A_p$.}
\label{fig:umap-min-2}
\end{figure*}

Traditionally approaches such as Principal Components Analysis (PCA) have
been used to visualise reduced dimensional representations of
high-dimensional data. PCA is a linear approach, which limits its power. In
this section we use a modern nonlinear algorithm, {\it Uniform Manifold
  Approximation and Projections (UMAP)} \citep{McIHeaMel18}, to arrange the
reduced, 4-dimensional model parameter vectors on a two-dimensional
space. This allows us to visualise the smooth changes in model parameters
over the space, as shown in Fig.~\ref{fig:umap-min-1} and helps us
associate them with qualitative changes in pointing behaviour, as shown in
Fig.~\ref{fig:umap-min-2}. This gives an impression of the locations of
specific users in parameter space for a given condition, and the impact of
different parameters on time-series variability. In general, the individual
users have behaviours which are spread widely around the space, with only
extreme behaviours (e.g. P2) in more distinct distributions.

In Fig.~\ref{fig:umap-min-1}, the resulting two-dimensional embedding
obtained using UMAP is displayed, while its samples are coloured according
to the value of the two first state-feedback gains $\kkk_1$ and $\kkk_2$ (A
and B), as well as the threshold $q$ and mismatch gain $\ma A_p$ (C and
D). This gives insight on how the relevant parameters change in the 2D
space. For instance, the samples for $\kkk_1$ and $\kkk_2$ have higher
values in general towards the top left corner of the embedding. The
threshold $q$ is generally low except for a patch of samples on the top
left corner, and $\ma A_p$ shows the lowest values on the left half of the
embedding, specially at the bottom.

From the parameter mappings in Fig.~\ref{fig:umap-min-1}, we generated
Fig.~\ref{fig:umap-min-2} which displays the type of phase planes that IC
would generate for a specific condition and participant, while looking at
the position of the associated samples over the space. The two-dimensional
UMAP embedding is shown in the centre, and relevant samples of different
colours are overlapped. The corresponding phase planes are shown on the
left and right columns relating the pointer position against velocity.

Participant 2 (A, blue) is clearly clustered in the upper left corner of
the embedding for an ID of 2, which coincides roughly with the high gain
$\kkk$, high threshold $q$ and low $\ma A_p$ region of the space. The phase
plane for participant 2 shows trajectories of high velocity (vertical
axis), giving rise to a more round phase plane. If this is compared with
participant 5 in pink (B), for the same ID and distance, the phase plane
does not show the same level of velocity as participant 2 and the
associated samples lie mostly on the bottom left region of the embedding
that corresponds to low levels for all of the parameters. This suggests
that the state-feedback gains have an effect on the velocity profile of the
participants.

In similar fashion, participant 4 (C, green) and 9 (D, light purple) are
compared for an ID of 6 on the right column of Fig.~\ref{fig:umap-min-2}. A
more precise trajectory is observed for both participants since the targets
are smaller for this condition, but if we look at their samples on the
embedding we can see that participant 4 has samples on both the top right
and bottom left sections of the embedding, whereas participant 9 is more
localised around the bottom right section. These two participants have a
considerable amount of samples in the high mismatch gain $\ma A_p$ regions
of the embedding.

\begin{figure*}[!htbp]
\centering
\includegraphics[width=1\linewidth]{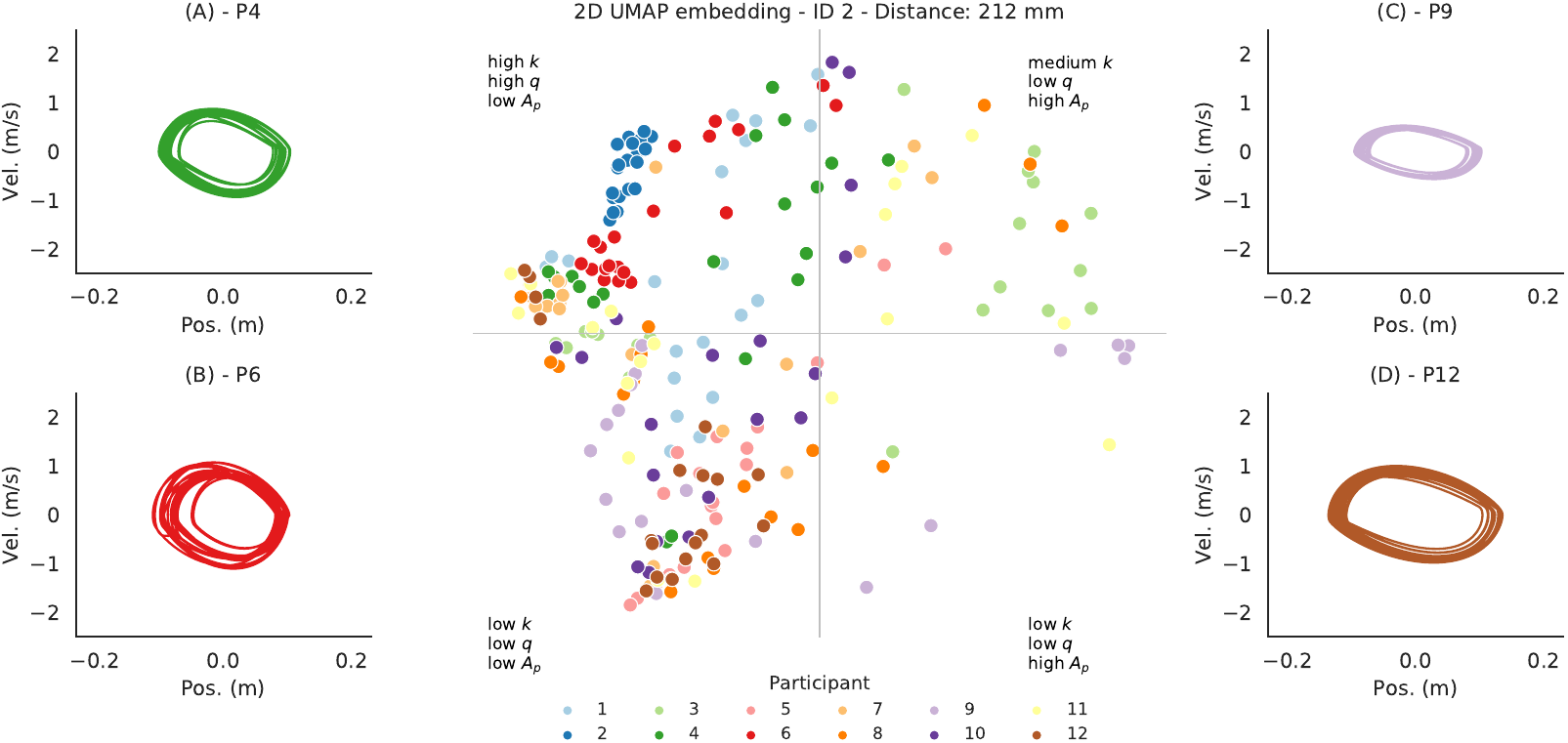}
\caption[UMAP 2D embedding for 212 mm and ID 2 for the reduced parameter
set]{Samples of a 2D UMAP embedding corresponding to all participants, an
  ID of 2, and a distance of 212 mm. Each participant is shown in a
  different colour. The embedding is classified roughly in four regions
  with approximate features of the optimised parameters, such as high and
  low values of $\kkk$, $q$ and $\ma A_p$ as in
  Fig.~\ref{fig:umap-min-2}. Participants 2 (blue), 4 (green), and 6 (red)
  stand out with most samples in the areas of high state-feedback gains
  $\kkk$ and threshold $q$, located in the top left region of the
  embedding. Participants 9 (light purple) and 12 (brown) have most samples
  in the bottom left region where the values for all parameters are
  relatively low. The phase planes for participants 4, 6, 9 and 12 are
  shown in A, B, C and D respectively.}
\label{fig:umap-min-3}
\end{figure*}

The easiest condition, i.e., ID 2, has greater degrees of freedom compared
to the rest and separates out participants the most in terms of their
position in the 2D embedding. This is shown in Fig.~\ref{fig:umap-min-3},
where the samples for all participants are displayed for ID 2 and a
distance of 212 mm. Participants 2 (blue), 4 (green), and 6 (red) tend to
have clusters of samples in the top left region of the embedding, while
participants 9 (light purple) and 12 (brown) tend to be in the lower
half. The corresponding phase planes for participants 4, 6, 9 and 12 are
shown in Fig.~\ref{fig:umap-min-3}A, B, C and D, respectively. The samples
for all participants are more mixed for the rest of the conditions
\footnote{Fig.~\ref{fig:umap-min-4} shows the 2D UMAP embedding for all
  participants and conditions, and a distance of 212 mm.}

\section{Discussion}
\label{sec:discuss}
\subsection{Key conceptual advantages of this model}
Our results show that intermittent control provides a viable model of mouse
movements.  In particular, we show that existing continuous control models
of mouse movements (e.g., \citep{MulOulMur17}), while a significant advance
on previous models with no dynamics, provide a highly simplified view of
movement dynamics in HCI tasks.  In particular, these previous models
assume that humans can react to computer feedback continuously. On a high
level, humans can appear to react to feedback continuously, such that
continuous control models can be viewed as a (rough) approximation of human
behavior. However, newer research suggests that humans may not be able to
react truly continuously because of the psychological refractory period
\citep{Loram2014}. Instead, intermittent control can masquerade as
continuous control, such that in some cases both behaviors cannot be
distinguished. However, intermittent control is a more general, and more
physiologically plausible, explanation of behavior. In particular, it can
explain phenomena that can not be explained by continuous control,
including the following:

\noindent {\bf Variability: } The parametric, continuous control models
proposed in \citep{MulOulMur17} are unable to replicate human variability.
Instead of replicating the variance of human movements, these models always
recreate the same movement given the same initial conditions and
task. Continuous control models could be augmented by injecting
sensorimotor noise into the control loop to replicate human
variability. However, intermittency of control is a physiologically
plausible additional noise source \citep{Gollee2017} that, coupled with the
model's internal dynamics, can explain the variability in pointing
movements well.

\noindent {\bf Submovements: } One practical consequence of the
simplification associated with continuous control models is their
fundamental inability to explain or replicate submovements. While humans
demonstrate identifiable submovements in difficult aimed movements, e.g.,
in mouse pointing, continuous control models predict a single smooth
movement towards the target. Submovements are a characteristic feature of
aimed movements, and the inability to replicate these represent a
fundamental shortcoming of continuous control models. However, this
shortcoming is solved through the intermittent control models proposed in
this paper.  Intermittent control provides a plausible explanation of the
physiological mechanisms underlying the existence of submovements, and our
work shows that it is also able to replicate movements with submovements
empirically. The threshold functions used in this paper are simple fixed
levels, but we anticipate that future work can capture more structure by
having more complex classifiers which are a function of both the error and
the state.

\noindent {\bf Predictive models in users: }
One particularly interesting aspect of intermittent control models is that
they posit that users maintain an internal model of the computer interface
dynamics and predict the feedback they will receive from the computer
interface in the future. They only change their control when the received
feedback deviates from their internal prediction. This maintenance of an
internal model is in line with developments in neuroscience, e.g.
\citep{wolpert2000computational}.  More awareness among HCI designers of the
fact that control is intermittent and that humans control according to an
internal model of the interface dynamics has the potential to fundamentally
change the design of the feedback of computer interfaces.  Computer
interfaces evolve according to an internal state, and aspects of this state
are communicated as feedback to the user.  The ability of the user to
construct an internal model of the state dynamics, as well as to observe
the state from feedback is crucial for their ability to close the
intermittent control loop more rarely, potentially freeing cognitive
resources.  Designers of feedback of computer interfaces should pay more
attention to making as much as possible of the full state of the computer
interface observable, e.g., by visualizing not only the positions but also
the velocities of virtual objects. The congruence of feedback received from
the interface and the feedback predicted by the user is important not only
for the control performance, but also for psychological phenomena such as
the agency perceived by the user over the interface \citep{seinfeld20}.

Furthermore, IC models previously assumed that the system matched hold was
a perfect approximation of the controlled system. In this paper we
introduce the concept of the mismatch gain, $\ma A_p$, to account for the
difference between the internal model (which is used in the hold and in the
predictor) and the real system (which is subject to disturbances). This
results in the ability of our model to represent the undershooting and
overshooting trajectories which are characteristic of this task. We also
extend our identification approach to be suitable for a reference tracking
task rather than compensatory control, as used previously
\citep{Gollee2017}.

\subsection{Limitations of the model}
This paper presents initial work building on experimental data for
one-dimensional pointing tasks of 12 test users in an artificial lab
setting, with constant mouse gain, no visual distractors, and a limited
range of ID and distance. In particular, our current model formulation
assumes linear user interface dynamics, e.g. pointing with constant mouse
gain. However, the model can be extended to the case of non-linear user
interface dynamics, e.g., non-linear pointing transfer functions, via
effectively local linearisation of a nonlinear observer model. The case of
a static non-linearity can be addressed by using its inverse, so that the
overall plant appears to be linear. Further directions of future work
involve generalising the model to everyday mouse interactions with typical
levels of visual distractors on screen, and multivariable movement tasks
(e.g. 2D targets of different shapes). However, we anticipate that the
nature of the Intermittent Control approach is naturally well-suited to
multivariate control models, allowing a flexible expansion to
higher-dimensional interaction tasks. Similarly, the dynamic systems roots
make it well suited for dynamic tasks such as {\it Steering law} tasks
\citep{Accot1997}, allowing us to better model user behaviour in tunnel or
trajectory following tasks, tracking of moving targets \citep{Pou74} and
crossing-based interfaces \citep{Accot2002}. It will also be of use in
general gesture recognition models which can often be of significantly
higher dimension.

\section{Conclusion}
\label{sec:conclusions}
We proposed intermittent control models as a model for
movement in Human-Computer Interaction, investigating intermittent
control of pointing movements with a computer mouse.

We identified parameters of our model from data of a reciprocal pointing
experiment.  Simulation of our intermittent control model shows that it can
replicate human pointing movements well, given a small number of
parameters. A significant difference from our previous work on continuous
control models in \citep{MulOulMur17} is that IC models can both reproduce
the empirical variability of human pointing movements and provide a
physiologically plausible explanation for the variability, in the system
dynamics coupled with the variability of triggering the sampling of
feedback information. Intermittent control models are also inherently able
to explain and predict submovements. We conclude that for the
one-dimensional pointing task, intermittent control models are
physiologically plausible and have stronger empirical support than current
continuous control models of movement in HCI.

The availability of identifiable, dynamic human control models, which also
accurately capture the rich variability of human behaviour, will have an
important role in the future design, testing, and analysis of
human--computer interaction. While point-and-click interfaces have
dominated interaction with computers for decades, they exploit only a tiny
fraction of the richness of motion that humans are able to
produce. Interfaces that only interpret the movement endpoint at the time
of click throw away all information that was generated in the process of
movement. This can include using the movement dynamics to give the system
predictive information about what user is trying to achieve, what their
emotional state is, or user identification, for security or personalisation
purposes (the dynamics of movement are much more difficult to counterfeit
than, say, the static image of a signature). In the future, interfaces that
accumulate information over the whole interaction might go far beyond that.
Future interfaces are also tightly integrated in the physical world and
equipped with multitudes of sensors such as gyroscopes, accelerometers and
cameras.  This would enable the interaction to generate richer reactions to
the environment and more complex, engaging, dynamics, such as during
physical scrolling or layout adjustments, especially in the case of Virtual
Reality based on physics-based environment simulations.

The control perspective on interaction provides a unifying theoretical and
modelling framework for the description, analysis and model-based design of
interfaces.  They enable the simulation of user behavior in interaction,
reducing the number of user studies necessary in the design phase. Such
simulations allow us to gain insight into processes that are difficult to
observe because they are internal to the user, and to design and optimize
for them.

\section*{Funding disclosure}

All authors acknowledge funding support from EPSRC grant EP/R018634/1, {\it
  Closed-loop Data Science.}

\bibliographystyle{apalike}
\bibliography{biblio.bib}

\appendix
\section*{APPENDIX: Supplementary content}

This appendix contains the reference Kullback-Leibler values used
in~\ref{subsec:kullback-leibler} as a table, followed by the low
dimensional visualisations obtained via uniform manifold approximations,
for a distance of 212 mm in all conditions, which complements
section~\ref{sec:umap}.

\subsection*{Kullback-Leibler divergence table}

This section contains a table with the Kullback-Leibler divergence values
used in Fig.~\ref{fig:kl-values}.

\begin{table}[ht]
\centering
\begin{threeparttable}
\caption{Kullback-Leibler divergence values of IC and 2ol controllers\tnote{*}}\label{tab:kl-values}
\begin{tabular}{ccccccccc}
\hline
\toprule
Participant & IC (ID2) & 2ol (ID2) & IC (ID4) & 2ol (ID4) & IC (ID6) & 2ol (ID6) & IC (ID8) & 2ol (ID8) \\
\midrule
1 & 0.161 & 0.745 & 0.619 & 2.567 & 0.987 & 3.319 & 2.149 & 2.867 \\
\midrule
2 & 0.941 & 0.879 & 0.680 & 2.881 & 1.092 & 3.834 & 1.950 & 3.409 \\
\midrule
3 & 0.594 & 2.085 & 0.796 & 2.867 & 1.504 & 3.700 & 2.127 & 3.606 \\
\midrule
4 & 0.814 & 0.425 & 0.568 & 2.255 & 1.264 & 3.921 & 1.806 & 3.183 \\
\midrule
5 & 0.556 & 0.709 & 0.773 & 3.267 & 1.742 & 3.980 & 2.523 & 3.869 \\
\midrule
6 & 0.461 & 1.134 & 1.121 & 2.893 & 1.086 & 4.103 & 1.883 & 3.539 \\
\midrule
7 & 0.514 & 1.415 & 0.716 & 3.158 & 1.134 & 3.798 & 2.050 & 3.414 \\
\midrule
8 & 1.120 & 1.412 & 0.730 & 2.510 & 1.877 & 3.880 & 1.406 & 3.510 \\
\midrule
9 & 0.754 & 1.944 & 1.652 & 4.824 & 1.563 & 4.801 & 2.687 & 5.242 \\
\midrule
10 & 0.550 & 1.829 & 0.368 & 3.090 & 0.824 & 3.954 & 1.609 & 3.460 \\
\midrule
11 & 0.480 & 1.974 & 0.828 & 3.204 & 1.436 & 3.506 & 1.853 & 3.469 \\
\midrule
12 & 0.994 & 1.077 & 0.759 & 3.497 & 1.055 & 3.734 & 1.141 & 3.118 \\
\bottomrule
\hline
\end{tabular}
\begin{tablenotes}
\item[*]Divergence values for all participants in conditions ID 2, ID 4, ID
  6, and ID 8 are shown. Each row corresponds to a participant and the
  values for both IC and 2ol are included. For IC, individual divergence
  values were registered from 20 different simulation runs, from which a
  mean value was calculated. The mean is reported in this table.
\end{tablenotes}
\end{threeparttable}
\end{table}

\clearpage

\subsection*{Uniform manifold approximations for a distance of 212 mm}

\begin{figure*}[!htbp]
\centering
\includegraphics[width=1\linewidth]{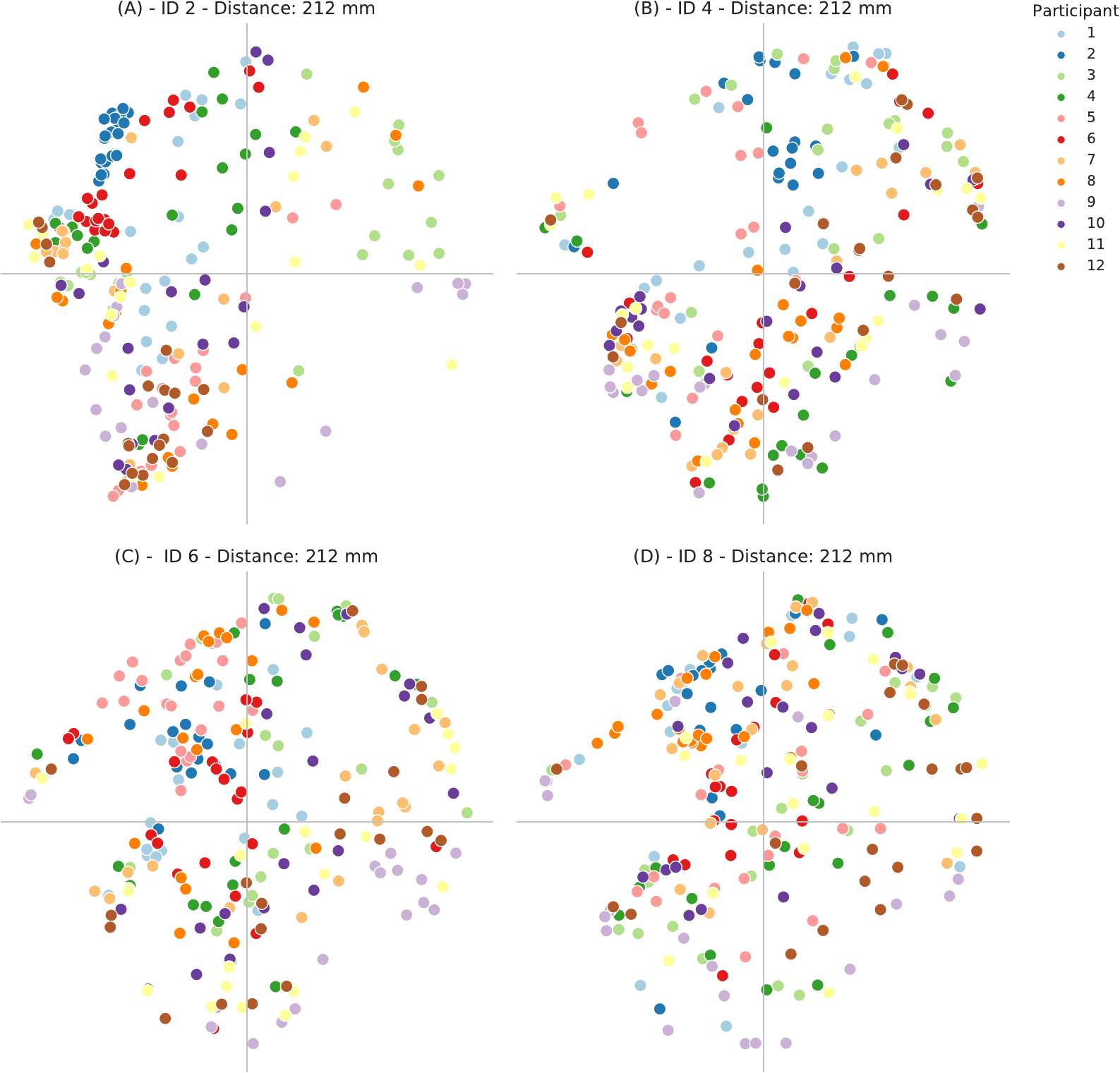}
\caption[UMAP 2D embedding for 212 mm and all conditions - reduced
parameter set]{Samples of a 2D UMAP embedding corresponding to all
  participants and conditions for the 212 mm distance between targets. Each
  participant is shown in a different colour. The embedding is classified
  roughly in four regions with approximate features of the optimised
  parameters, such as high and low values of $\kkk$, $q$ and $\ma A_p$ as
  in Fig.~\ref{fig:umap-min-2}. (A) and (B) at the top contain the samples
  for ID 2 and 4 respectively, (C) and (D) show the samples for ID 6 and
  8. ID 2 stands out with small clusters for participant 2 (blue) and 6
  (red). For the rest of the conditions, the samples tend to be more
  mixed.}
\label{fig:umap-min-4}
\end{figure*}

Fig.~\ref{fig:umap-min-4} shows the 2D UMAP embedding for all IDs, a
distance between targets of 212 mm, and when the reduced parameter set is
used for the optimisation.

\subsection*{Further data visualisations}

The following code repository was created as an online appendix of this paper:

\url{https://bitbucket.org/6Albert/ic_point/src/master/}.

\end{document}